\documentclass[structabstract]{aa}  
\usepackage{graphicx}
\usepackage{aalongtable}
\usepackage{times}
\usepackage[lowtilde]{url}
\usepackage{natbib}
\bibpunct{(}{)}{;}{a}{}{,}
\newcommand{\nodata}{$\cdots$}
\newcommand{\cmsq}{cm$^{-2}$}
\newcommand{\cmcc}{cm$^{-3}$}

\newcommand{\mic}{$\mu$m}

\newcommand{\Mjysr}{MJy~sr$^{-1}$}
\newcommand{\msunyr}{$M_\odot$~yr$^{-1}$}

\newcommand{\ergps}{ergs~s$^{-1}$}
\newcommand{\ergcms}{ergs~cm$^{-2}$~s$^{-1}$}

\newcommand{\cxc}{\textit{Chandra}}
\newcommand{\xmm}{\textit{XMM-Newton}}
\newcommand{\spitz}{\textit{Spitzer}}
\newcommand\bv{\mbox{$B\!-\!V$}}
\newcommand\jh{\mbox{$J\!-\!H$}}
\newcommand\hk{\mbox{$H\!-\!K$}}
\begin{document}


\title{A multi-wavelength study of the young star V1118 Orionis in outburst\thanks{Tables~\ref{tab:smartsmag} and \ref{tab:villmag}, and Figure~\ref{fig:irsconv} are only available in electronic format at http://www.aanda.org}}

\author{M.~Audard\inst{1,2},  G.~S.~Stringfellow\inst{3}, M.~G\"udel\inst{4}, S.~L.~Skinner\inst{3},
F.~M.~Walter\inst{5}, E.~F.~Guinan\inst{6}, R.~T.~Hamilton\inst{6,7}, K.~R.~Briggs\inst{4}, C.~Baldovin-Saavedra\inst{1,2}}

\authorrunning{M. Audard et al.}
\titlerunning{The 2005 Outburst of V1118 Ori}

   \offprints{\email{Marc.Audard@unige.ch}}

\institute{ISDC Data Center for Astrophysics, University of Geneva, Ch. d'Ecogia 16, CH-1290 Versoix, Switzerland
 \and
 Observatoire de Gen\`eve, University of Geneva, Ch. des Maillettes 51, 1290 Versoix, 
Switzerland
 \and
Center for Astrophysics and Space Astronomy, University of Colorado, Boulder, CO 80309-0389, USA
 \and
Institut f\"ur Astronomie, ETH Z\"urich, 8093 Z\"urich, Switzerland
 \and
Department of Physics and Astronomy, Stony Brook University, Stony Brook, NY 11794-3800, USA
 \and
Department of Astronomy and Astrophysics, Villanova University, Villanova 19085, PA, USA
\and
Department of Astronomy, New Mexico State University, 320 East Union Ave, Apt. 1434, Las Cruces, NM 88001, USA
}
   \date{Received 2009 July 31; accepted 2009 December 16}

 
  \abstract
   {{The accretion} history of low-mass young stars is not smooth but shows spikes of accretion that can last from months and years to decades and centuries.}
   {{Observations of young stars in outbursts can help us understand the temporal evolution of accreting stars and the interplay between the accretion disk and the stellar magnetosphere.}}
   {{The young late-type star V1118 Orionis was in outburst from 2005 to 2006. We followed the outburst } with optical and near-infrared photometry; the X-ray emission was further probed with observations
   taken with \xmm\ and \cxc\ during and after the outburst. In addition, we obtained mid-infrared photometry and spectroscopy with \spitz\ at the peak of the outburst and in the post-outburst phase.}
   {The spectral energy distribution of V1118 Ori varied significantly over the course of the outburst. The optical flux showed the largest variations, most likely due to enhanced emission by a hot spot. The latter dominated
   the optical and near-infrared emission at the peak of the outburst, while the disk emission dominated in the mid-infrared. 
   The emission silicate feature in V1118 Ori is flat and does not vary in shape, but was slightly brighter at the peak of the outburst compared to the post-outburst spectrum. 
   The X-ray flux correlated with the optical and infrared fluxes, indicating that accretion affected the magnetically active corona and the stellar magnetosphere. The thermal structure of the corona was variable with some indication of a cooling of the coronal temperature in the early phase of the outburst with a gradual return to normal values. Color--color diagrams in the optical and infrared showed variations during the outburst, with no obvious signature of reddening due to circumstellar matter. Using Monte-Carlo realizations of star+disk+hotspot models to fit the spectral energy distributions in ``quiescence''   and at the peak of the outburst, we determined that the mass accretion rate varied from about $2.5 \times 10^{-7}$~\msunyr\  to $1.0 \times 10^{-6}$~\msunyr; in addition the fractional area of the hotspot increased significantly as well.
   }
   {The multi-wavelength study of the V1118 Ori outburst helped us to understand the variations in spectral energy distributions and demonstrated the interplay between the disk and the stellar magnetosphere in a young, strongly accreting star.}


   \maketitle

\section{Introduction}
\label{sect:intro}

The star formation process involves strong accretion of circumstellar matter
onto the protostar. The time evolution of the mass accretion rate is of deep
interest to understand the timescale of stellar growth and lifetime of proto-planetary disks.
While the mass accretion rate in young stars overall decreases with increasing stellar age \citep{hartmann98}, 
it can also show significant and rapid changes over time \citep[e.g.,][]{hartmann93}.
A handful of accreting young stars display powerful eruptive events
with flux increases in the optical regime of a few magnitudes. Two classes have
emerged: FUors, which display outbursts of 4 magnitudes and more, last
several decades and, therefore, show a low recurrence rate. EXors (named
after the prototype EX Lup), in contrast, show somewhat smaller outbursts
($\Delta V=2-3$~mag) on a much shorter timescale, from a few months to a few
years, and may occur repeatedly (see review by \citealt{hartmann96}; also \citealt{herbig08}).
Such outbursts are believed to originate during  a rapid increase of the disk accretion 
rate over a short period of time, from values of $10^{-7}$~$M_\odot$~yr$^{-1}$ 
to $10^{-4}$~$M_\odot$~yr$^{-1}$, although the underlying cause of such increase in mass accretion rate
is unclear (thermal disk instabilities, \citealt{lin85,bell94}; close companions, \citealt{bonnell92}; cluster-induced encounters,  \citealt{pfalzner08}; giant planets in the disk, \citealt{clarke96,lodato04}; 
a combination of gravitational instability and the triggering of the magnetorotational instability, \citealt{armitage01}; accretion of
clumps in a gravitationally unstable disk, \citealt{vorobyov05,vorobyov06}). 
The limited number of eruptive  young stars and the 
long recurrence time (especially for FUor-type objects) make it difficult 
to test models. It is, therefore, important to study in as much detail as possible the 
evolution of outbursts and to understand the place of outbursting young stars
in the evolutionary scheme from highly accreting, embedded protostars to
``normal'' accreting classical T Tauri stars (CTTS).

The optically revealed CTTS can make excellent comparison stars to outbursting young stars:
they have been studied extensively in the past: optical and infrared observations have given
clues about the presence of accretion disks, on the mass accretion rates, and disk winds, while
millimeter observations provided constraints on the dust disk mass and on CO outflows. In the
X-rays, the origin of emission in young, accreting stars has gained significant attention in the last few years.
\citet{kastner02} obtained the first X-ray grating spectrum of TW Hya, a
moderately accreting CTTS, and observed unexpected new
spectral features: the O~{\sc vii} and Ne~{\sc ix} He-like triplets showed very low
forbidden-to-intercombination line ratios and indicated high electron densities
of the order of $10^{12}-10^{13}$ cm$^{-3}$. The effect of UV photoexcitation on the
observed line ratio was deemed negligible. In addition, the X-ray spectrum was
consistent with a quasi-isothermal plasma of about 3 MK. 
\citet{kastner02} concluded that the X-ray emission in TW Hya was due
to accretion, a result further supported by \citet{stelzer04} and \citet{ness05}. 
Additional high-resolution X-ray spectra of CTTS also indicated that
accretion may play a significant role for X-rays in accreting stars
\citep{schmitt05,guenther06,robrade06,argiroffi07,guenther07,sacco08}. 
Soft X-ray emission from shocks in jets may also be detected \citep{guedel05,guedel08,kastner05,robrade07,schneider08,guenther09}.
\citet{telleschi07} and \citet{guedel07} showed that CTTS display
a soft X-ray excess from plasma at low temperature. Such a plasma (mostly detected in
the O~VII lines) is difficult to detect with CCD spectroscopy but can easily
be revealed with high-resolution X-ray spectra when the absorbing column density
is not too high. The soft X-ray excess can co-exist with the hot plasma
observed in the vast majority of young stars, which is due to 
scaled-up solar-like magnetic activity \citep{smith05,audard05a,preibisch05,telleschi07}. 
The origin of such an excess is unclear, but it could be
due in part to accretion onto the stellar photosphere. It is crucial to understand how
the X-ray emission in young stars, in particular those accreting matter actively,
can be influenced by the accretion process. FUor and EXor stars are, therefore, 
the ideal cases to understand the physical mechanism common to both CTTS and
outbursting stars.

In the recent years, several young low-mass stars were observed in outburst, in the optical and infrared, but also in the X-rays:
V1647 Ori in 2003--2005 (e.g., \citealt{kastner04}, \citealt{grosso05}, \citealt{kastner06a}; see also 
\citealt{aspin08a} and references therein) and recently in 2008--2009 \citep{itagaki08,aspin08b,venkat08,aspin09},
V1118 Ori in 2005-2006 \citep{audard05b,lorenzetti06,lorenzetti07,herbig08}.
\citet{kastner06b} also observed with \cxc\ a young stellar object in eruption in 
LDN 1415 but failed to detect it. EX Lup also showed an extreme outburst in early 2008 \citep{jones08a,jones08b,kospal08}. \citet{abraham09} detected
crystalline features in the silicate feature of EX Lup in outburst which were not present in the pre-outburst spectrum, suggesting crystallization
by thermal annealing in the surface layer of the inner disk.  X-ray observations were obtained with \cxc\  (PI: Weintraub) and \textit{Swift} (PI: Stringfellow) 
but are not yet published.  We note also that the FUor objects FU Ori, V1735 Cyg, and Z CMa were also detected in X-rays \citep{skinner06,skinner09a,stelzer09}.

The V1647 Ori campaign showed an increase of the X-ray flux up to a factor of 200 from
its pre-outburst flux, in line with the flux increase in the infrared. The X-ray flux then followed
the optical outburst flux and returned to its pre-outburst level after the outburst
ended \citep{kastner06a}. The initial observations of V1118 Ori in X-rays showed a 
different behavior, with little flux enhancement \citep{audard05b}, but also indicated that  the rapid increase of 
accretion rate in outbursts can impact the X-ray emission of young accreting stars.
The present paper aims to present the remainder of the X-ray data of V1118 Ori taken
during and after the 2005--2006 outburst, together with contemporaneous optical and infrared data.

\section{The young erupting star V1118 Ori}
\label{sect:v1118ori}
The outburst of \object{V1118 Ori}, a young low-mass M1e star in the Orion Nebula ($d = 400$~pc; \citealt{muench08} for a discussion), was reported
in early January 2005 by \citet{williams05}.
\citet{hillenbrand97} and \citet{stassun99} provide details about its physical properties:
$M_\star = 0.41 M_\odot$; $R_\star=1.29 R_\odot$; $P_\mathrm{rot} = 2.23 \pm 0.04$~d; 
$L_\mathrm{bol}  \geq 0.25~L_\odot$; $\log T_\mathrm{eff} [K] = 3.562$; $\log t [yr] = 6.28$. 
Recently, \citet{reipurth07} resolved V1118 Ori into a close binary separated by $0 \farcs 18$ with
a position angle of $329\degr$ and a magnitude difference of $\Delta m = 0.4$~mag in the H$\alpha$-band image.
The observation was obtained on 2004 Jan 29, thus before the 2005--2006 outburst (see Fig.~\ref{fig:optXlc}).
The binarity of V1118 Ori leaves unclear which star actually erupted. However, we note that the small magnitude
difference indicates that the components are similar, suggesting similar effective temperatures. We may also assume
that they show similar disk evolution. While the binarity of V1118 Ori complicates the interpretation of the combined
photometry and spectroscopy in quiescence, this should have little impact on the interpretation of the outburst data, as
only one star+disk component dominates the emission. In this paper, we have used the above stellar properties that
assumed a single star.

V1118 Ori has shown frequent outbursts (e.g., 1983-84, 1988-90, 1992-94, 1997-98; see \citealt{garcia00} and \citealt{herbig08} for details). 
In fact, after returning in quiescence in mid-2006, V1118 Ori had another outburst in late 2007 \citep{garcia08}.
We focus here on our monitoring campaign to study the 2005--2006 outburst of V1118 Ori in the X-rays,
optical and infrared.
Additional properties in the optical and infrared were obtained independently during and after the outburst by \citet{lorenzetti06,lorenzetti07} and \citet{herbig08}.
Using wind models, \citet{lorenzetti06} derived a mass loss rate of $4 \times 10^{-8}~M_\odot$~yr$^{-1}$ from the \ion{H}{i} recombination line, and
$(3-8) \times 10^{-7}$~\msunyr\ from the CO emission at 2.3~\mic\ for the neutral molecular gas; they
also found no evidence of infrared cooling from a collimated jet or outflow. \citet{lorenzetti07} also found
evidence of intrinsic polarization in the $I$ band, and irregular fluctuations during the outburst. 
\citet{herbig08} obtained Keck/HIRES spectra of V1118 Ori in the decaying phase of the outburst and in the
post-outburst phase. He noted the detection of \ion{Li}{i} $\lambda 6707$ in emission during the outburst,
in contrast with the absorbed feature in CTTS spectra. The feature, however, returned in absorption after the outburst. A similar
behavior occurred in the \ion{K}{i} lines at $\lambda\lambda$ 7664 and 7698~\AA. A P Cyg profile was also found
in H$\alpha$ during the outburst, which disappeared thereafter. In the near-infrared, \citet{lorenzetti07} also found
that the  emission lines detected during the outburst (\ion{H}{i}, \ion{He}{i}, CO $2.3~\mu$m band, a few neutral metals) 
disappeared about a year later.

\begin{table*}[!ht]
\caption{X-ray observation log.\label{tab:obslog}}
\centering
\begin{tabular}{lcccccc}
\hline\hline
 \noalign{\vskip .8ex}%
\makebox[40mm][c]{Parameter} 		& Sep 2002 &	& Jan 2005 & Feb 2005 & Mar 2005 & Sep 2005\\
 \noalign{\vskip .8ex}%
\hline\\
 \noalign{\vskip -2ex}%
Satellite\dotfill				& \cxc				& & \cxc	  & \xmm		  & \xmm  & \xmm  \\
ObsID\dotfill					& 2548				& & 6204		  & 0212480301  	    & 0212480401		  & 0212481101 \\
Duration (ks)\dotfill				& 48 				& & 5			  & 20  		    & 20			  & 20  	  \\ 
Observation date\dotfill			& 2002 Sep 6--7			& & 2005 Jan 26 	  & 2005 Feb 18-19	    & 2005 Mar 21		  & 2005 Sep 8    \\ 
UT\dotfill					& 12:57--02:54			& & 03:07--05:04	  & 22:36--04:50	    & 16:20--22:08		  & 12:43--18:51  \\ 
Average JD - 2,450,000\dotfill			& 2,524.3			& & 3,396.7		  & 3,420.6		    & 3,451.3			  & 3,622.2	  \\ 
 \noalign{\vskip 1.5ex}%
 \hline\\
 \noalign{\vskip -1.5ex}%
{} 		         	               & Jan 2006 &  Feb 2006 &  Mar 2006 &  Apr 2006&  Jul 2006 & Dec 2007\\
 \noalign{\vskip .8ex}%
 \hline\\
 \noalign{\vskip -2ex}%
Satellite\dotfill				& \cxc		       & \cxc				& \xmm			& \cxc			& \cxc 		& \cxc 	        	      \\
ObsID\dotfill					& 6416 			& 6417   			& 0403200101 		&  6418 		& 6419 		& 8936 	        	      \\
Duration (ks)\dotfill			        & 30 		       & 30 			        & 100		        & 30 		        & 30 		& 37.5 	        	     \\ 
Observation date\dotfill		        & 2006 Jan 4--5        & 2006 Feb 23		        & 2006 Mar 2--3         & 2006 Apr 23	        & 2006 Jul 24	& 2007 Dec 14--15   	     \\ 
UT\dotfill				        & 18:33--03:09	       & 09:41--17:47		        & 19:02--21:06	        & 14:42--23:31	        & 07:50--16:49	& 14:49--01:13  	     \\ 
Average JD - 2,450,000\dotfill	        	& 3,740.5	       & 3,790.2		        & 3,797.8	        & 3,849.3	        & 3,941.0	& 4,449.3       	     \\ 
\hline
\end{tabular}
\end{table*}

{In the X-ray regime, the initial Jan-Mar 2005 data were published by \citet{audard05b}. In brief, the X-ray data of early 2005 
indicated that the X-ray flux and luminosity  stayed  similar within a factor of two during the outburst, and at the same level as in a pre-outburst 
observation in 2002. The fluxes in the optical and near-infrared varied more significantly, within factors of 
$2-10$. The hydrogen column density showed no evidence for variation from its modest pre-outburst value of  
$N_\mathrm{H}  \approx 3 \times 10^{21}$~\cmsq. However, there was evidence of a spectral change from a dominant 
hot plasma  ($\approx 25$~MK) in 2002  and in January 2005 to a cooler plasma ($\approx 8$~MK) in February 2005 and 
probably in March 2005. We hypothesized that the hot magnetic  loops high in the corona were disrupted by the closing in of the accretion disk due to 
the increased  accretion rate during the outburst, whereas the lower cooler loops were probably  less 
affected and became the dominant coronal component \citep{audard05b}. We argued that the cool component in V1118 Ori
could not originate from shocks because free-fall velocities of matter falling from the truncation radius are 
too low (see \citealt{audard05b} for further discussion). In a subsequent paper, \citet{lorenzetti06} independently analyzed our public data sets
of early 2005, and also included the September 2005 \xmm\ observation, together with their near-infrared data sets. The September 2005 observation
showed a decrease in X-ray flux at the start of the decay phase.

The initial paper by \citet{audard05b} presented and analyzed the outburst data through March 2005. We present here the \xmm\  and \cxc\ data obtained
from September 2005 on, which are analyzed in the context of our multiwavelength analysis.  We also include a post-outburst data set obtained by \cxc\ 
in December 2007. The latter was taken during a minor outburst detected in the optical \citep{garcia08}.
Extensive optical and near-infrared photometry are presented and analyzed, obtained from our team's imaging data, along with mid-infrared photometry and spectroscopy
with \spitz.}

\section{Observations and data reduction}
\label{sect:datared}

\subsection{X-ray}
Table~\ref{tab:obslog} provides the observation log of our 2005--2006 monitoring campaign of the outburst of V1118 Ori with \xmm\  \citep
{jansen01} and \cxc\ \citep{weisskopf96}. We also provide the information about the 2002 serendipitous observation of V1118 Ori (an observation in 2001 with \xmm\
was reported in \citealt{audard05b} but is not mentioned here since the star was not detected). A deep \xmm\  observation was
obtained in early 2006 that complements the short monitoring observations. Finally, we include the post-outburst \cxc\ observation 
in December 2007 as well. Note that the angular resolutions of both X-ray satellites were not high enough to separate the V1118 Ori binary.

The \xmm\ data were processed with SAS 7.0. Standard procedures were applied. We used an extraction circle of radius 20\arcsec\  
for the source and a nearby background circular region of 60\arcsec\ radius (40\arcsec\ for Sep 2005 and 35\arcsec\ for Mar 2006).
Event patterns lower than 4 and 12 were used only for the European Photon Imaging Cameras (EPIC) pn and MOS \citep{strueder01,turner01}, respectively.
The background flux levels were high during all \xmm\  observations, and in particular in March and September 2005. As described in \citet{audard05b}, we used only the MOS1 and MOS2 data for the March 2005 observation.
The September 2005 observation was so affected by the background that we lost about half the exposure time in the EPIC pn, while there
was no MOS data available. The deep March 2006 observation was affected by a system failure at the Mission Operations Centre and the EPIC
pn experienced full scientific buffer in the last part of the observation, explaining the reduced exposure time in the EPIC pn (78ks) compared to the
EPIC MOS (90 ks). No Optical Monitor data were taken with the \xmm\  X-ray observation due to the presence of the nearby bright Trapezium stars.

The \cxc\  data
were processed with CIAO 3.3 and CALDB 3.2.3\footnote{We used the pipeline data for the December 2007 observation, which was calibrated with CALDB 3.4.2.}. 
The task \textit{psextract} 
was used to extract the spectra
for V1118 Ori and the nearby background. We used a circle of 2\arcsec\  radius ($\approx 4$~pixels) for the star and an annulus 
centered at the position of the star but with radii of 10 and 60 pixels for the background; our background area was, therefore, 212 times larger
than the source extraction area. Our extraction radius for the star includes 95\% of the encircled energy at 1.5~keV and 90\% at 4.5 keV.
Note that for the January 2005 observation in sub-array mode, we used an outer radius of 40 pixels for the background annulus.
For the 2002 observation, we used two circles of radii of 15\arcsec\  and 45\arcsec\  for the source and the background, respectively (see \citealt{audard05b}).

\begin{table*}[!t]
\caption{\textit{Spitzer} IRAC and MIPS flux density measurements (units of mJy).\label{tab:spitzer}}
\centering
\begin{tabular}{lrrrrr}
\hline\hline
 \noalign{\vskip .8ex}%
\makebox[40mm][c]{Date} 		& $3.6~\mu$m & $4.5~\mu$m & $5.8~\mu$m & $8.0~\mu$m & $24~\mu$m\\
 \noalign{\vskip .8ex}%
\hline\\
 \noalign{\vskip -2ex}%
2004 Mar 09\dotfill		&$28.6 \pm 0.1$ & $36.3 \pm 0.1$ & $39.2 \pm 1.3$ & $54.8 \pm 0.5$ & \multicolumn{1}{c}{\nodata}\\
2004 Mar 20\dotfill		& \multicolumn{1}{c}{\nodata} & \multicolumn{1}{c}{\nodata} &  \multicolumn{1}{c}{\nodata} & \multicolumn{1}{c}{\nodata} & $74.2 \pm 0.5$ \\
2004 Oct 12\dotfill		&$46.5 \pm 0.2$ & $50.6 \pm 0.2$ & $53.3 \pm 0.3$ & $55.9 \pm 0.5$ & \multicolumn{1}{c}{\nodata}\\
2004 Oct 27\dotfill		& \multicolumn{1}{c}{\nodata} & $43.4 \pm 0.1$ &  \multicolumn{1}{c}{\nodata} & $57.5 \pm 0.3$ &  \multicolumn{1}{c}{\nodata}\\
2005 Feb 20\dotfill		& $143.3 \pm 1.7$ & $169.7 \pm 2.0$ & $148.2 \pm 4.9$ & $144.1 \pm 1.9$ &  \multicolumn{1}{c}{\nodata}\\
2005 Mar 28\dotfill		& $173.7 \pm 1.8$ & $203.9 \pm 2.0$ & $194.1 \pm 5.0$ & $167.6 \pm 1.9$ & \multicolumn{1}{c}{\nodata}\\
2007 Oct 21\dotfill		& $42.4 \pm 1.7$ & $48.7 \pm 1.9$ & $52.2 \pm 4.9$ & $53.1 \pm 1.8$ &  \multicolumn{1}{c}{\nodata}\\
\hline
\end{tabular}
\end{table*}

\subsection{Optical and near-infrared}

\subsubsection{SMARTS}
We observed V1118~Ori with the ANDICAM dual-channel imager on the SMARTS/CTIO\footnote{SMARTS, the Small and Medium Aperture Research Telescope Facility, is
a consortium of universities and research institutions that operate the
small telescopes at Cerro Tololo under contract with AURA.}
1.3m telescope on 115 nights. Such dense coverage is possible by the service
mode operations employed by the SMARTS consortium. 
We have 5 pre-outburst observations (2004 Feb 2 through 2004 Apr 8) that can
be used to set the quiescent flux levels.
The intensive monitoring began 2005 Jan 10, shortly after the outburst was
reported. There were no observations from 2005 Apr 4 through 2005 Jul 31
due to proximity to the Sun. We observed with cadences between 1 per day
and 1 per week, with observations generally every 3-4 days.
Our intensive monitoring terminated on 2006 May 6.
We also followed the decay of the second outburst from 2007 Dec 27
through 2008 Mar 27. Details about the data reduction are given in {Appendix \ref{app:smarts}}.
Table~\ref{tab:smartsmag} (available online only) gives the nightly average photometry for SMARTS.

\subsubsection{Villanova}
Photometric coverage in the optical (standard Bessel $VRI$) was obtained with 
{the Celestron 14" optical tube assembly with a Paramount ME German equatorial mount at the Villanova University Observatory located on the campus near Philadelphia, PA. Observations
were carried out with a SBIG ST7-XME detector thermoelectrically cooled to -25 degrees Celsius. Dark and flat field frames were collected at the end of each night's observations. }
A standard error of 0.05 mag  was estimated from the signal-to-noise ratio and seeing 
conditions. Table~\ref{tab:villmag}  (available online only) lists the magnitudes obtained at Villanova.

\subsubsection{Additional data}
We have used published optical and near-infrared photometric data from \citet{lorenzetti07} ($IJHK$), \citet{garcia06} and \citet{garcia08} ($V$).

\subsection{\textit{Spitzer}}

\begin{figure}[!b]
\centering
\includegraphics[angle=0,width=\linewidth]{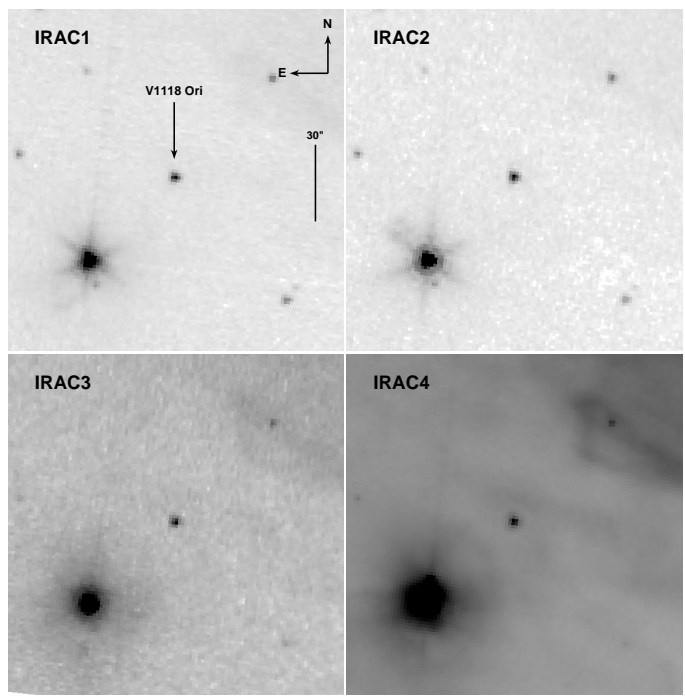}
\caption{\spitz\ IRAC images centered on V1118 Ori, taken pre-outburst in March 2004. The images are shown with a linear scale from 0 to 200~\Mjysr. The Herbig Ae star V372 Ori is seen near V1118 Ori.
\label{fig:irac}}
\end{figure}

In addition to our observations of V1118 Ori in outburst (program ID 3716, PI: G.~Stringfellow) and in post-outburst (program ID 41019, PI: M.~Audard), V1118 Ori was serendipitously observed with the InfraRed Array Camera (IRAC; \citealt{fazio04}) by the \textit{Spitzer} Space Telescope \citep{werner04}  in March 2004 and twice in October  2004 (program IDs 43 and 50, PI: G.~Fazio). We show the IRAC images taken before the outburst in March 2004 centered on V1118 Ori (Fig.~\ref{fig:irac}). 
We provide the IRAC fluxes for all observations in Table~\ref{tab:spitzer} {(see also \citealt{lorenzetti07} for the IRAC data of October 27, 2004). Details about the \spitz\ IRAC data reduction are given in Appendix~\ref{app:spitz}.}

\begin{figure*}[!t]
\centering
\includegraphics[angle=0,width=\linewidth]{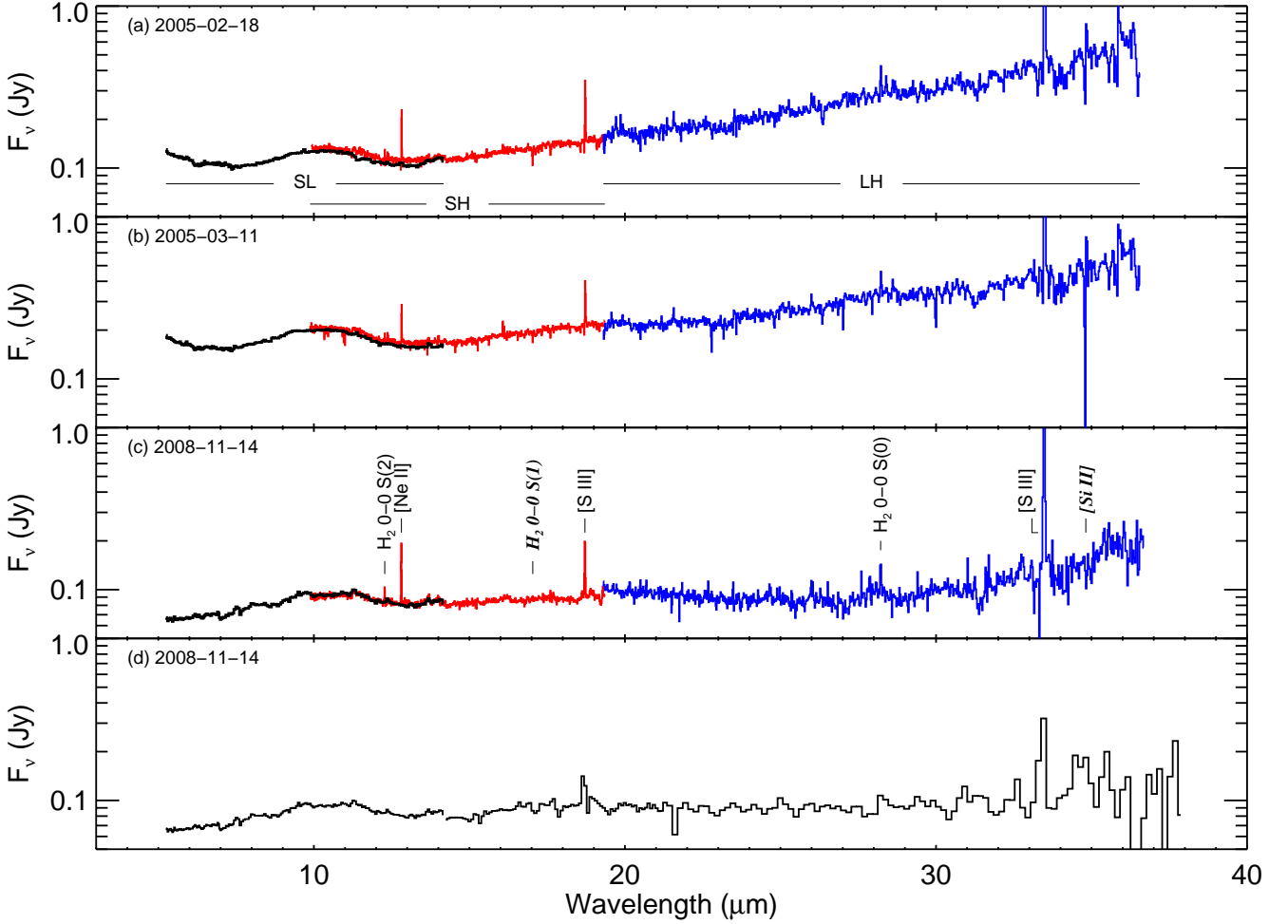}
\caption{\spitz\ IRS background-subtracted spectra obtained near the peak of the outburst (panels {\it a} and {\it b}) and after the outburst (panels {\it c} and {\it d}). Panel d shows the SL and LL module data. Panels {\it a}, {\it b}, and {\it c} show the
SL spectra as thick black lines, whereas the SH and LH spectra are shown as red and blue lines. Detected emission lines are labeled (while other lines well-subtracted by the background spectrum are shown in italics).
\label{fig:irsspec}}
\end{figure*}

\begin{table*}
\centering
\begin{minipage}[t]{1.5\columnwidth}
\caption{{Tentative} \textit{Spitzer} IRS line fluxes {derived from the SH and LH spectra}.\label{tab:irsline}}
\centering
\begin{tabular}{lcrrr}
\hline\hline
 \noalign{\vskip .8ex}%
Line 		& $\lambda$ (\mic) & \multicolumn{3}{c}{Flux ($10^{-18}$~W~m$^{-2}$)}\\
{}						&					& Feb 2005 & Mar 2005 & Nov 2008\\
 \noalign{\vskip .8ex}%
\hline\\
 \noalign{\vskip -2ex}%
 H$_2$ $0-0$ $S(2)$\dotfill						&	12.2786	& \nodata			& \nodata	&	$7.8 \pm 0.8$ \\
$[\ion{Ne}{ii}]$	$^2P_{1/2}$ -- $^2P_{3/2}$\dotfill 		&	12.8136	& $41.1 \pm 0.7$	& $39.7 \pm 1.1$	&	$42.7 \pm 0.9$ \\
$[\ion{S}{iii}]$ $^3P_2$ -- $^3P_1$\dotfill			&	18.7130	& $46.5 \pm 1.4$	& $42.5 \pm 1.5$	&	$34.6 \pm 1.2$ \\
H$_2$ $0-0$ $S(0)$\dotfill						&	28.2188	& $15.2 \pm 2.2$	& $18.9 \pm 3.4$	&	$9.5 \pm 1.6$ \\
$[\ion{S}{iii}]$ $^3P_1$ -- $^3P_0$\dotfill				&	33.4810	& $463.3 \pm 9.6$	& $374.4 \pm 7.9$	&	$183.4 \pm 4.7$\\
\hline
\end{tabular}
\end{minipage}
\end{table*}

\begin{figure*}[!t]
\centering
\includegraphics[angle=0,width=\linewidth]{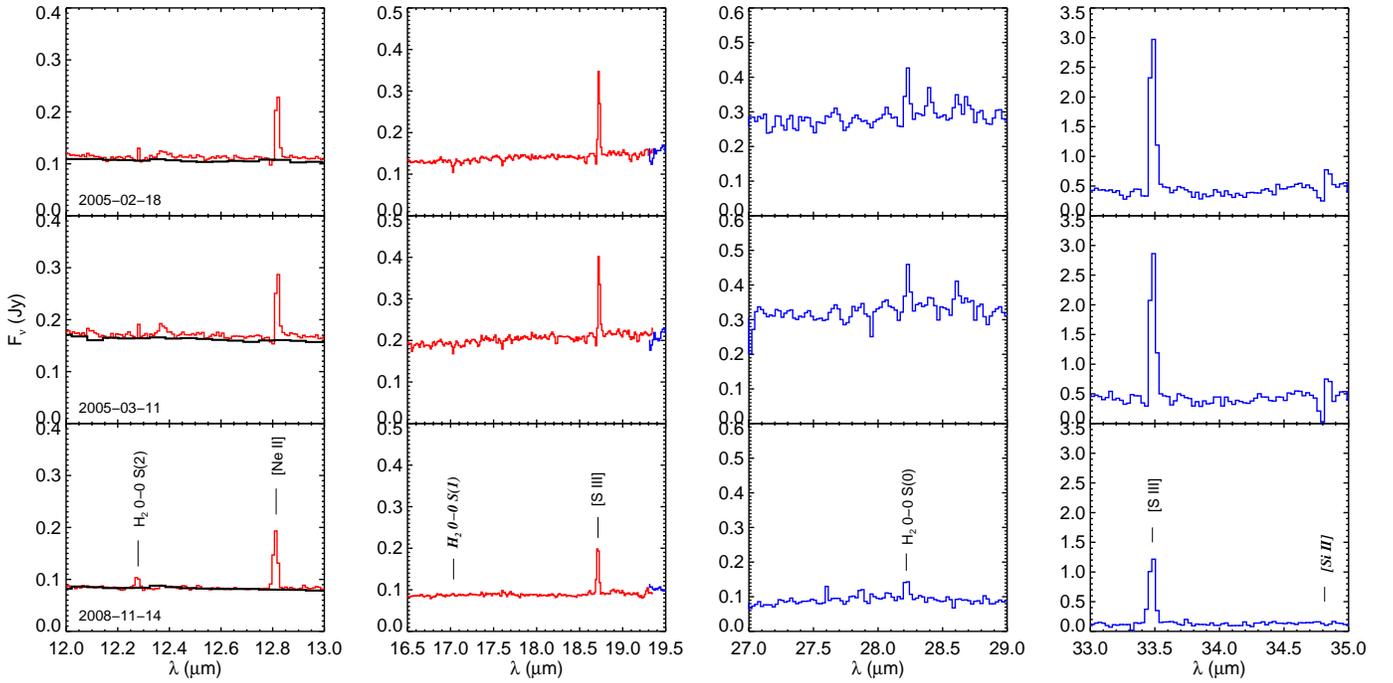}
\caption{Zoomed in regions of the \spitz\ IRS background-subtracted spectra. The post-outburst spectra are shown in the bottom panel, while the top two panels are for the outburst spectra. In the leftmost panels, we show both the high-resolution SH (thin, red) and the low-resolution SL (thick, black) data. Detected emission lines are labeled in the bottom panels (while other lines well-subtracted by the background spectrum are shown in italics). {Observation dates applicable to each horizontal set of spectra are listed in the left panels.}
\label{fig:irsline}}
\end{figure*}

\begin{figure}[!ht]
\centering
\includegraphics[angle=0,width=\linewidth]{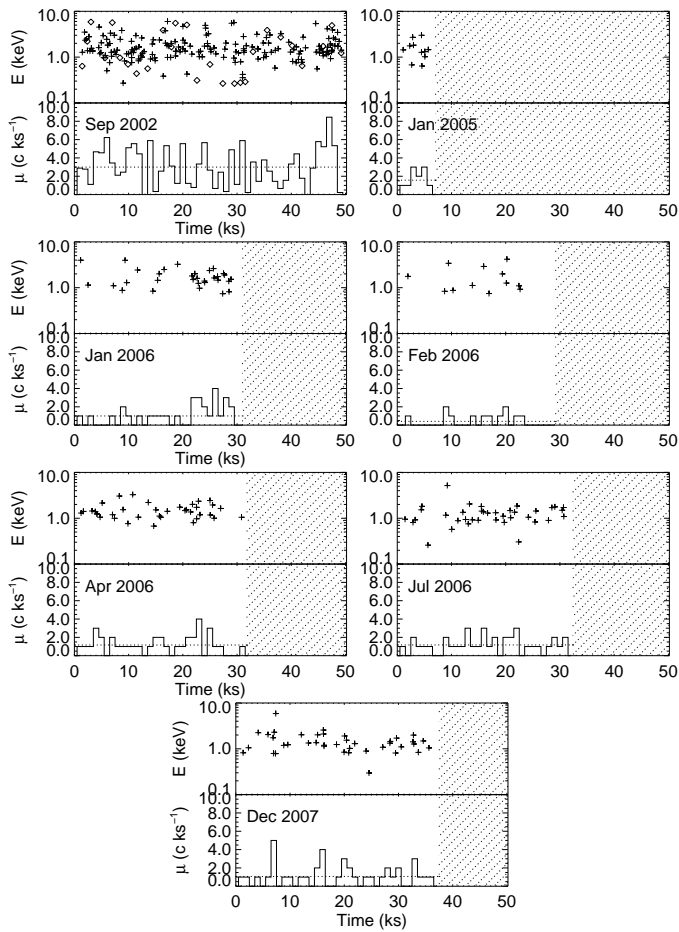}
\caption{Light curves for the \cxc\  observations  (0.4--6.0 keV range). The upper panels show the CCD energy of the events as a function
of time after the start of the observations, whereas the lower panels show count rate light curves ($\mu$) with a bin size of 1 ksec. The average count rates 
are also shown as horizontal dotted lines. The time span for each panel was kept similar and equal to the time span of the Sep 2002 observation
($\approx 50$~ksec). Since the other observations were shorter than 50~ksec, the rest of the time span is marked as hashed regions. 
{See text for a detailed description}.
\label{fig:lcchandra}}
\end{figure}

\begin{figure}[!ht]
\centering
\includegraphics[angle=0,width=\linewidth]{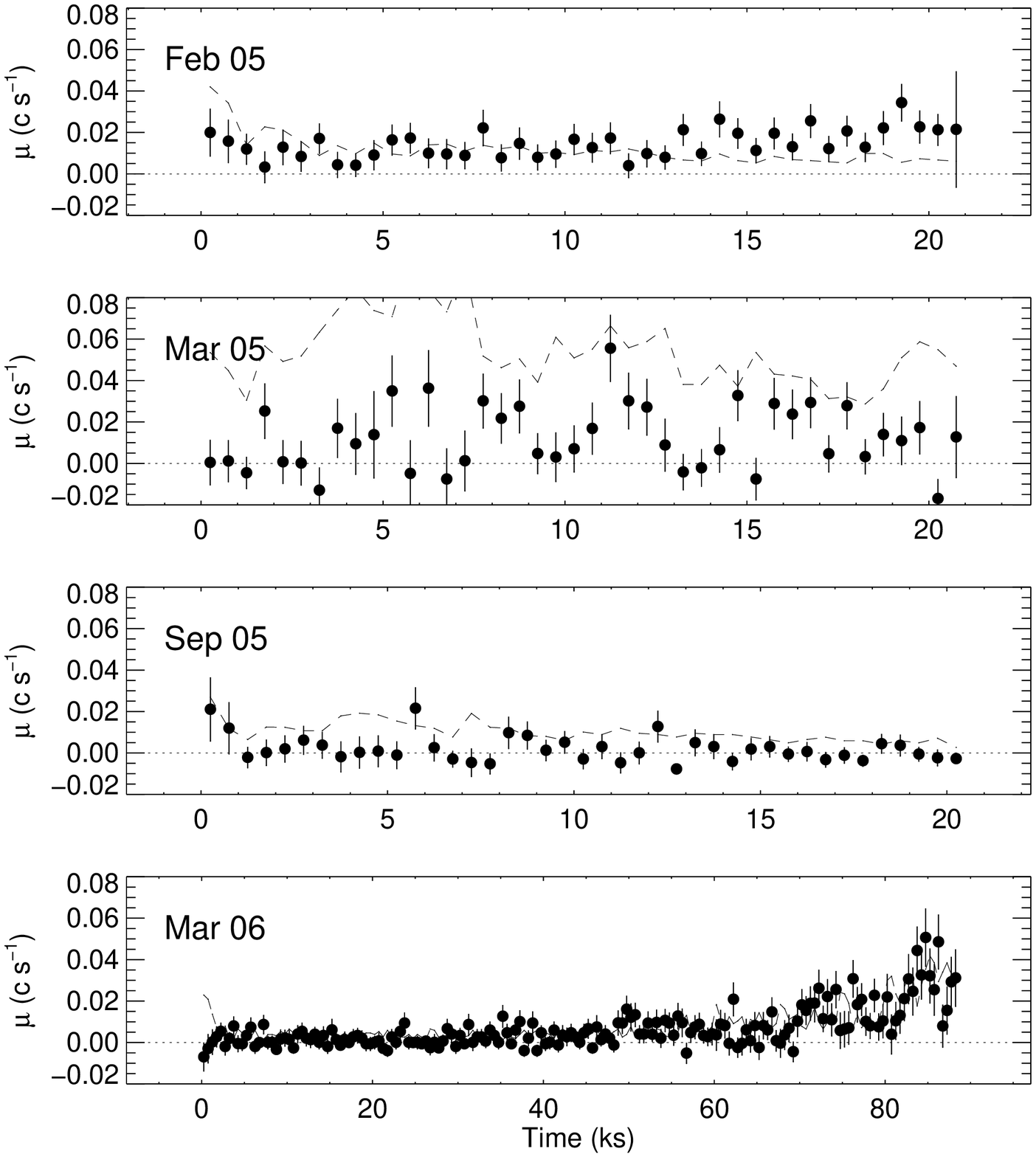}
\caption{Background-subtracted count rate light curves ($\mu$) for the \xmm\  observations, with bin sizes of 500~s and in the 0.4--6.0 keV range. 
The EPIC pn data are shown, except in March 2005 for which we summed the data of both MOS detectors. The (scaled) 
background light curves are shown as dashed curves.
\label{fig:lcxmm}}
\end{figure}

V1118 Ori was also observed with the Multiband Imaging Photometer for \spitz\  (MIPS; \citealt{rieke04}) at 24~$\mu$m before the outburst, on March 20, 2004 (program ID 58, PI: G.~Rieke). {Again, Appendix \ref{app:spitz} provides the details of the data reduction.} No MIPS photometry is available for V1118 Ori during or after the outburst. Our programs (3716 and 41019) included MIPS spectral energy distribution data ($R \approx 20$) in the 70~$\mu$m band; however, the on-time exposure (180~s) did not allow us to detect V1118 Ori, even during the outburst. Indeed, the background level (due to diffuse emission in the Orion nebula) produced a much higher flux (of order $40$~Jy at 70~$\mu$m) than the expected signal from V1118 Ori (of the order 0.1 Jy if extrapolating the SED at 70~$\mu$m, see Fig.~\ref{fig:sed}).

\textit{Spitzer} also observed V1118 Ori with the InfraRed Spectrograph (IRS; \citealt{houck04}) twice during the outburst (PID 3716) with the Short-Low  (SL: 5.2-14.7~$\mu$m, $R=\lambda/\Delta\lambda \approx 64$), Short-High (SH: 9.9-19.6~$\mu$m, $R=\lambda/\Delta\lambda \approx 600$), and Long-High (LH: 18.7-37.2~$\mu$m, $R=\lambda/\Delta\lambda \approx 600$) modules (2005 February 18 and 2005 March 11). No background observations were taken with the high-resolution modules. On the other hand, the post-outburst data (PID 41019; 2008 November 14) were taken with background spectra for the high-resolution modules and also included the Long-Low (LL: 14.0-38.0~$\mu$m, $R=\lambda/\Delta\lambda \approx 64$) module.

{Figure~\ref{fig:irsspec} shows all IRS spectra after background subtraction. We describe in Appendix  \ref{app:spitz} our methodology to derive the background-subtracted spectra of V1118 Ori. 
We also discuss in the Appendix the validity to use the post-outburst background observation for our outburst IRS SH spectra. In brief, the post-outburst, background-subtracted high-resolution spectrum is well-subtracted
for the continuum, as the flux is consistent with the low-resolution spectrum. In the case of the outburst SH spectra, the continuum is also accurate below 14~\mic\  since they are consistent with the low-resolution SL spectra.
However, we prefer to stay on the safe side and claim that the observed increase in continuum flux for $\lambda > 14$~\mic\   during the outburst is unreliable, since we have no MIPS photometry or low-resolution spectra to confirm the increase. We also urge caution with regards to the detection of lines in the background-subtracted SH and LH spectra. The strong background line emission of the Orion nebula, and its inhomogeneity (see Fig.~\ref{fig:mipsirs}) make the background subtraction difficult, although not impossible (e.g., a very strong PAH emission at 11.3~\mic\ is well-subtracted). The[\ion{S}{iii}] emission lines at 18.71 and 33.48~\mic\  are likely robust detections as their detection is consistent with the low-resolution SL and LL data. The [\ion{Si}{ii}] is also generally well subtracted. Interestingly, the rotational H$_2$ S(0) line is detected during the outburst and in post-outburst. However the S(1) line is not detected (it is is even slightly oversubtracted in the outburst spectra), and the S(2) line is only detected in the post-outburst spectrum (a faint excess in the outburst spectra is present only in one bin, likely due to slight differences in the wavelength scales of the source and background spectra). It is likely that the H$_2$ emitting conditions vary spatially in the Orion nebula, and that the background spectrum did not accurately reflect the H$_2$ background line fluxes near V1118 Ori, creating an excess in line emission observed in the post-outburst spectrum. Alternatively, if the H$_2$ emission lines are real, at least in the post-outburst spectra, the physical conditions near V1118 Ori are such that the para (odd quantum number $J$) lines dominate the emission. A similar case of difficult background line subtraction arises in the  [\ion{Ne}{ii}] line at 12.81~\mic: its detection in the SH spectrum is also subject to caution, as it is not detected in the low-resolution spectra (although it should have been, see the discussion in the Appendix). In any case, we provide in Table~\ref{tab:irsline} the tentative, measured line fluxes from the high-resolution module spectra, while Figure~\ref{fig:irsline} shows zoomed in regions of the IRS spectra of the detected lines.}

\section{X-ray analysis}
\label{sect:xray}

\subsection{Light curves}

Figure~\ref{fig:lcchandra} presents the X-ray light curves for the \cxc\  observations. 
The lower panels show the binned light curves, whereas the upper panels show the photon CCD energy
as a function of time.  Except for Sep 2002, the background is not subtracted in the light curves and not shown in the upper panels,
since it contributed to less than 1 count to the \cxc\  total (source + background) number of counts
over the total observing duration. Since V1118 Ori was observed serendipitously in the Sep 2002 observation
and was placed at the edge of the ACIS-I camera, the background contributed more significantly; therefore, the
light curve shown in the bottom panel is background-subtracted, and the CCD energy vs time background events
are shown together with the CCD energy vs. time total events (for clarity, every ninth background event was
plotted, which corresponds approximately to the area-scaled background contribution).

Small-scale variability is observed in the \cxc\  light curves; however, no strong flares were detected during
the observations. The January 2006 observation displayed nevertheless a significant increase in the count
rate in the last 10 ks, probably due to a moderate flare. The average count rates are 3.14, 2.31, 1.10, 0.39, 1.23, 1.23, and 1.09
ct~ks$^{-1}$ ($0.2-4.0$~keV, except for Sep 2002: $0.2-6.0$~keV) for the Sep 2002, Jan 2005, Jan-Jul 2006, and Dec 2007 observations, respectively. While the effective areas in all 
pointed observations are similar, the Sep 2002 effective area was about 40-50\% lower (due to the off-axis position of V1118 Ori). Therefore, to compare the
above Sep 2002 count rate with the later \cxc\  count rates, one needs to multiply by a factor of about 2, i.e.,
6.00~ct~ks$^{-1}$. Clearly, assuming only a change in emission measure, the X-ray flux dropped by a factor of 4--6,
and even 15 in Feb 2006, during the 2005 outburst compared to 2002. It is important to emphasize that the serendipitous
observation did not show evidence of strong flaring, and V1118 Ori was likely caught in ``quiescence''.

Figure~\ref{fig:lcxmm} presents the X-ray light curves for the \xmm\  observations. In comparison with the \cxc\ data,
the \xmm\  data were heavily impacted by background radiation. The (scaled) background contributions generally were
similar or even higher than the net source contribution. In particular, during the March 2005 observation, the background
completely overwhelmed the EPIC pn data which could not be used. No obvious large flare was observed during the \xmm\  observations,
like in the \cxc\ observations. However, in March 2006, V1118 Ori showed a significant increase in X-ray flux in
the last 35--40~ks of the observation, probably due to flares or the onset of an active region. This increase
is observed in all EPIC cameras, comforting us that this behavior is not due to an increase in background flux that
was improperly subtracted in the EPIC pn data.

The light curves {in Figures~\ref{fig:lcchandra} and \ref{fig:lcxmm}} suggest that the spectral fits represent snapshots of the thermal emission measure distribution of V1118 Ori's corona observed serendipitously before the 2005 outburst, and over the course of the optical/infrared outburst.

\subsection{X-ray and optical/near-infrared flux correlations}

\begin{figure*}
\centering
\includegraphics[angle=0,width=\linewidth]{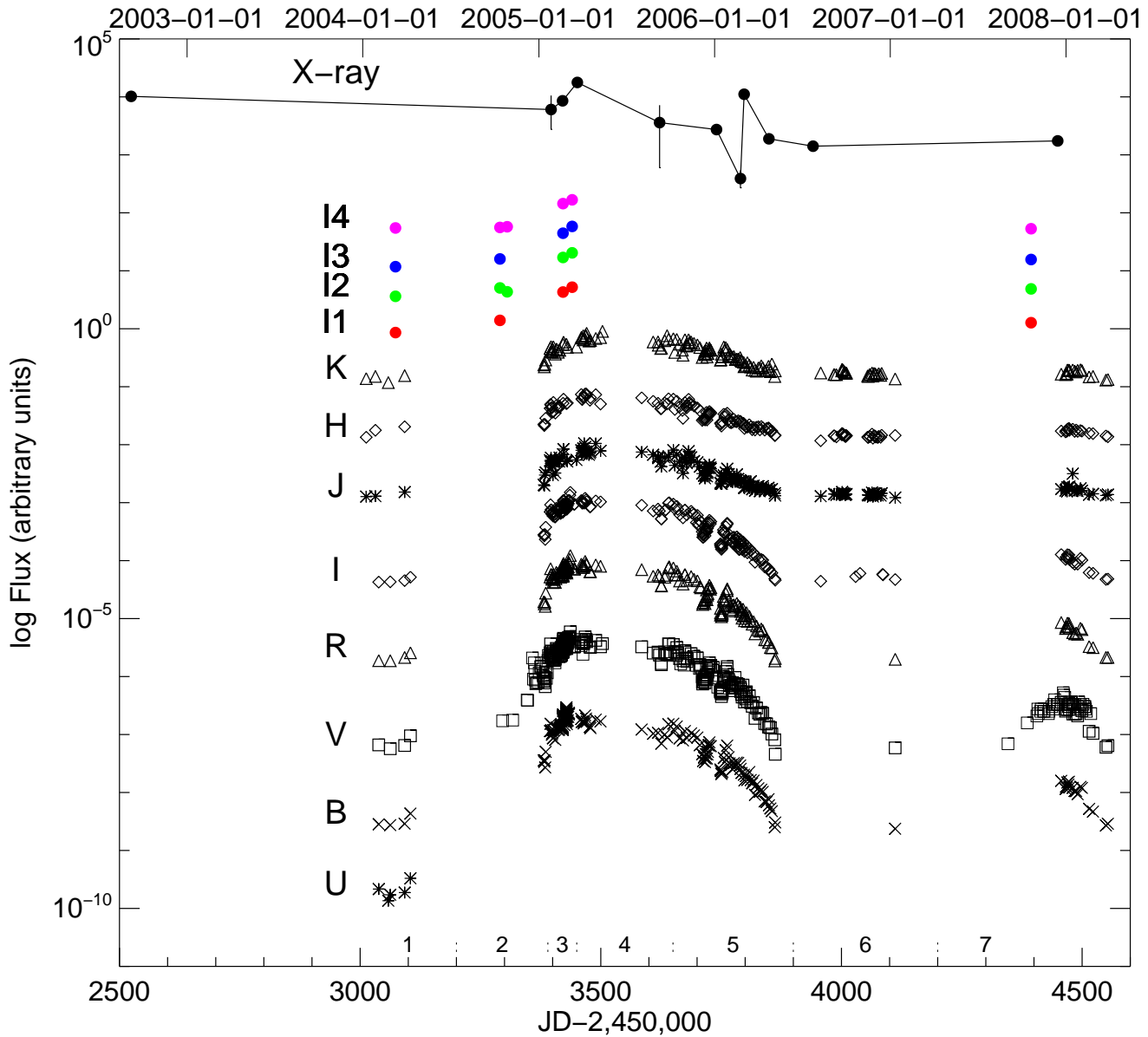}
\caption{Optical ($UBVRI$),  near-infrared ($JHK$), mid-infrared (IRAC bands at 3.6, 4.5, 5.8, 8.0~$\mu$m), and X-ray light curves versus Julian date. Arbitrary scales were applied to help visualize the light curves.  See text for the definition of the time intervals labeled at the bottom. Data from the SMARTS and Villanova campaigns, from \citet{lorenzetti07}, \citet{garcia06} and \citet{garcia08} are included.\label{fig:optXlc}}
\end{figure*}

\begin{figure*}[!ht]
\centering
\includegraphics[angle=0,width=\linewidth]{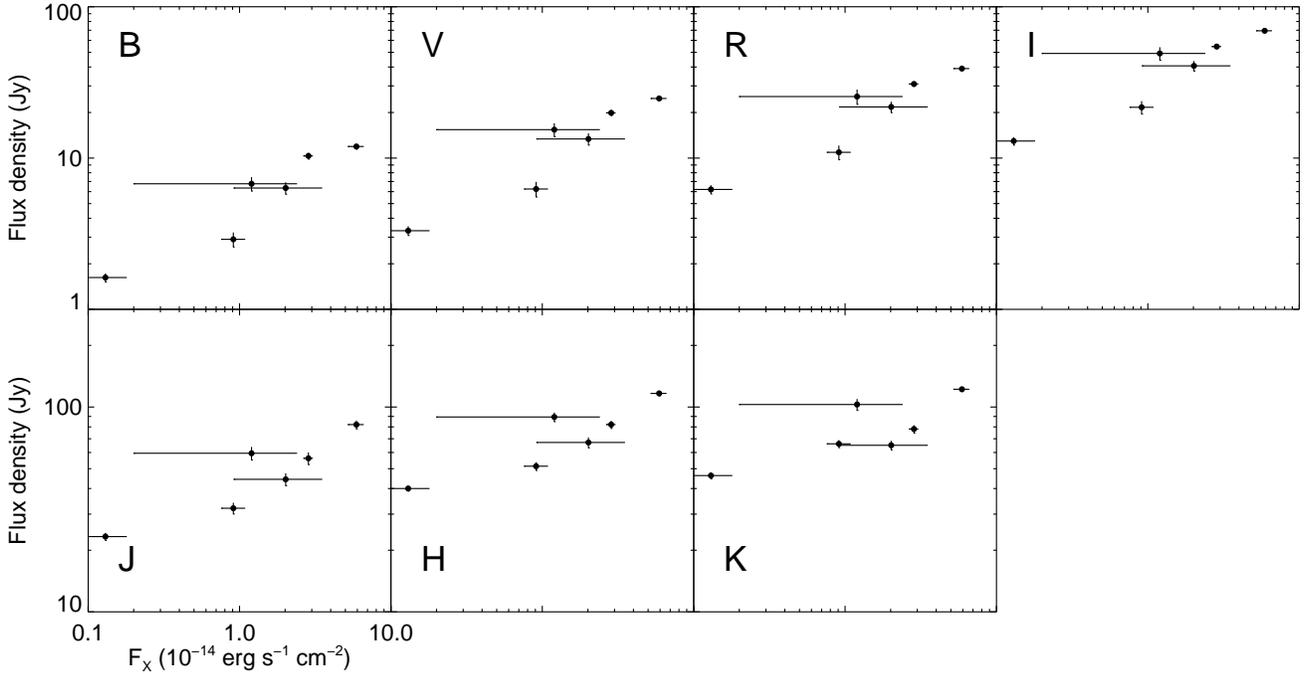}
\caption{Optical and near-infrared flux densities (averaged within $\pm 20$ days of the X-ray observation) as a function of the observed X-ray flux during the outburst (Jan 2005 to Feb 2006). Notice the different flux axis scales for the optical and near-infrared bands.
\label{fig:Xoptir}}
\end{figure*}

Figure~\ref{fig:optXlc} shows the long-term light curve of V1118 Ori in the X-rays, optical, and infrared, from its pre-outburst detection in September 2002, through its outburst in 2005 and 2006 to the post-outburst phase at the end of December 2007. We have looked into correlations between the X-ray and optical and infrared photometry.
Since we obtained { a few sequential} optical and infrared photometry points every night of our campaign, it was not possible to determine short-term variations to compare with the X-ray light curves. Nevertheless, to quantify further the correlation between the optical and near-infrared flux densities with the X-ray flux, we have calculated average optical and near-infrared flux densities within  $\pm 20$~days of the 6 X-ray observations between January 2005 and February 2006 (we did not include the March 2006 observation, as this one might have been contaminated by a flare). The optical and near-infrared flux densities are well-correlated with the X-ray fluxes measured at Earth (Fig.~\ref{fig:Xoptir}) as indicated by  Pearson's correlation coefficients, which are in the range of { $\rho = 0.8-0.95$}. The sparsity of X-ray data points does not allow to measure any clear delay between the X-ray and optical/near-infrared flux densities. We have also done cross-correlation studies with  the optical and near-infrared data (including only the data points for which we had simultaneous measurements). {The onset of the outburst is unclear in most light curves (except perhaps in the $V$ band), but the end of the outburst occurs at a similar epoch for all bands (around MJD 53860). The shape of the light curves also differ: the fluxes vary most at short wavelengths ($BVRI$), while the near-infrared fluxes display shallower variations and they reach their peak some time later (e.g., $K$ peaks around MJD 53470 while $V$ peaks around MJD 53433). The overall shapes of the light curves suggest that the mechanism dominating in the optical and near-infrared bands may be the same (e.g., due to a hot spot), although the different peak times suggest contamination of another mechanism longward of $1~\mu$m, probably disk thermal emission (see below).}

\subsection{Spectral fits}

\begin{table*}
\caption{Spectral fits and quantile properties.\label{tab:fits}}
\centering
{\tiny 
\begin{tabular}{lcccccc}
\hline\hline
 \noalign{\vskip .8ex}%
\makebox[40mm][c]{Parameter} 		& Sep 2002 & \makebox[5mm][c]{}& Jan 2005 & Feb 2005 & Mar 2005 & Sep 2005\\
 \noalign{\vskip .8ex}%
\hline
 \noalign{\vskip 0.5ex}%
 Satellite				&	\cxc	& &  \cxc	&	\xmm	&	\xmm	&\xmm\\
 \noalign{\vskip .5ex}%
\multicolumn{7}{l}{Spectral analysis:}\\
 \noalign{\vskip .5ex}%
Net counts$^{a}$\dotfill			& $150.9$			& & $10.8$			  & $285.5/105.4/103.2$ 	  & $(47.1)/190.4/108.0$	  & $(12.4)$	  \\
Scaled background counts$^{a}$\dotfill		& $27.1$			& & $0.2$			  & $224.5/56.6/52.8$		  & $(698.9)/325.6/331.0$	  & $(227.6)$	  \\
Exposure (ks)$^{a}$\dotfill			& $47.85$			& & $4.66$			  & $18.34/21.87/21.91$ 	  & $(7.83)/19.96/19.22$	  & $(13.03)$	  \\ 
$N_\mathrm{H}$ ($10^{21}$~\cmsq)\dotfill	& $2.7^{+1.2}_{-0.9}$		& & $1.4^{+3.6}_{-1.4}$ 	  & $4.3^{+1.3}_{-1.1}$ 	  & $1.4^{+0.8}_{-0.5}$ 	  & \nodata \\
$T$ (MK)\dotfill				& $25.1^{+6.3}_{-4.8}$		& & $46^{+\infty}_{-28}$	  & $7.7^{+1.3}_{-0.8}$ 	  & $15.9^{+3.9}_{-2.0}$	  & \nodata \\
EM ($10^{53}$~\cmcc)\dotfill			& $1.0^{+0.3}_{-0.2}$		& & $0.4^{+0.4}_{-0.1}$ 	  & $2.9^{+1.0}_{-1.0}$ 	  & $2.1^{+0.2}_{-0.2}$ 	  & \nodata \\
$Z/Z_\odot$\dotfill				& $:= 0.17$			& & $:=0.17$			  & $0.17^{+0.20}_{-0.06}$	  & $:=0.17$			  & \nodata	\\
$F_\mathrm{X}$ ($10^{-14}$~\ergcms)\dotfill	& $3.4^{+0.3}_{-0.3}$		& & $2.0^{+1.5}_{-1.1}$ 	  & $2.9^{+0.2}_{-0.2}$ 	  & $5.9^{+0.7}_{-0.7}$ 	  & $1.2^{+1.2}_{-1.0}$$^{b}$	  \\
$L_\mathrm{X}$ ($10^{30}$~\ergps)\dotfill	& $1.2^{+0.1}_{-0.1}$		& & $0.5^{+0.4}_{-0.3}$ 	  & $2.5^{+0.2}_{-0.2}$ 	  & $2.1^{+0.2}_{-0.2}$ 	  & $0.04-0.12$$^{b}$	  \\
 \noalign{\vskip 1.5ex}%
 \multicolumn{7}{l}{Quantile analysis:}\\
 \noalign{\vskip .5ex}%
Energy range\dotfill				& $0.2 - 6.0$			& & $0.2 - 4.0$ 		  & $0.2 - 4.0$ 		  & $0.2 - 4.0$ 		  & $0.2 - 4.0$\\
Net counts$^{c}$\dotfill 			& $150.44$			& & $10.79$			  & $286.45$			  & $196.34$			  & $9.75$\\
Scaled background counts$^{c}$\dotfill		& $26.56$			& & $0.21$			  & $222.55$			  & $425.66$			  & $228.25$\\
$Q_\mathrm{25\%}$\dotfill			& $0.17 \pm 0.01$		& & $0.24 \pm 0.07$		  & $0.15 \pm 0.01$		  & $0.15 \pm 0.02$		  & $0.0 \pm 0.5$ \\
$Q_\mathrm{50\%}$\dotfill			& $0.22 \pm 0.01$		& & $0.33 \pm 0.07$		  & $0.20 \pm 0.01$		  & $0.25 \pm 0.02$		  & $0.3 \pm 1.5$ \\
$Q_\mathrm{75\%}$\dotfill			& $0.31 \pm 0.03$		& & $0.42 \pm 0.14$		  & $0.29 \pm 0.01$		  & $0.34 \pm 0.03$		  & $0.9 \pm 1.3$ \\
$E_\mathrm{50\%}$~(keV)\dotfill			& $1.45 \pm 0.06$		& & $1.44 \pm 0.25$		  & $0.95 \pm 0.04$		  & $1.14 \pm 0.07$		  & $1.2 \pm 4.0$ \\
$x = \log[Q_{\rm 50\%}/(1-Q_{\rm 50\%})]$\dotfill& $-0.56_{-0.03}^{+0.03}$ 	& & $-0.31_{-0.14}^{+0.12}$	& $-0.61_{-0.03}^{+0.03}$	  & $-0.48_{-0.04}^{+0.04}$	  & $-0.4_{-\infty}^{+\infty}$\\
$y = 3 \times Q_\mathrm{25\%}/Q_\mathrm{75\%}$\dotfill& $1.63_{-0.13}^{+0.13}$  & & $1.70_{-0.61}^{+0.61}$	& $1.54_{-0.07}^{+0.07}$  	  & $1.33_{-0.19}^{+0.19}$	  & $0.05_{-0.05}^{+1.42}$\\
$T_{xy}$~(MK)\dotfill				& $\approx 20$			& & $\approx 12$			  & $\approx 12$			  & $\approx 20$			  & \nodata \\
$N_{\mathrm{H},xy}$~($10^{21}$~\cmsq)\dotfill	& $\approx 3$			& & $\approx 8$			  & $\approx 1$			  & $\approx 0.8$  		  & \nodata \\
 \noalign{\vskip 1.5ex}%
 \hline\\
 \noalign{\vskip -1.5ex}%
 {} 		                        & {Jan 2006} 		&  {Feb 2006} 		&  {Mar 2006} 				&  {Apr 2006}		&  {Jul 2006}           &  {Dec 2007}\\
 \noalign{\vskip .8ex}%
 \hline
 \noalign{\vskip .5ex}%
 Satellite				&	\cxc	& \cxc	&	\xmm	&	\cxc	&\cxc	& \cxc\\
 \noalign{\vskip 0.5ex}%
 \multicolumn{6}{l}{Spectral analysis:}\\
 \noalign{\vskip .5ex}%
Net counts$^{a}$\dotfill			& $30.5$			& $10.5$			& $623.4/183.8/232.4$			  	& $36.4$			& $36.3$  			& $37.9$  		 \\
Scaled background counts$^{a}$\dotfill		& $0.5$				& $0.5$				& $998.6/215.2/209.6$			  	& $0.6$				& $0.7$				& $1.1$			 \\
Exposure (ks)$^{a}$\dotfill			& $27.82$	       		& $27.11$			& $78.38/90.90/90.95$	  			& $29.57$	       		& $29.57$     			& $34.73$     		 \\ 
$N_\mathrm{H}$ ($10^{21}$~\cmsq)\dotfill	& $3.1^{+2.4}_{-1.9}$		& $3.6^{+4.3}_{-3.1}$		& $4.2^{+0.3}_{-1.1}$		  		& $4.9^{+5.2}_{-2.0}$		& $7.3^{+2.2}_{-2.4}$		& $2.5^{+1.6}_{-1.0}$	 \\
$T$ (MK)\dotfill				& $46.8^{+\infty}_{-26.0}$	& $8.7^{+9.7}_{-3.5}$		& $11.3^{+3.5}_{-2.8}~/~89^{+32}_{-43}$		& $15.9^{+8.6}_{-7.0}$		& $6.5^{+2.3}_{-2.1}$		& $19.9^{+7.4}_{-4.7}$	\\ 
EM ($10^{53}$~\cmcc)\dotfill			& $0.18^{+0.12}_{-0.05}$	& $0.10^{+0.34}_{-0.10}$	& $0.22^{+0.12}_{-0.13}~/~0.51^{+0.08}_{-0.07}$	& $0.35^{+0.63}_{-0.11}$	& $1.0^{+1.9}_{-1.0}$		& $0.20^{+0.08}_{-0.05}$	\\ 
$Z/Z_\odot$\dotfill				& $:=0.17$			& $:=0.17$			& $:=0.17$		 			& $:=0.17$			& $:=0.17$			& $:=0.17$		 \\ 
$F_\mathrm{X}$ ($10^{-14}$~\ergcms)\dotfill	& $0.91^{+0.18}_{-0.15}$	& $0.13^{+0.05}_{-0.04}$	& $3.7^{+0.2}_{-0.2}$		  		& $0.63^{+0.11}_{-0.10}$	& $0.47^{+0.08}_{-0.07}$	& $0.58^{+0.10}_{-0.09}$ \\ 
$L_\mathrm{X}$ ($10^{30}$~\ergps)\dotfill	& $0.27^{+0.05}_{-0.05}$	& $0.10^{+0.03}_{-0.03}$	& $1.16^{+0.07}_{-0.07}$		 		& $0.36^{+0.06}_{-0.06}$	& $0.9^{+0.2}_{-0.1}$		& $0.20^{+0.03}_{-0.03}$	\\ 
 \noalign{\vskip 1.5ex}%
 \multicolumn{7}{l}{Quantile analysis:}\\
 \noalign{\vskip .5ex}%
Energy range\dotfill				& $0.2 - 4.0$			& $0.2 - 4.0$			& $0.3 - 10.0$					& $0.2 - 4.0$			& $0.2 - 4.0$			& $0.2 - 4.0$		 \\
Net counts$^{c}$\dotfill			& $30.48$			& $10.53$			& $626.9$				        & $36.37$		        & $36.32$			& $37.91$		 \\
Scaled background counts$^{c}$\dotfill		& $0.52$			& $0.47$			& $985.1$				        & $0.63$ 		        & $0.68$			& $1.09$		 \\
$Q_\mathrm{25\%}$\dotfill			& $0.28 \pm 0.04$ 	      	& $0.18 \pm 0.03$		& $0.081 \pm 0.010$			        & $0.23 \pm 0.02$	        & $0.18 \pm 0.01$ 		& $0.22 \pm 0.03$ 	 \\
$Q_\mathrm{50\%}$\dotfill			& $0.35 \pm 0.04$ 	      	& $0.24 \pm 0.09$		& $0.145 \pm 0.006$			        & $0.31 \pm 0.03$	        & $0.24 \pm 0.04$ 		& $0.28 \pm 0.02$ 	 \\
$Q_\mathrm{75\%}$\dotfill			& $0.48 \pm 0.08$ 	      	& $0.46 \pm 0.19$		& $0.255 \pm 0.019$			        & $0.40 \pm 0.06$	        & $0.34 \pm 0.03$ 		& $0.40 \pm 0.06$ 	 \\
$E_\mathrm{50\%}$~(keV)\dotfill			& $1.52 \pm 0.15$ 	      	& $1.12 \pm 0.35$		& $1.709 \pm 0.06$				& $1.39 \pm 0.09$		& $1.11 \pm 0.14$ 		& $1.28 \pm 0.09$ 	 \\
$x = \log[Q_\mathrm{50\%}/(1-Q_\mathrm{50\%})]$\dotfill	& $-0.27_{-0.08}^{+0.07}$ & $-0.50_{-0.26}^{+0.20}$	& $-0.77_{-0.02}^{+0.02}$			& $-0.34_{-0.05}^{+0.05}$	& $-0.50_{-0.09}^{+0.08}$	& $-0.40_{-0.05}^{+0.05}$\\
$y = 3 \times Q_\mathrm{25\%}/Q_\mathrm{75\%}$\dotfill	& $1.74_{-0.29}^{+0.29}$  & $1.19_{-0.42}^{+0.42}$	& $0.95_{-0.11}^{+0.11}$			& $1.68_{-0.22}^{+0.22}$	& $1.65_{-0.15}^{+0.15}$	& $1.63_{-0.25}^{+0.25}$ \\
$T_{xy}$~(MK)\dotfill				& $\approx 13$			& $\approx 35$			& $\approx 90$					& $\approx 10$			& $\approx 15$			& $\approx 16$		 \\
$N_{\mathrm{H},xy}$~($10^{21}$~\cmsq)\dotfill	& $\approx 10$			& $\approx 0.2$			& $\approx 2.5$					& $\approx 9.0$			& $\approx 1.5$			& $\approx 3.5$		 \\
\hline
\end{tabular}
\begin{list}{}{}
\item The uncertainties are based on 68\% Bayesian confidence ranges. X-ray luminosity and absorbed X-ray flux in the $0.1-10$~keV range, assuming $d=400$~pc.
\item[$^{\mathrm{a}}$] For \xmm\ observations, values refer to the EPIC pn, MOS1, and MOS2, respectively (pn only for September 2005). The values in parentheses refer
to detectors not used in the spectral fits.
\item[$^{\mathrm{b}}$] Estimates based on models for February 2005 and March 2006.
\item[$^{\mathrm{c}}$] Based on pn data in general, except for March 2005 for which MOS1 data only were used (similar values and results were obtained for 
MOS2 data).
\end{list}
}
\end{table*}

\citet{audard05b} presented spectral fits of the V1118 Ori data taken through March 2005. \citet{lorenzetti06} repeated
the analysis for the February 2005 data and added the September 2005 data, noting a decrease in the X-ray flux of V1118 Ori
at that period compared to earlier in the outburst. For this paper, we have reprocessed the 2005 data with the latest calibration 
(see Sect.~\ref{sect:datared}).

We used XSPEC 11 \citep{arnaud96} to fit the \cxc\ ACIS and the \xmm\  EPIC spectra for each epoch. For February 2005 and March 2006, we fitted the EPIC
pn, MOS1, and MOS2 spectra simultaneously, while we used the EPIC MOS only for March 2005, and the EPIC pn only for 
September 2005 (no MOS data were available). In general, we used a 1-$T$ collisional ionization equilibrium (CIE) model (\textit{apec};
\citealt{smith01}) with a photoelectric absorption model, except for the deep March 2006 observation for which the signal-to-noise was
large enough to use a 2-$T$ CIE model. {We have fitted the plasma metallicity in the high signal-to-noise ratio spectra of February 2005 and used the best-fit
value, $Z=0.17Z_\odot$, for the other epochs.
The coronal abundances are relative to the solar photospheric standard set of \citet{grevesse98}. The best-fit metallicity is in line with the values measured for Fe
in the coronae of young stars in Orion (Table~3 in \citealt{maggio07}).} Since the spectral fits of the February and March 2005 spectra are similar
to those reported in \citet{audard05b}, we provide in this paper their best-fit values (but adapt the emission measure and luminosities to the adopted
distance of 400~pc). Note that the
September 2005 data are heavily contaminated by the high background level during the observation. Contrary to \citet{lorenzetti06},
we preferred not to provide spectral fits for this observation; however, we provide estimates for the observed X-ray flux at Earth
and absorption-corrected X-ray luminosity, based on the previous February 2005 and the posterior March 2006 \xmm\ observations.  
The results of our spectral fits are given in Table~\ref{tab:fits}.

\subsection{Quantiles}

\begin{figure*}[ht]
\centering
\includegraphics[bb=85 360 550 700,angle=0,width=0.30\linewidth]{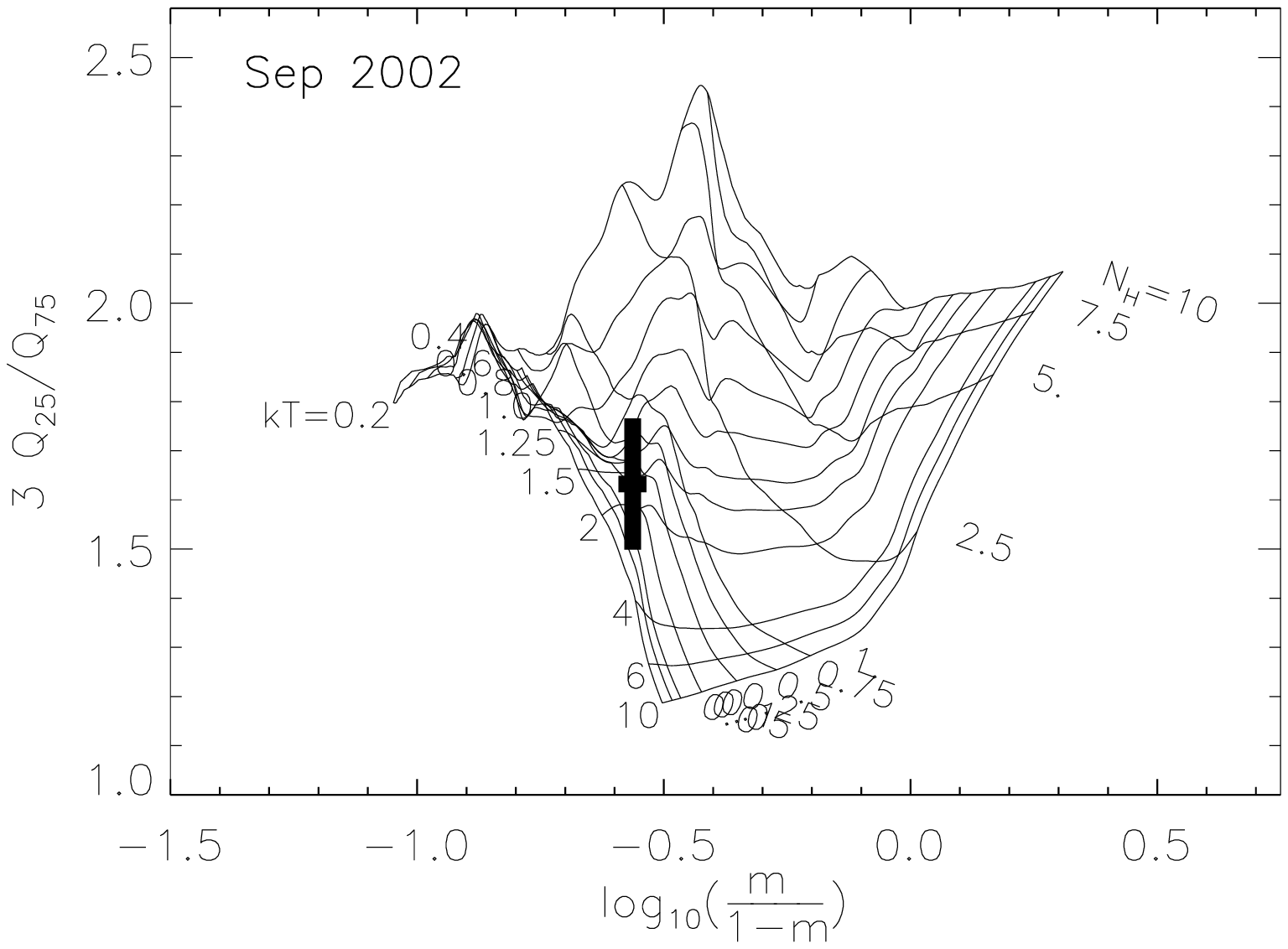}\\
\includegraphics[bb=85 360 550 700,angle=0,width=0.30\textwidth]{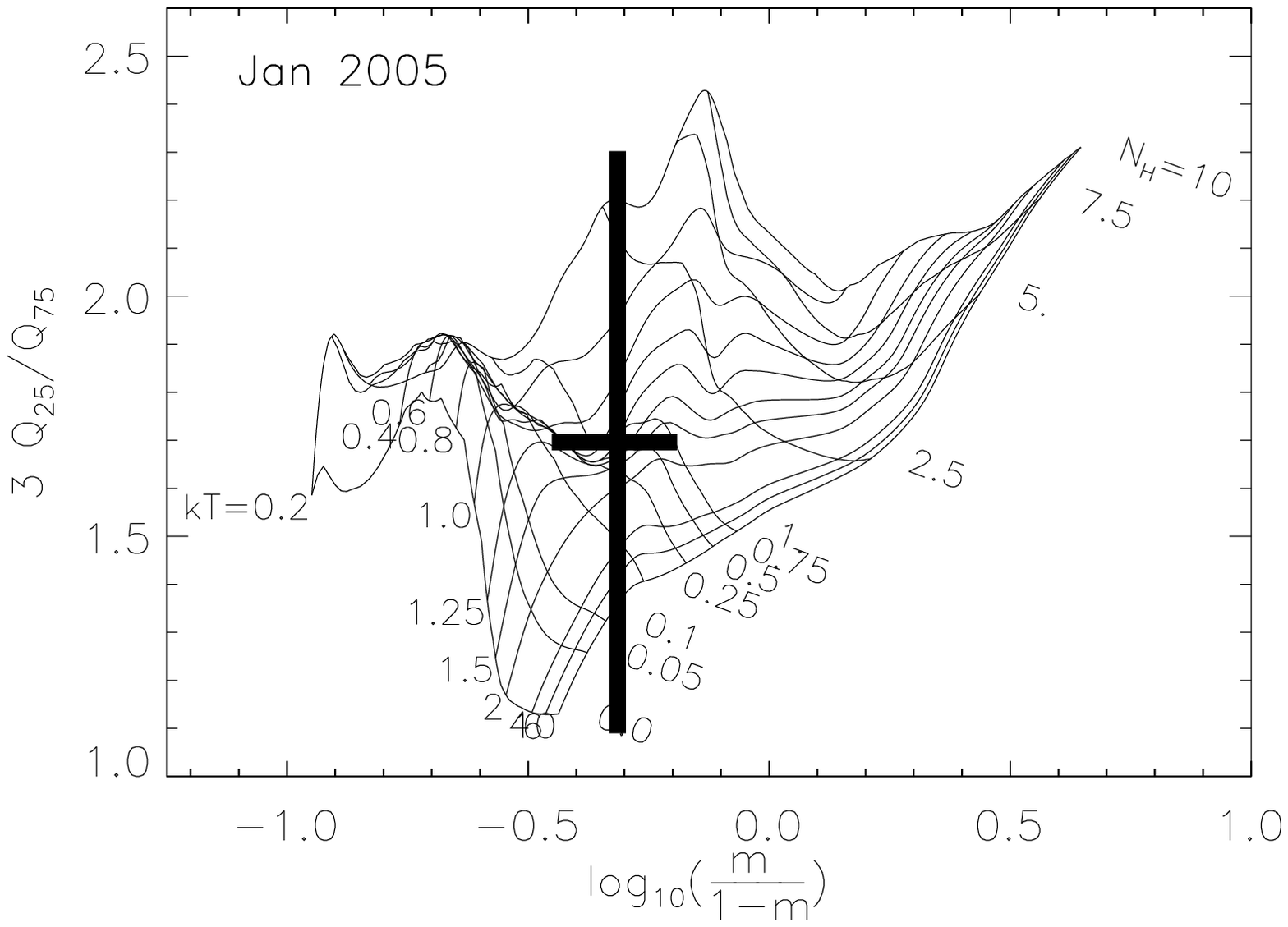}\hfill
\includegraphics[bb=85 360 550 700,angle=0,width=0.30\textwidth]{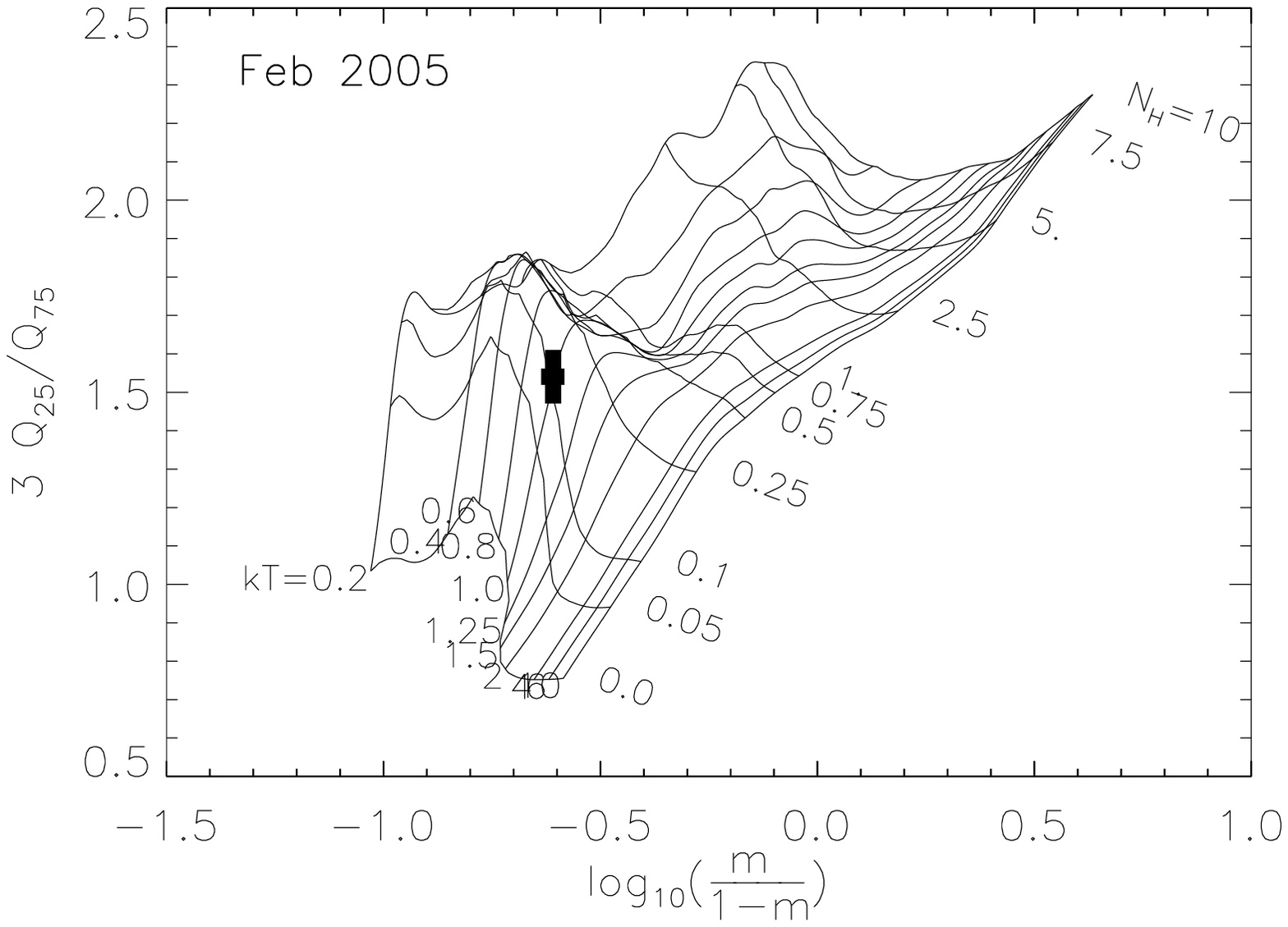}\hfill
\includegraphics[bb=85 360 550 700,angle=0,width=0.30\textwidth]{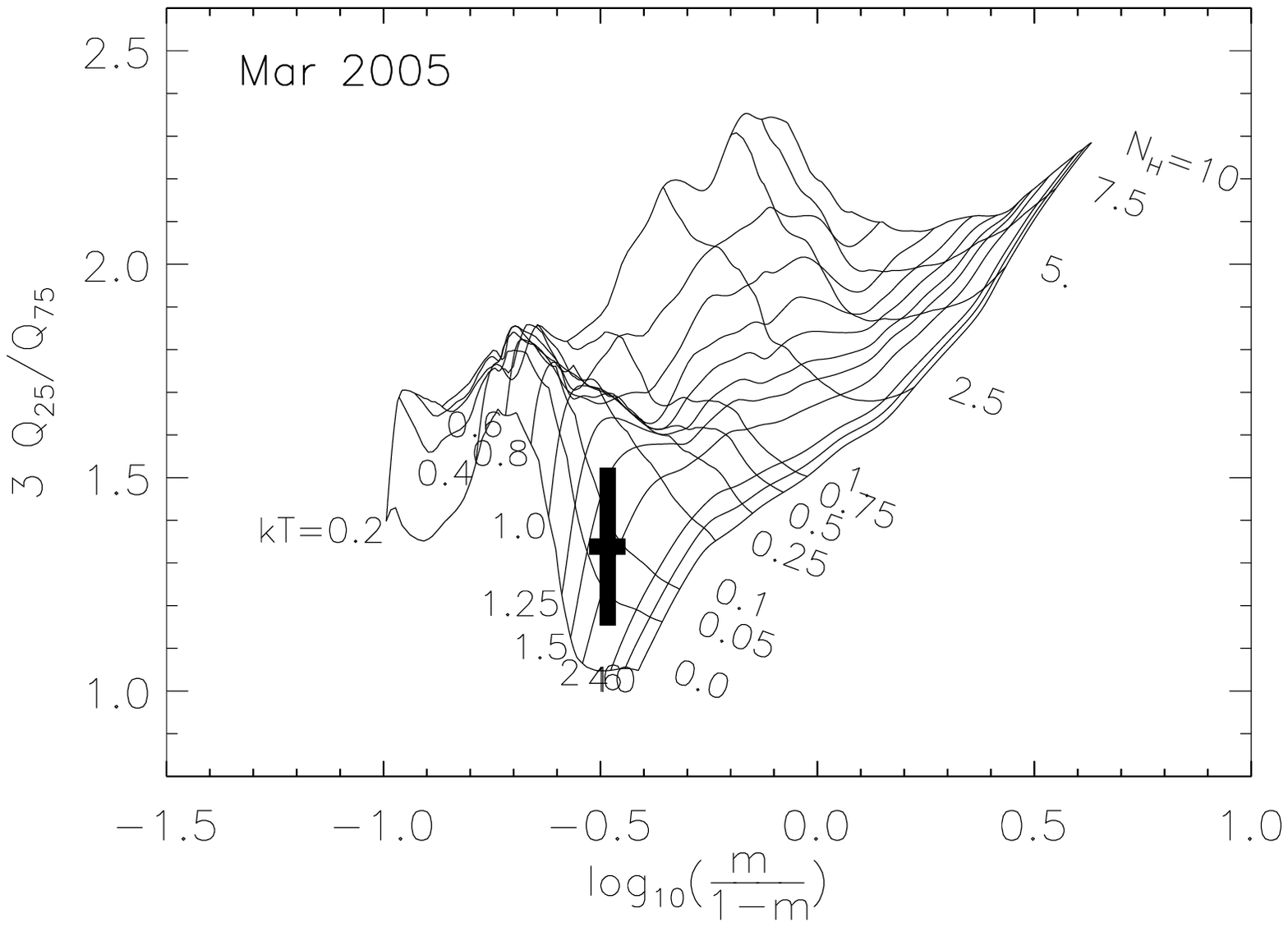}\\
\includegraphics[bb=85 360 550 700,angle=0,width=0.30\textwidth]{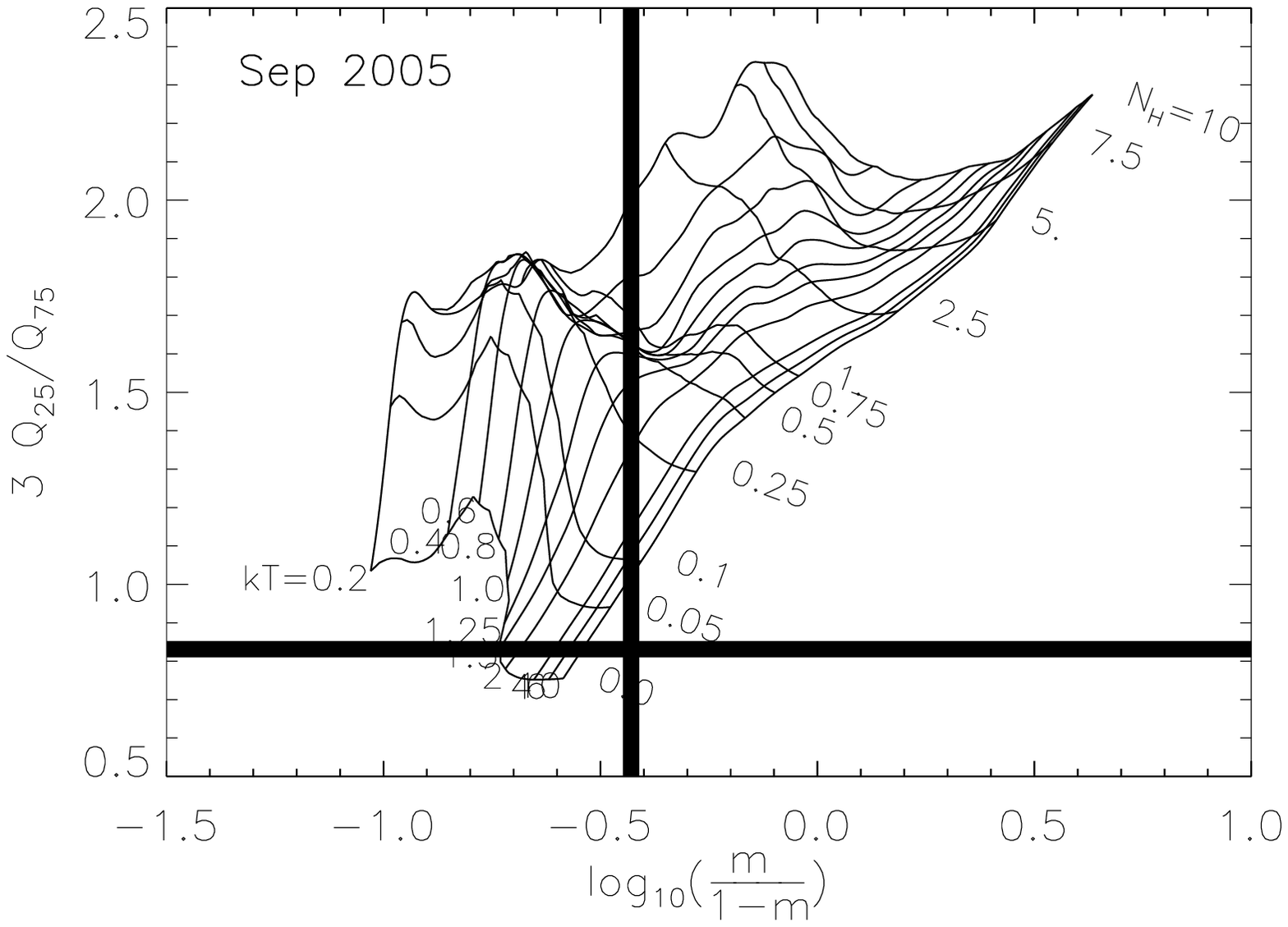}\hfill
\includegraphics[bb=85 360 550 700,angle=0,width=0.30\textwidth]{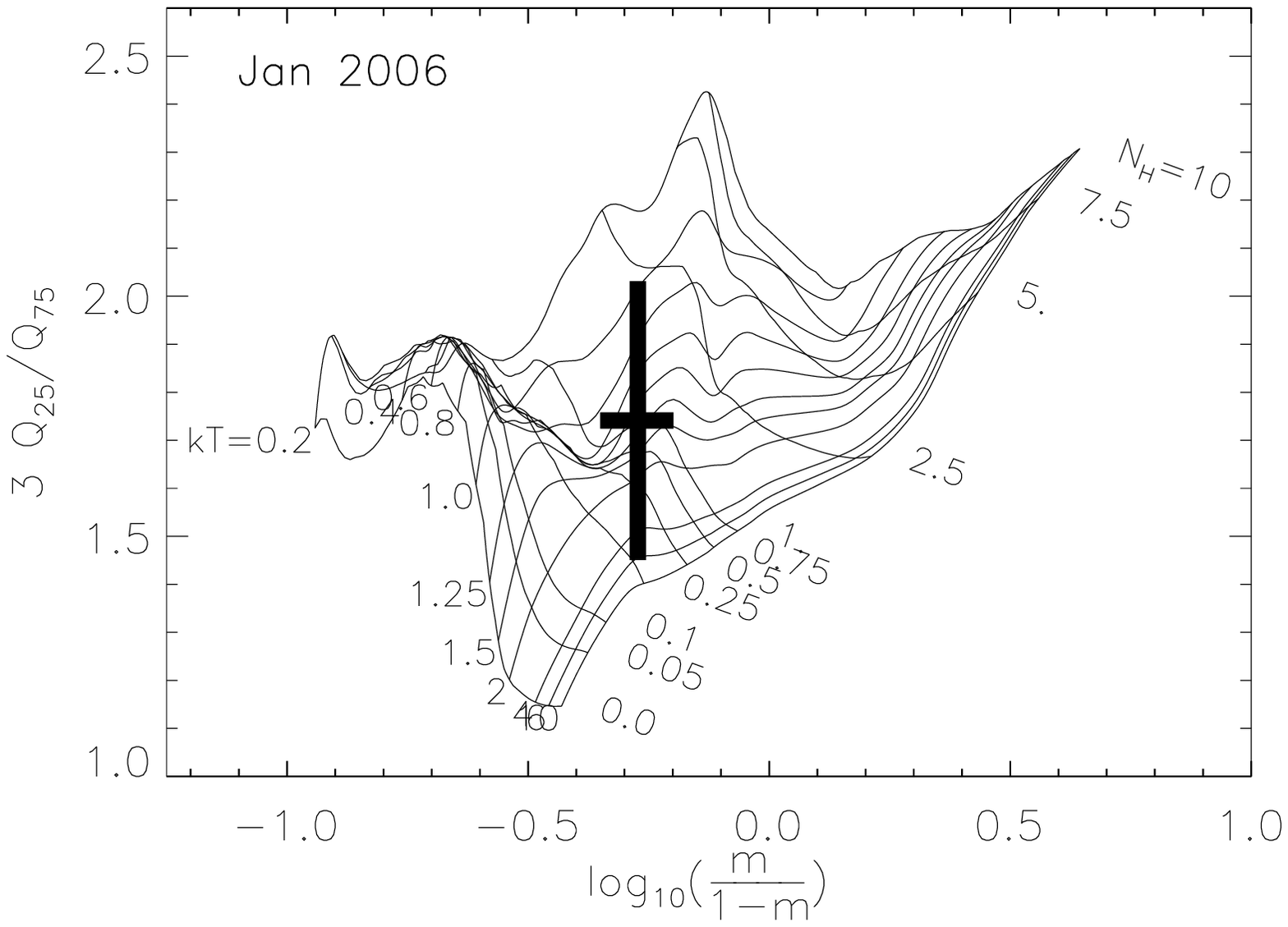}\hfill
\includegraphics[bb=85 360 550 700,angle=0,width=0.30\textwidth]{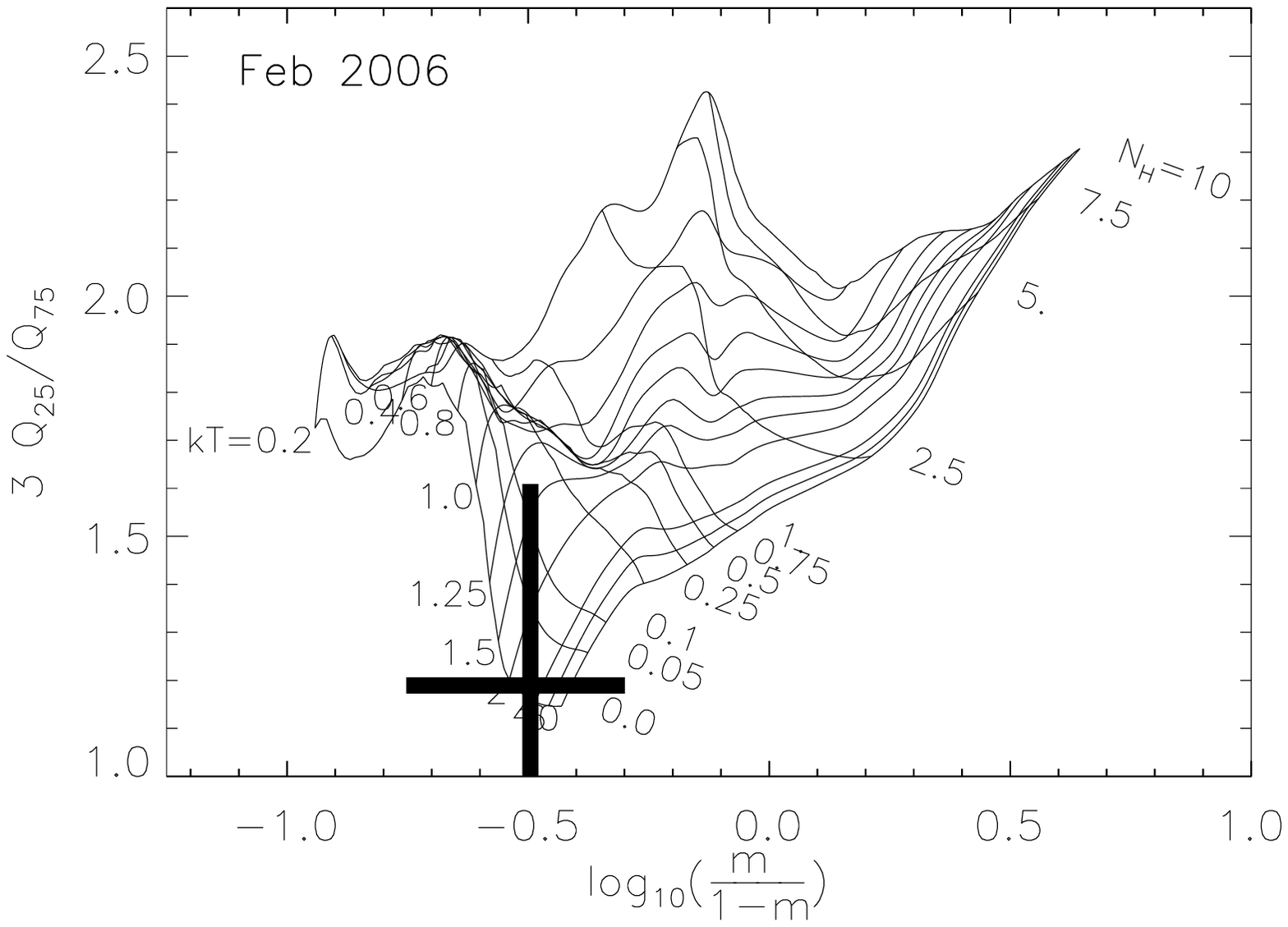}\\
\includegraphics[bb=85 360 550 700,angle=0,width=0.30\textwidth]{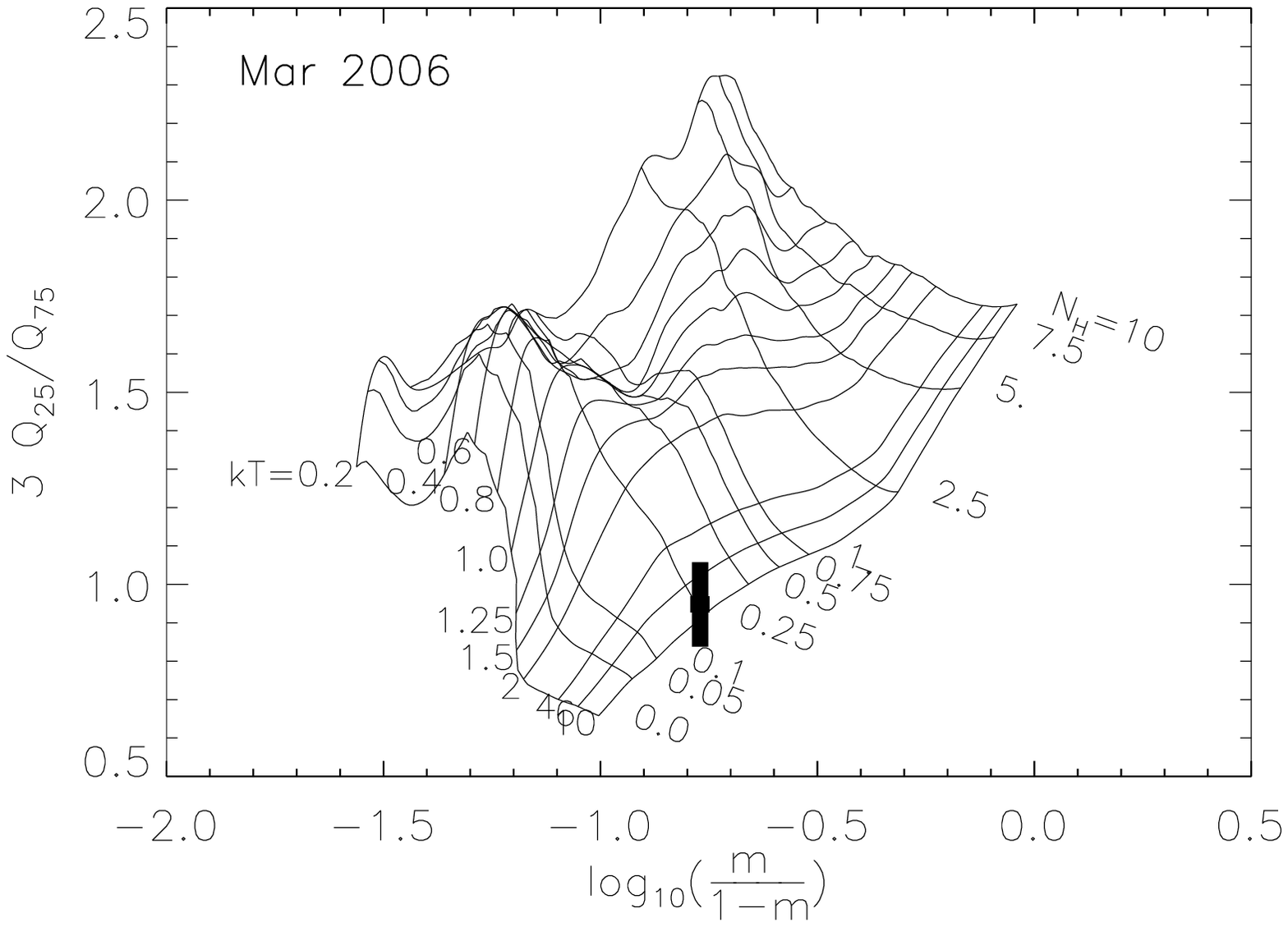}\hfill
\includegraphics[bb=85 360 550 700,angle=0,width=0.30\textwidth]{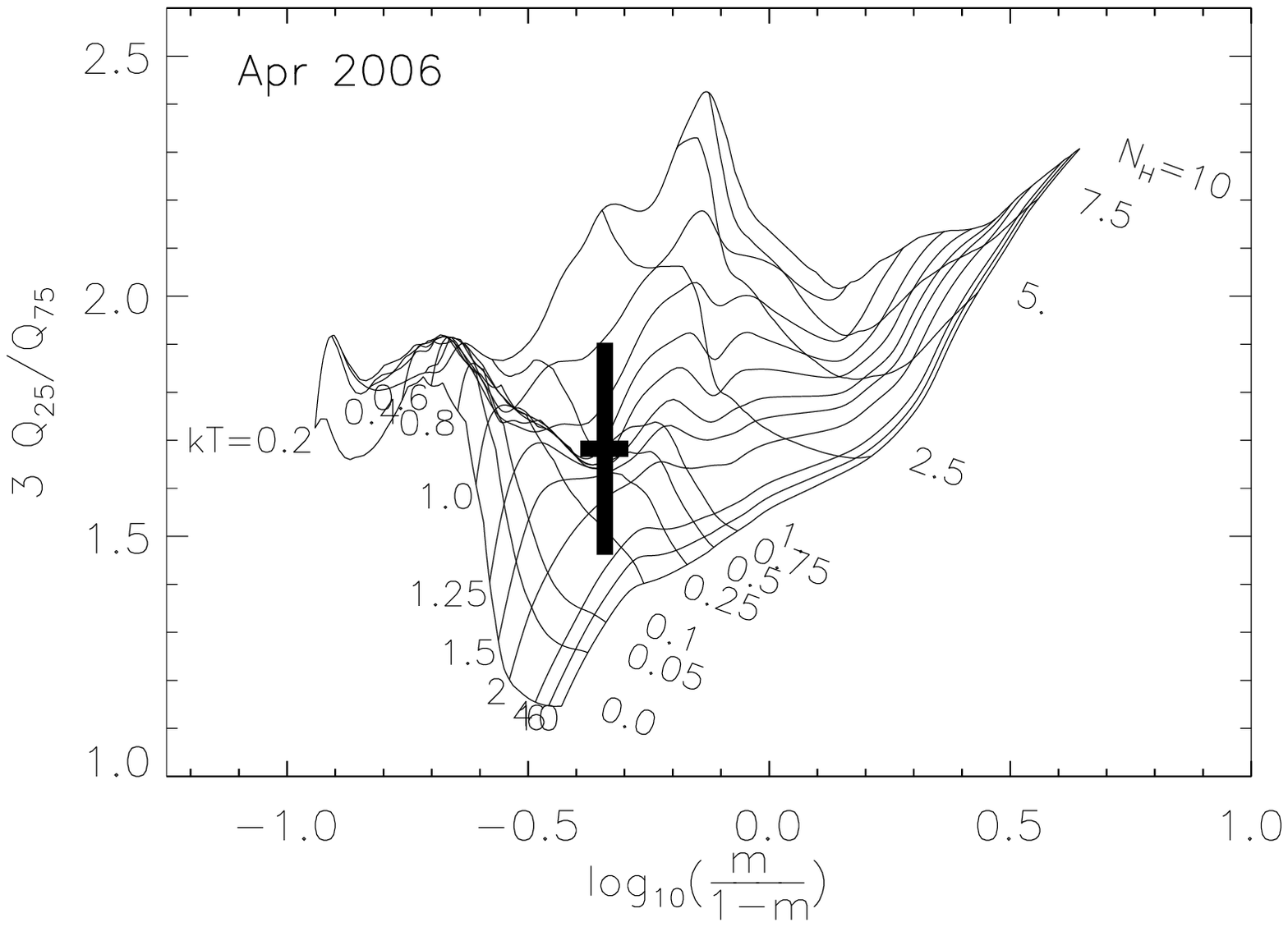}\hfill
\includegraphics[bb=85 360 550 700,angle=0,width=0.30\textwidth]{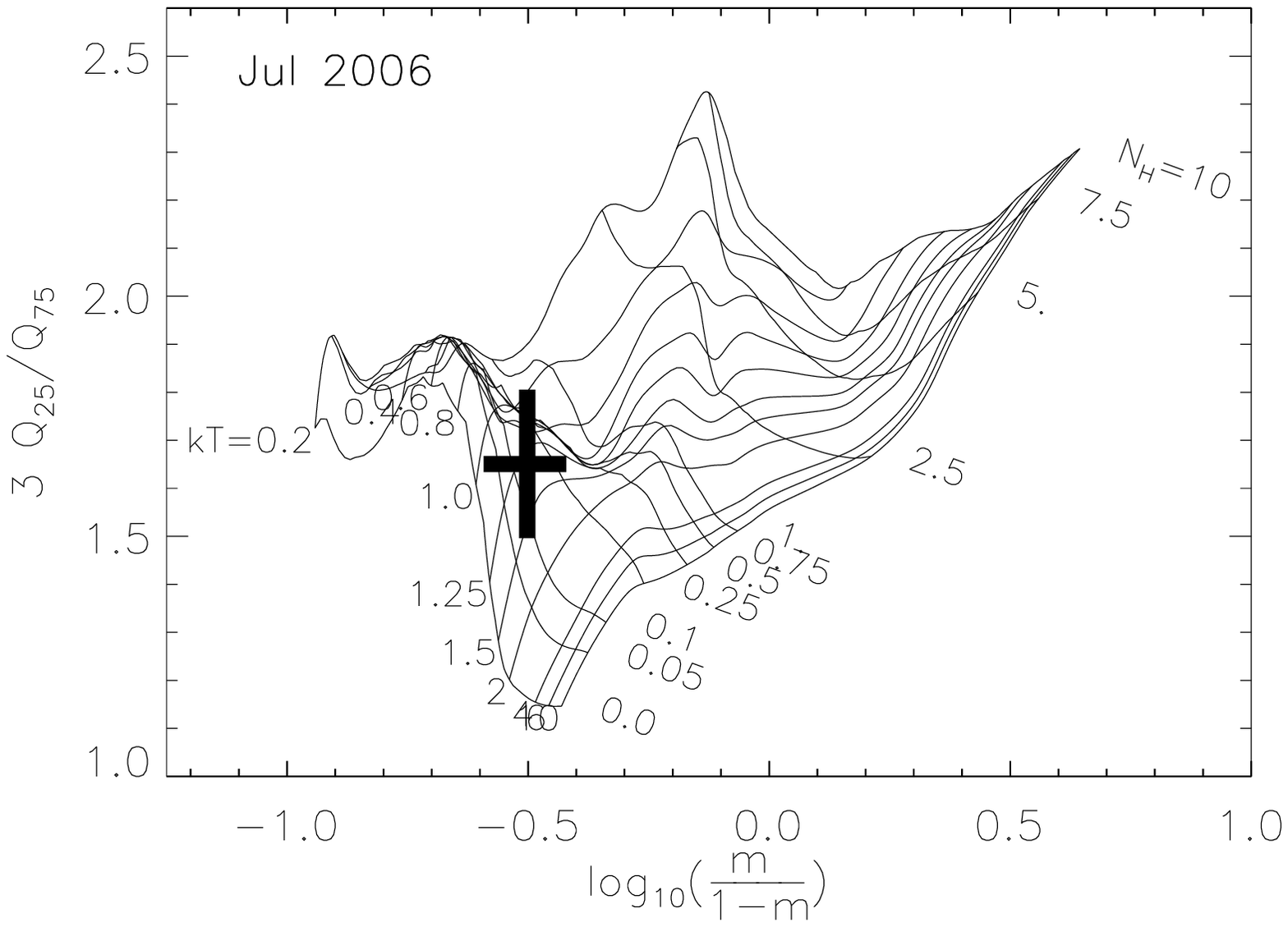}\\
\includegraphics[bb=85 360 550 700,angle=0,width=0.30\textwidth]{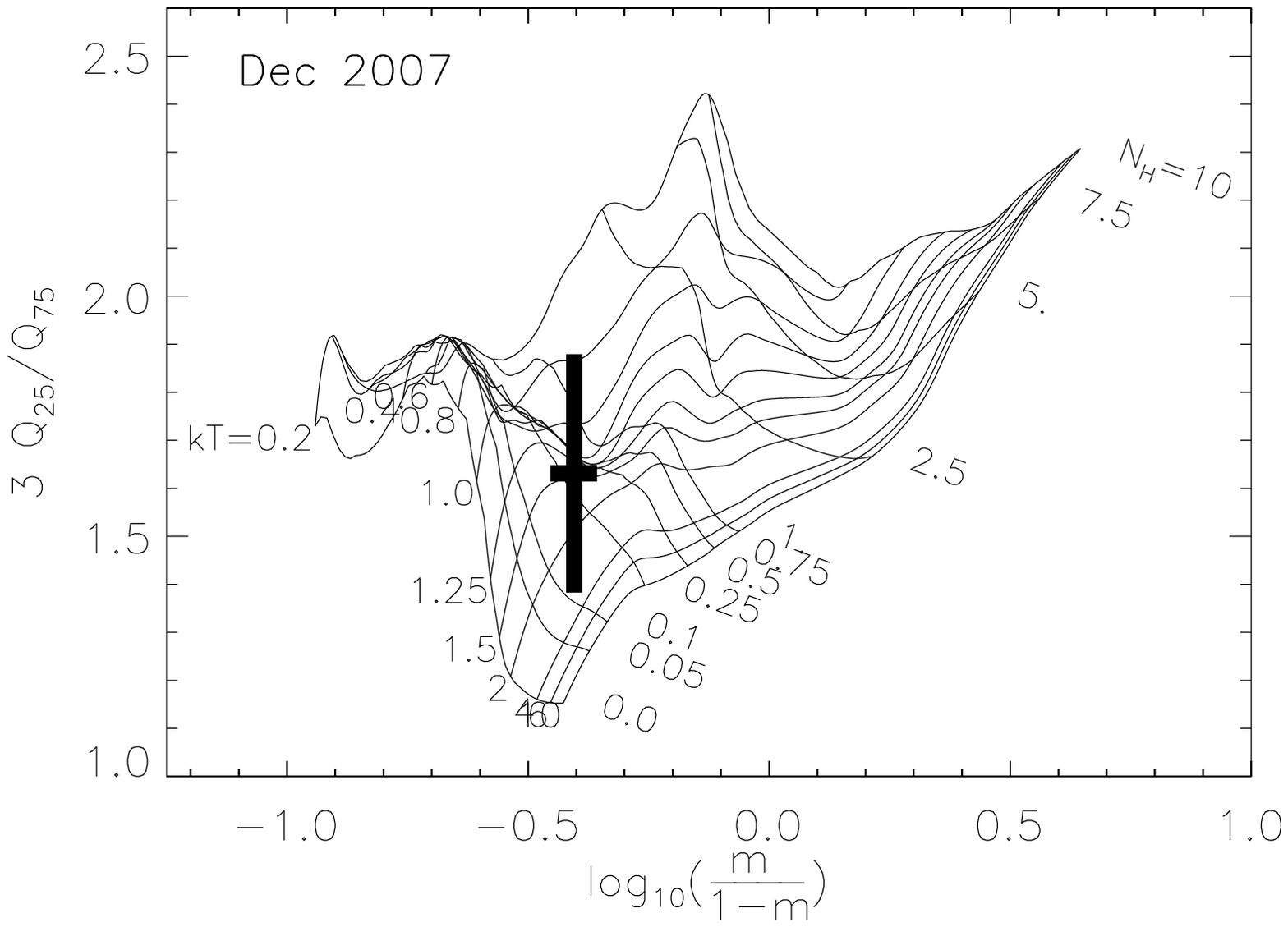}
\caption{Color-color diagrams based on the $Q_{50\%}$ ($=m$), $Q_{25\%}$, and $Q_{75\%}$ quantiles. The grid is in plasma temperature (kT in keV) and hydrogen column density ($N_\mathrm{H}$ in $10^{22}$~cm$^{-2}$). The serendipitous,
pre-outburst observation of September 2002 is shown at the top, the post-outburst version of December 2007 is shown at the bottom. 
The 9 outburst observations in between are ordered in time from left to right and top to bottom, 
with January 2005 at the top left and July 2006 at the bottom right.
\label{fig:quantiles}}
\end{figure*}

In view of the low observed count rates with \xmm\ and \cxc\  during the outburst, we have explored an alternative method to determine
spectral properties for low-count X-ray spectra and have used the quantile analysis presented by \citet{hong04}\footnote{{The code is }available at \url{http://hea-www.harvard.edu/ChaMPlane/quantile}. We used version 1.7.}. This type of analysis was, in particular, 
successfully used by \citet{grosso05} for their analysis of the spectral properties of V1647 Ori (see also \citealt{skinner09b}). In brief, this method makes
direct use of the event energy values, and determines spectral properties based on quantiles
of the total number of counts (e.g., median 50\% quantile, and quartiles, 25\% and 75\%),  instead of spectral energy bins. Such quantiles are then used
as indicators for the X-ray color of the source. In particular, we make use of a diagram with the $x =  \log_{10} (m / (1-m) ) $ index in the $x$-axis, where
$m$ corresponds to the median quantile $Q_{50}$, and the index $y = 3 \times Q_{25}/Q_{75}$ in the $y$-axis. \citet{hong04} define 
the $\alpha$\% quantile, $Q_\alpha$, as

\begin{displaymath}
Q_\alpha = \frac{E_{\alpha\%} - E_{\rm lo}}{E_{\rm up} - E_{\rm lo}},
\end{displaymath}

where $E_{up}$ and $E_{\rm lo}$ are the lower and upper boundaries of the used energy band, and $E_{\alpha\%}$ is the energy below which the net counts is
$\alpha$\% of the total number of counts. Note that this method is able to take the distribution of background event energies into account. We refer
the reader to \citet{hong04} for more details. 

The total (source + background) and background extraction regions used in the quantile analysis were the same as those for spectral fits.
The energy bands used to determine the extraction of events are reported in Table~\ref{tab:fits} together with the 25\%, 50\%, and 75\% quantiles,
the $E_{50\%}$ energy, and the $x$ and $y$ indices. We also provide estimated values for the hydrogen column density and plasma temperature 
derived from the $x$ and $y$ indices. Finally, Figure~\ref{fig:quantiles} shows the color-color diagrams for all observations.

\begin{figure*}[!ht]
\includegraphics[angle=0,width=\linewidth]{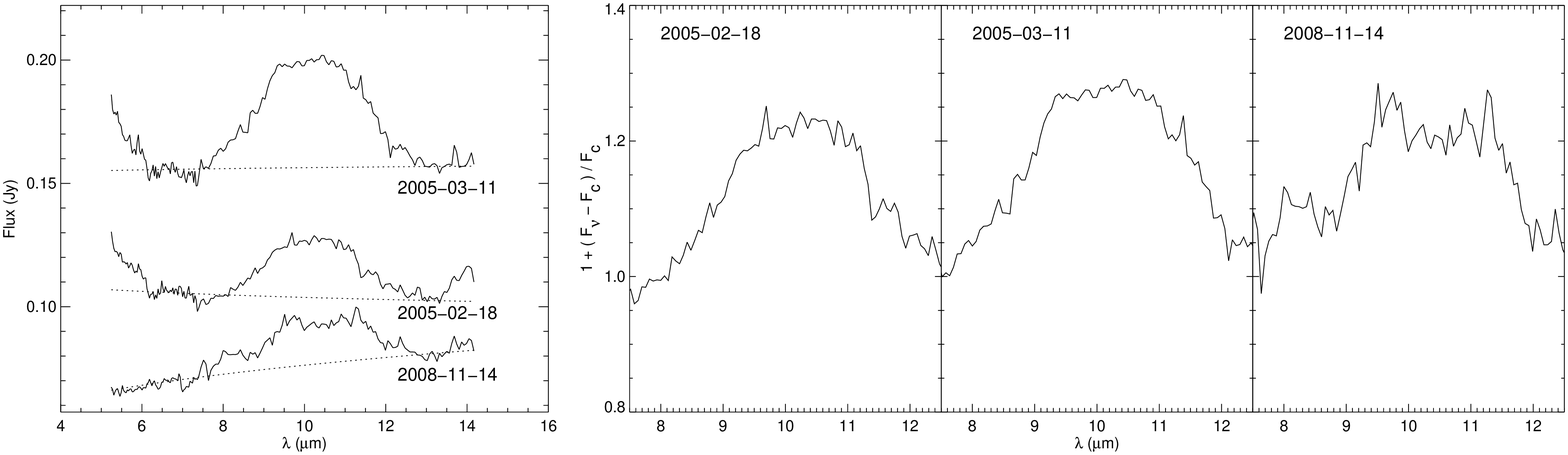}
\caption{\spitz\ IRS SL spectra of V1118 Ori for the three epochs (\textit{left} panel). The dotted line shows the fit to the underlying continuum in the [6-8,13-14.5]~\mic\ range. The right panels show the normalized flux at the three epochs.
\label{fig:sil}}
\end{figure*}

There exists a certain degree of degeneracy in the $x$ and $y$ indices \citep[e.g.,][]{grosso05}: indeed, the same color indices can sometimes be described by
different combinations of column density $N_{\rm H}$ and plasma temperature $T$. However, these degeneracies are limited to certain
sets of $T$ and $N_{\rm H}$: for example, in the September 2002 observation, color indices of $x = -0.8$ and $y=1.8$ can represent
any plasma temperature from about $kT \approx 0.2$ to about $kT \approx 0.8$ keV and $N_{\rm H}$ can range from $0$ to $0.75 \times 10^{22}$~\cmsq.
At higher temperatures, the degeneracy breaks down, and the observed $x/y$ values of V1118 Ori can be better constrained to $kT \approx 1.2-2.5$~keV 
and $N_{\rm H} \approx (1-8) \times 10^{21}$~\cmsq. The degeneracies differ slightly depending on the response matrix generated in the observation.
For example, with our \cxc\ monitoring data, the degeneracies occur around $(x,y)=(-0.7,1.85)$ and at $(-0.4,1.65)$. The first set again
indicate a degeneracy for $kT < 0.8$~keV and $N_{\rm H} < 1 \times 10^{22}$~\cmsq. The second set shows that the plasma properties cannot be disentangled
between $0.8$ and $1.5$~keV for a column density of $(3-8) \times 10^{21}$~\cmsq.

In contrast to the temperatures obtained from spectral fits, the temperature derived from the color indices are more stable and vary
little from $10-20$~MK, except in February 2006 ($T_{xy} \approx 35$~MK) and March 2006 ($T_{xy} \approx 90$~MK). In the case of March 2006, the 
quantile temperature is in fact close to the second temperature component (which also has the higher emission measure) obtained with the 2-$T$ fit. 
In the case of February 2006, the {low number of counts did not allow a good constraint on the temperature.} 
There also appears to have some kind of degeneracy: while the spectral fit gives a low plasma temperature and
``high'' column density, the quantile analysis gives the inverse. 
It remains unclear which result is more robust, but the quantile diagram does not show
evidence of a degeneracy. It is then likely that the spectral fit reached a local minimum at a low temperature and could not find a
combination of high plasma temperature and low $N_{\rm H}$.  The February 2005 quantile analysis and spectral fit both confirm that the temperature
of the coronal plasma was cooler than in September 2002 and March 2005, as initially report in \citet{audard05b}.
The error bars of the color indices in January 2005 do not allow us to constrain the temperature of the coronal plasma, while
the spectral fit with XSPEC indicates no evidence of cool plasma at the start of the outburst.

\subsection{X-ray parameter correlations}

We have attempted to search for correlations between X-ray properties ($N_\mathrm{H}$, $T$, $N_{\mathrm{H},xy}$, $T_{xy}$, $\mathrm{EM}$, $L_\mathrm{X}$, $E_{50\%}$), but found generally little or no correlation based on the Pearson correlation coefficient, $\rho$. Indeed, its absolute value amounted to values typically below $0.5$. One exception was a mild correlation between the average median photon energy, $E_{50\%}$, and the absorbing column density derived from the quantile analysis, $N_{\mathrm{H},xy}$, which showed $\rho = 0.57$. This is expected, as larger column densities will absorb low-energy photons. However, we emphasize that we have only used the estimated  $N_{\mathrm{H},xy}$ based on the location of the $xy$ values in the quantile color-color diagrams, and that we have not taken into account the (relatively) large range of column densities covered by the quantile errors (typically, $1-5 \times 10^{21}$~\cmsq, similar to values derived from spectral fitting). Another exception was a strong correlation between the emission measure and the X-ray luminosity, with $\rho = 0.96$. Together with the above lack of correlation between the plasma temperature and $L_\mathrm{X}$, and $N_\mathrm{H}$ and $L_\mathrm{X}$, this is a strong indication that the X-ray variability observed during the outburst of V1118 Ori is only due to the amount of material in the corona. It is, thus, probably related to the amount of mass falling from the disk into the stellar magnetosphere, i.e., to the mass accretion rate.

\section{Optical and infrared analysis}

\begin{figure*}[!ht]
\centering
\includegraphics[angle=0,width=\linewidth]{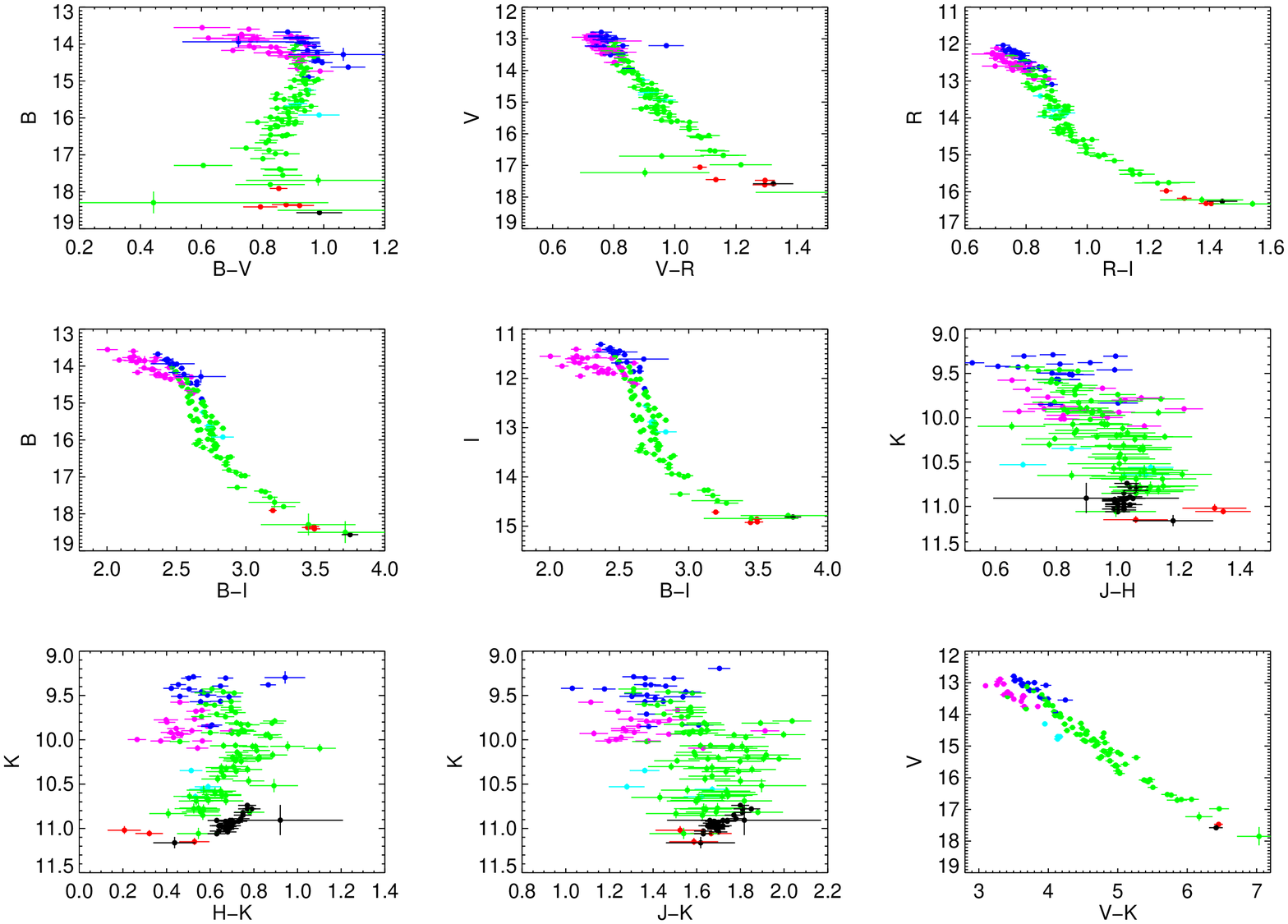}
\caption{Magnitude versus color indices for the different time intervals (red: \#1, lime: \#2, violet: \#3, blue: \#4, green: \#5, black: \#6).
\label{fig:colindmag}}
\end{figure*}

\begin{figure*}[!ht]
\centering
\includegraphics[angle=0,width=\linewidth]{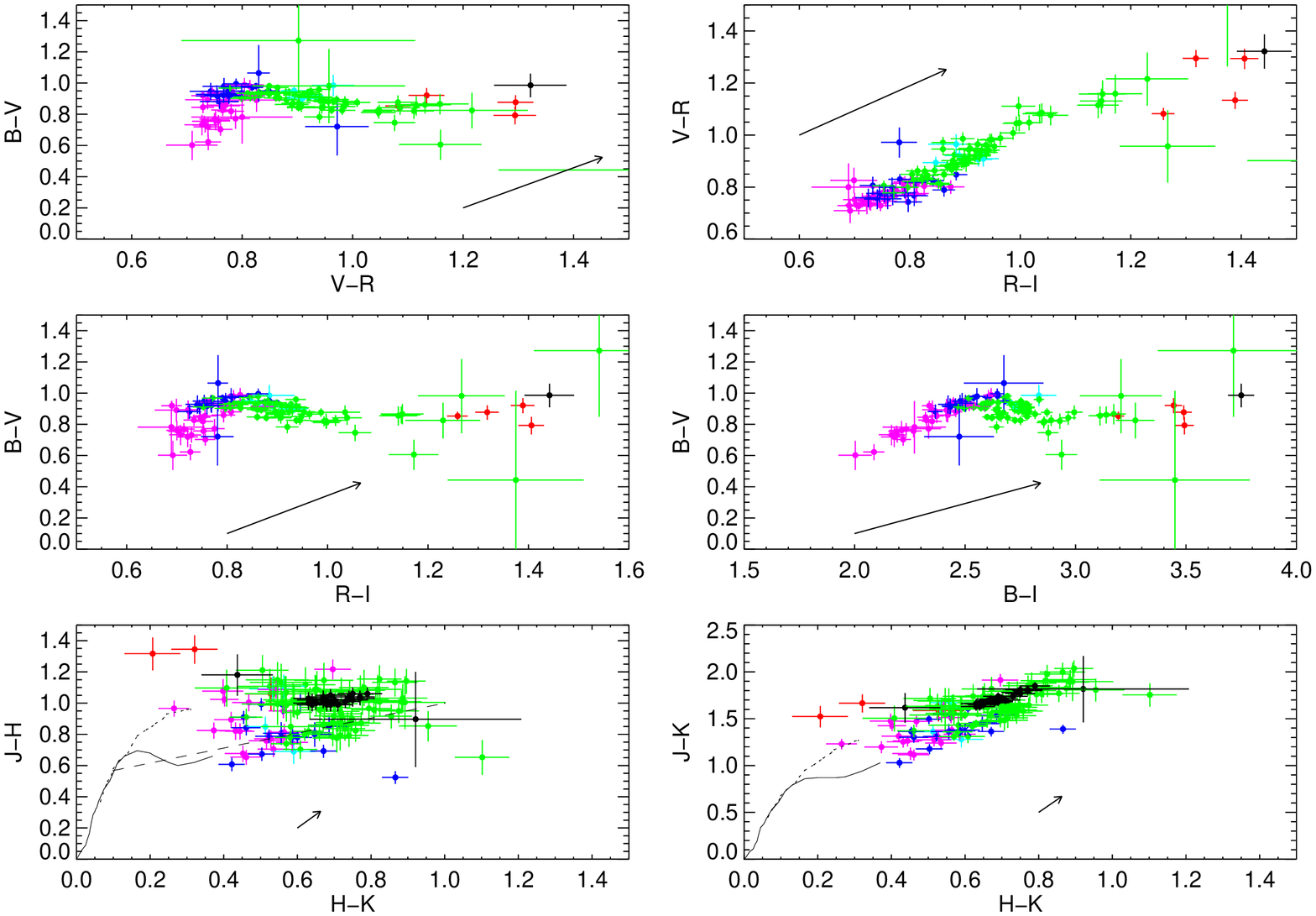}
\caption{Color index diagrams for the same time intervals as in Fig.~\ref{fig:colindmag}. An $A_V=1$~mag vector is added using the reddening law of \citet{rieke85}. The loci of unreddened main sequence and giant stars \citep{bessell88} are shown for the near-infrared panels as solid and dotted curves, respectively. Finally the loci of classical T Tauri stars \citep{meyer97} are shown as
a dashed line in the (\jh) vs. (\hk) diagram.
\label{fig:colindcolind}}
\end{figure*}

\subsection{Silicate feature}
\label{sect:sil}

Figure~\ref{fig:sil} shows the silicate feature obtained at the three different epochs and its normalized version ($S_\nu = 1 + F_\nu/F_C$). We have use a linear fit to the SL spectra in the $\log \lambda - \log F_\nu$ plane, using only wavelength ranges from 6.0--8.0~\mic\ and 13.0--13.5~\mic\   to determine the underlying continuum flux level, $F_C$. The shapes of the silicate feature are compatible for all three observations, but the Mar 2005 silicate feature appears to be brighter than the Feb 2005 or post-outburst (Nov 2008) normalized fluxes, i.e., more flux coming from the less optically thick disk upper layers, likely due to hotter temperature of such layers. The heating of the upper layers is probably {due to the strong irradiation}, e.g., by a hot spot (see below). The ratio $S_{11.3}/S_{9.8}$ is equal to $0.95-1.0$, suggesting grain growth, and the ratio $(S_{11.3}/S_{9.8})/(S_{\rm peak}^{10~\mu m})$ is $0.74-0.79$, evidence of a flat silicate feature. 

In the outburst spectrum of the EXor prototype EX Lup, \citet{abraham09} observed a silicate feature with a crystalline peak, in contrast to the pre-outburst silicate feature which was similar to the interstellar medium spectrum dominated by amorphous grains. They concluded that thermal annealing in the surface layer of the proto-planetary disk was the mechanism for this change in spectral features. Unfortunately, we have no pre-outburst spectrum of V1118 Ori before its outburst to compare with the outburst spectra, only a {spectrum taken 2 years after the end of the outburst and 3.5 years after our last outburst spectrum in March 2005}. If the pre-outburst spectrum was indeed typical of amorphous silicates, it means that the time scale for disappearance of the crystalline features must be longer than 2 years. \citet{abraham09} proposed that similar crystallization must have occurred during the previous 1955-1956 outburst of EX Lup, and since the 2005 quiescent spectrum did not show crystalline features, the removal time scale must have been less than 50 years. On the other hand, in V1118 Ori, it is also possible that the unchanging silicate feature in the three spectra is typical of the quiescent silicate feature in the young star, perhaps an indication that the disk is more evolved. In any case, M-type young stars typically show flat silicate features  \citep[e.g.,][]{kessler05,kessler06}. 
Finally, we note that the IRS data do not show strong evidence of the forsterite crystalline feature at 16~\mic, in contrast with the feature observed in EX Lup \citep{abraham09}. This is consistent with the lack of a sharp peak at 10~\mic\  and of a shoulder at 11.3~\mic.

\subsection{Color indices}
\label{sect:colorind}

\begin{figure*}[!ht]
\centering
\includegraphics[angle=0,width=\linewidth]{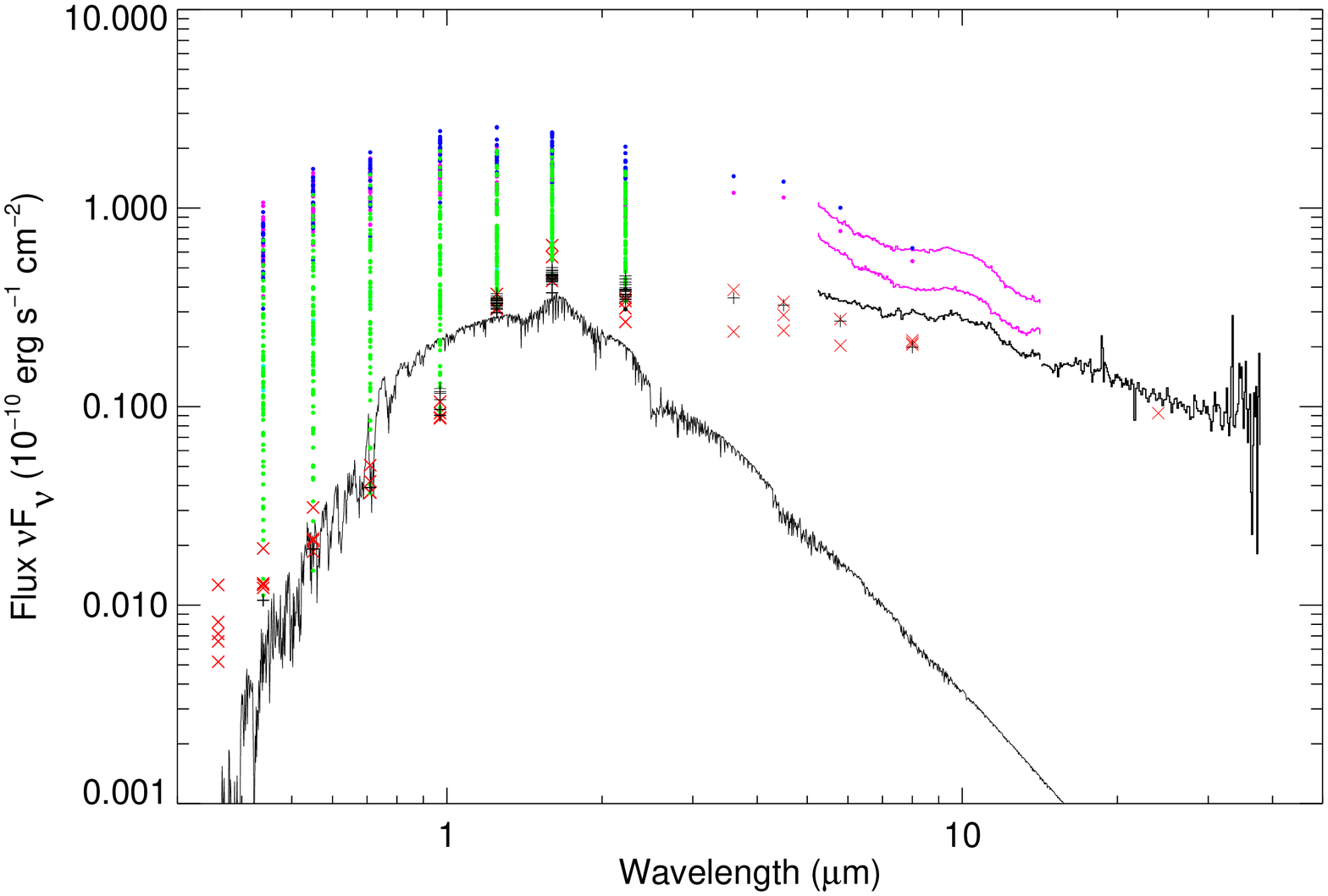}
\caption{Compilation of the optical and infrared photometry taken before, during, and just after the outburst (see text and caption of Fig.~\ref{fig:colindmag} for details about colors). The IRS SL and LL spectra and a 
reddened ($A_V = 1.5$~mag) stellar atmosphere model are also shown.\label{fig:sed}}
\end{figure*}

We have calculated color indices from our optical and near-infrared data, and looked into their evolution during the outburst. Time intervals
were defined in the pre-outburst epoch ($1$; JD $<2453200$), initial phase of rise ($2$; $2453200 < $JD $< 2453390$), second phase of rise 
($3$; $2453390 < $JD $< 2453450$), plateau ($4$; $2453450 < $JD $< 2453650$), decay ($5$; $2453650 < $JD $< 2453900$), and post-outburst ($6$; $2453900 < $JD $< 2454200 $).

Figure~\ref{fig:colindmag} shows the color-magnitude diagrams, whereas Figure~\ref{fig:colindcolind} shows the color-color diagrams, together with an
$A_V = 1$~mag reddening vector using the reddening law of \citet{rieke85},  the loci of the unreddened main sequence and giant stars \citep{bessell88}, and the loci of classical T Tauri stars
\citep{meyer97}.  Compared to the pre-outburst color indices of V1118 Ori, the optical color indices are \emph{bluer} in the early phases of the outburst (except for \bv\ which shows a peculiar pattern, 
probably because the $B$ magnitude is very sensitive to the evolution of hotspot emission), and gradually return to the pre-outburst values in the decaying phase of the outburst. 
In the near infrared, an apparent blueing occurs as well in the initial phases of the outburst with a return
to ``normal'' colors in the decay phase. {The above results are consistent with the hypothesis of \citet{lorenzetti07} that blueing must have occurred due to a hotter temperature component during the initial phase of the outburst (their data were taken only in the decaying phase and showed reddening). The variations in the color-color diagrams also show no trend related to extinction variations, as suggested by \citet{lorenzetti07} and \citet{lorenzetti09}.} There is, however, a larger scatter in the infrared color-color diagrams, especially in \jh\ versus \hk.   Since the stellar photospheric emission does not vary during the outburst, we need to understand
the variations in color indices and magnitudes during the outburst in the context of a star + disk + hot spot model.  { \citet{lorenzetti07} suggested that the polarization found in the $I$ band of the V1118 Ori outburst was intrinsic and that it was probably due to the spotted and magnetized stellar photosphere. As we will show below,} the photometric changes can {indeed} be explained by
changes  in the fluxes of the disk thermal emission and from a hot spot as the amount of mass falling from the disk onto the stellar photosphere increases during the outburst. We note also that during the outburst, 
any increase in hotspot emission will further increase the irradiation of the disk. The following section aims to understand the multi-wavelength photometry and spectroscopy observed during the outburst of V1118 Ori 
by comparing the data with young star disk models.

\subsection{Spectral energy distribution}
\label{sect:sed}

Figure~\ref{fig:sed} shows the observed SED for V1118 Ori. We have shown as small colored dots the data for the different phases (as defined at the beginning of section \ref{sect:colorind} and in the caption of Fig.~\ref{fig:colindmag}), and the pre- and post-outburst photometry as red crosses and black stars, respectively. 
The IRS low-resolution spectra are also shown for the outburst phases (violet) and post-outburst phase (black). For clarity, we do not show the high-resolution spectra. To emphasize the expected stellar photospheric contribution, we have added a stellar atmosphere model from the PHOENIX database\footnote{\url{http://www.hs.uni-hamburg.de/EN/For/ThA/phoenix/index.html}} \citep[e.g.,][]{hauschildt99} for a star with $T_\mathrm{eff} = 3600$~K, $\log g = 4.0$, and $Z=0$ (STARdusty2000 models), and reddened it with $A_V =1.5$~mag and the reddening law of \citet{fitzpatrick99}. Indeed, the hydrogen column density derived from X-rays is $2.7^{+1.2}_{-0.9} \times 10^{21}$~\cmsq\  in 2002, and does not change much during the outburst, generally staying in the range $1-5 \times 10^{21}$~\cmsq, which  corresponds to $A_V$ in the range $A_V = 1.71^{+0.76}_{-0.57}$~mag for $R_V = 3.1$ (Galactic value) or  $A_V = 1.44^{+0.64}_{-0.48}$~mag for $R_V = 5.5$ (dusty environment). 

\begin{figure}[!t]
\centering
\includegraphics[angle=0,width=\linewidth]{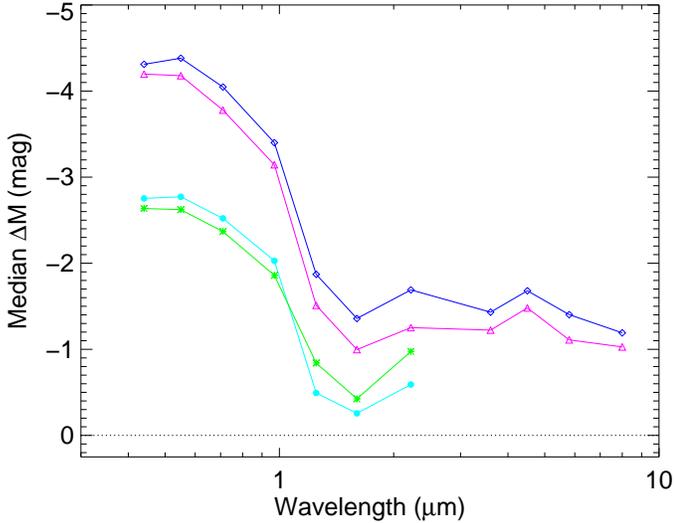}
\caption{Median magnitude difference relative to quiescence as a function of wavelength. The same time intervals as in Fig.~\ref{fig:colindmag} are used for the color coding (lime filled circles: \#2, violet triangles: \#3, blue diamonds: \#4, green stars: \#5). In the initial phase of the outburst, a strong optical component developed, with a peak in the $V$ band. 
\label{fig:pseudospec2}}
\end{figure}

We calculated the median magnitude differences from 0.44 to 8.0~$\mu$m (we have no $U$ photometry during the outburst), relative to the median magnitude in quiescence (we used both pre-outburst and post-outburst magnitude;  
Fig.~\ref{fig:pseudospec2}). There is clear evidence that the optical emission brightened much more than the infrared emission, with a peak in the $V$ band. 
We also observe that the $BVRI$ decayed faster than the $JHK$ bands, probably indicating a mixture of different emission components in the near-infrared bands (e.g., disk and hot spot). At long wavelengths ($\lambda \ge 2$~\mic), the increase in flux was similar at all wavelengths.

To quantify better the star+disk properties of V1118 Ori before and during its outburst, we have then used the code described in \citet{whitney03a,whitney03b}, version 20090224\footnote{\url{http://gemelli.colorado.edu/~bwhitney/codes/}}. We have run the code by fixing certain parameters and leaving others free to vary. First, the models did not include an envelope, but only an ambient medium with density of $10^{-22}$~g~cm$^{-3}$. The contribution of this medium is generally negligible. The exclusion of the envelope is motivated by the fact that it would contribute mostly at long wavelengths where we have no data. The IRS data also do not show strong evidence of an envelope (that could not be attributed to improper background subtraction).
The fixed parameters were the stellar radius ($R_\star = 1.29 R_\odot$), mass ($M_\star = 0.41 M_\odot$), and effective temperature ($T_\mathrm{eff} = 3600$~K; see \S\ref{sect:v1118ori}; we used as input a lower spectral resolution version of the same PHOENIX stellar atmosphere model as above), the disk
mass ($5 M_{\rm Jup}$) and outer radius ($100$~AU), the disk flaring power ($\propto R^{\beta}$, $\beta = 1.1$) and the disk scale height ($0.01R_\star$ at $R_\star$; the height is thus $2.6$~AU at 100 AU),  and the disk density profile ($\propto R^{-\alpha}$, $\alpha=2.1$). For the dust opacity, we used a model of grains as large as 1~mm for the denser regions of the disk (see \citealt{wood02} for HH 30), but used an interstellar medium grain model for the less dense regions of the disk (i.e., below a threshold set at $3.34 \times 10^{-14}$~\cmcc). {The disk mass and outer radius were not constrained from our optical and infrared data}
\footnote{We have obtained data with the IRAM 30m MAMBO2 bolometer at 1.2~mm in late November 2009. V1118 Ori is detected at $1.8 \pm$ 0.4 mJy. Assuming an optical thin regime, an opacity of $\kappa_\nu = 0.02$ cm$^2$~g$^{-1}$, using a gas-to-dust ratio of 100, we obtain a disk mass of $2.45 \pm 0.5 M_{\rm Jup}$ for a characteristic temperature of 20~K. The disk mass increases to $6.9 \pm 1.5 M_{\rm Jup}$ for $T_c = 10$~K. Although the determination of the exact disk mass is out of scope in this paper, the derived estimates indicate that the assumed disk mass in our model is realistic.}.

Changes in their values would also not affect much the models at optical-infrared wavelengths. We have left the mass accretion rate and inner disk radius (in units of the dust destruction radius, $R_\mathrm{sub}$) free to vary. The model also includes  the possibility for material to fall from the disk onto a hot spot by following the magnetic field lines.  We have kept the truncation radius to $5$~AU, but left the fractional area, $f_S$ {(ratio of the shock area to the star area; see \citealt{calvet98})}, free to vary. Finally, the code samples the model at 10 different inclination angles from $\cos i =0.05$ to $0.95$ every $0.10$, i.e., from $87\degr$ to $18\degr$. Thus, we have four different parameters that can vary ($[\cos i, \dot{M}, f_S, R_{\rm min}$]).

Note that the choice of $\alpha$ and $\beta$ is typical of those observed in T Tauri stars, and correspond to values that seem to fit the V1118 Ori data. 
The power law indices are related in the sense that if the disk surface density is $\Sigma \propto R^{p}$, then $p=\beta-\alpha$.  A value of $\beta = 1.1$ is close to the theoretical values for irradiated and  $\alpha$ disks ($\beta=9/8=1.125$), and, with $\alpha = 2.1$,  $p=\beta - \alpha = -1.0$, i.e., typical for a steady optically thick disk. 
Note that a fit to the IRAC and MIPS fluxes in quiescence gives $\lambda F_\lambda \propto \lambda^{-0.65}$, i.e., consistent with the commonly observed $\lambda F_\lambda \propto \lambda^{-2/3}$ for irradiated disks. Assuming that the temperature follows a power law $T \propto R^{-q}$, and if $\lambda F_\lambda \propto \lambda^{s}$, both indices are related by $s=2/q-4$. Thus, we have $q=0.6$, i.e., $T \propto R^{-3/5}$.

There is no pre-outburst IRS spectrum available to determine the shape of the silicate feature before the outburst (emission or absorption). However, we have used the post-outburst
IRS spectra, scaled by a factor of 0.7 to match the IRAC photometry (see Fig.~\ref{fig:sil}). For the photometry, we have used the median of the pre-outburst (interval \#1) and post-outburst (interval \#6) photometry. For the outburst data,  we have used the plateau (\#4) peak photometric data with the second epoch outburst IRS SL spectrum to characterize the SED at ``peak''. The IRS spectrum already matches well the 8~\mic\  flux without correction and shows the silicate feature in emission.

We ran the code with 5,000,000 ``photons'' or energy packets for each model. This results in SED models with reasonably good signal-to-noise ratios over the wavelength range covered by our data. 
The models were then reddened with $A_V = 1.5$~mag (assuming ISM opacities). For each realization we have calculated a reduced $\chi^2_\mathbf{p}$ value
\begin{equation}
\chi^2_\mathbf{p} = \frac{1}{N} \sum_{i=1}^{N} \left ( \frac{\log_{10} F_\nu(\lambda_i) - \log_{10} M_\nu(\lambda_i,\mathbf{p})}{\Delta (\log_{10} F_\nu(\lambda_i)) } \right )^2
\end{equation}
where $F_\nu(\lambda_i)$ are the data flux densities, $M_\nu(\lambda_i,\mathbf{p})$ are the model flux densities (for parameter vector $\mathbf{p} = [\cos i, \dot{M}, f_S, R_{\rm min}$]) interpolated at the wavelength of the data points 
($0.44, 0.55, 0.71, 0.97, 1.25, 1.60, 2.22, 3.6, 4.5, 5.8, 8.0, 24.0$~\mic\ for the photometry, and $5.5, 6.0, 6.5, \ldots, 14.0$~\mic\  for the IRS -- we only used the SL data).
For the photometric flux density uncertainties, we have used a fixed $\Delta (\log_{10} F_\nu(\lambda_i)) = 0.05$, i.e., about 10\% relative uncertainties for all data points (a value larger than many actual uncertainties on the measured values) to account for flux variations observed before and at the peak of the outburst. We used $\Delta (\log_{10} F_\nu(\lambda_i)) = 0.025$ instead for the IRS SL data to put more weight on the silicate feature.

\begin{figure}[!t]
\includegraphics[angle=0,width=\linewidth]{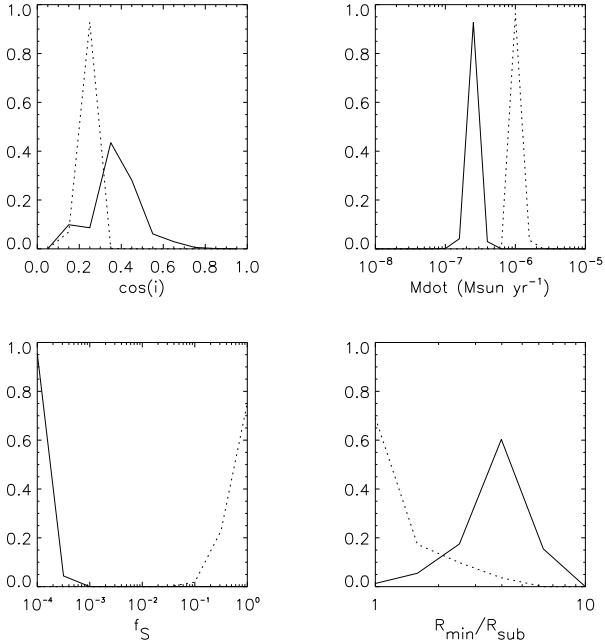}
\caption{Probability distribution of the parameters ($[\cos i, \dot{M}, f_S, R_{\rm min}$]) for the quiescent (solid) and outburst (dotted) cases.
\label{fig:sedprob}}
\end{figure}

Figure~\ref{fig:sedprob} shows the normalized model probabilities (calculated as $p_\mathbf{p} = e^{-\chi^2_\mathbf{p}}$) for the quiescent case (solid), and for the outburst case (dotted). The distributions of inclination angles are slightly different for the quiescent and peak of outburst cases, but the distributions overlap in the range $\cos i = 0.1 - 0.3$, with a maximum probability (obtained by multiplying the probability distributions) at $\cos i =0.25$, i.e., 
$i \approx 76^\circ$. The mass accretion rate distributions are sharply peaked: the most probable values of $\dot{M}$  are $10^{-6.6}$~\msunyr\  and $10^{-6.0}$~\msunyr\  in quiescence and and at the peak of the outburst, respectively, indicating an increase by a factor of 4. The coverage factor seems to be small ($f_S< 0.0001$) outside the outburst while the simulations indicate high values ($0.1 <f_S < 1$) at the peak of the outburst. Note that this parameter is essentially constrained by the optical and near-infrared photometry. The minimum disk radius has likely values of $R_{\rm min}  = (10^{0.4}-10^{0.8}) R_{\rm sub}$, with the most probable value at $R_{\rm min}  \approx 4 R_{\rm sub}$, whereas the values at the peak of the outburst are lower and most likely equal to the sublimation radius.
The best models give sublimation radii of about 0.10 AU and 0.20~AU in quiescence and at the peak of the outburst, implying inner disk radii of $0.25-0.6$~AU (most probable value of 0.4~AU) and 0.2~AU, respectively. This result suggests that the inner disk radius (determined from the infrared data arising from dust emission) probably diminished during the outburst, although we emphasize that quiescent values of 0.2~AU are possible, though less probable from our SED modeling.

\begin{figure*}[!t]
\centering
\includegraphics[angle=0,width=0.475\linewidth]{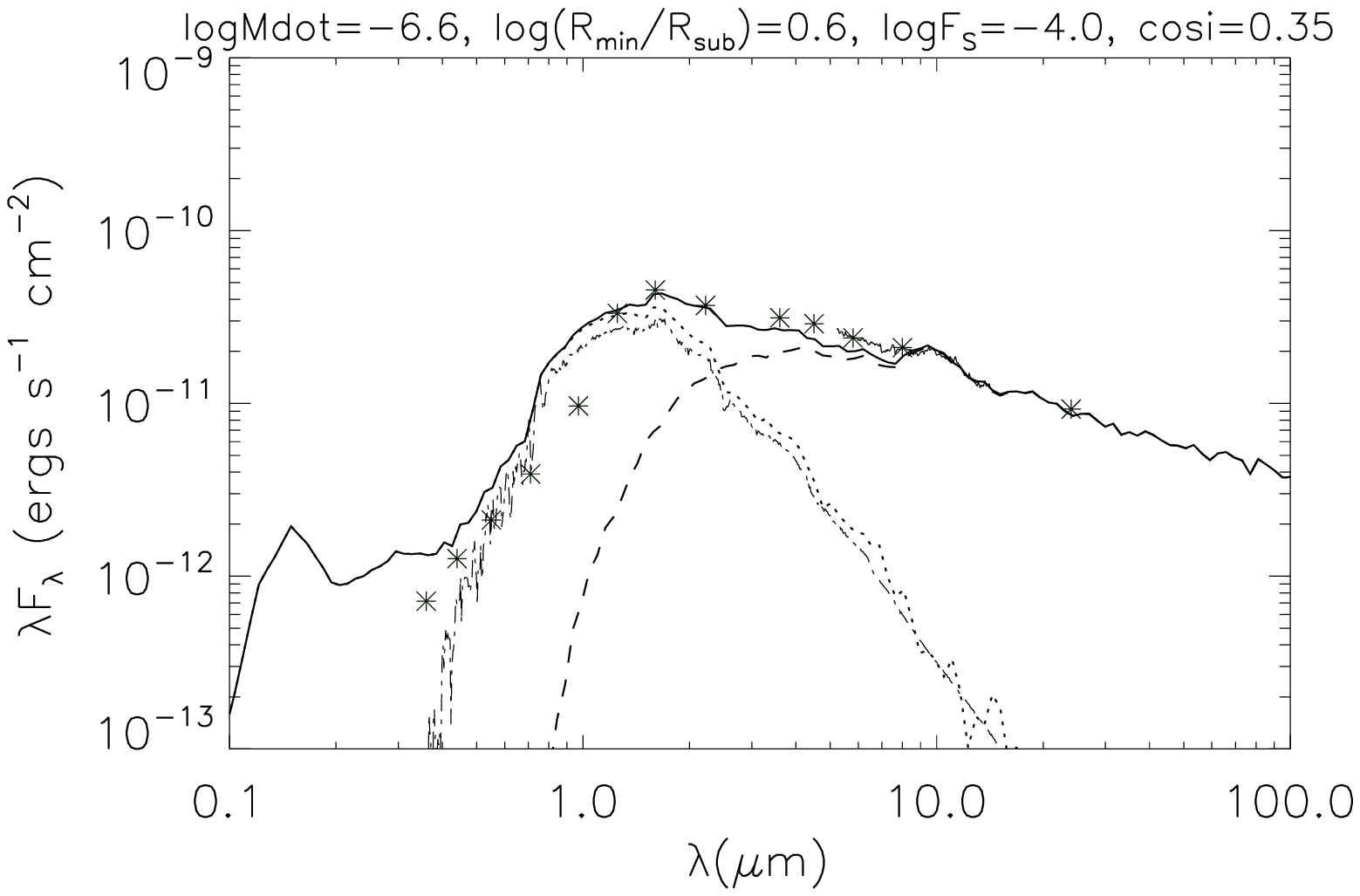}
\hfill
\includegraphics[angle=0,width=0.475\linewidth]{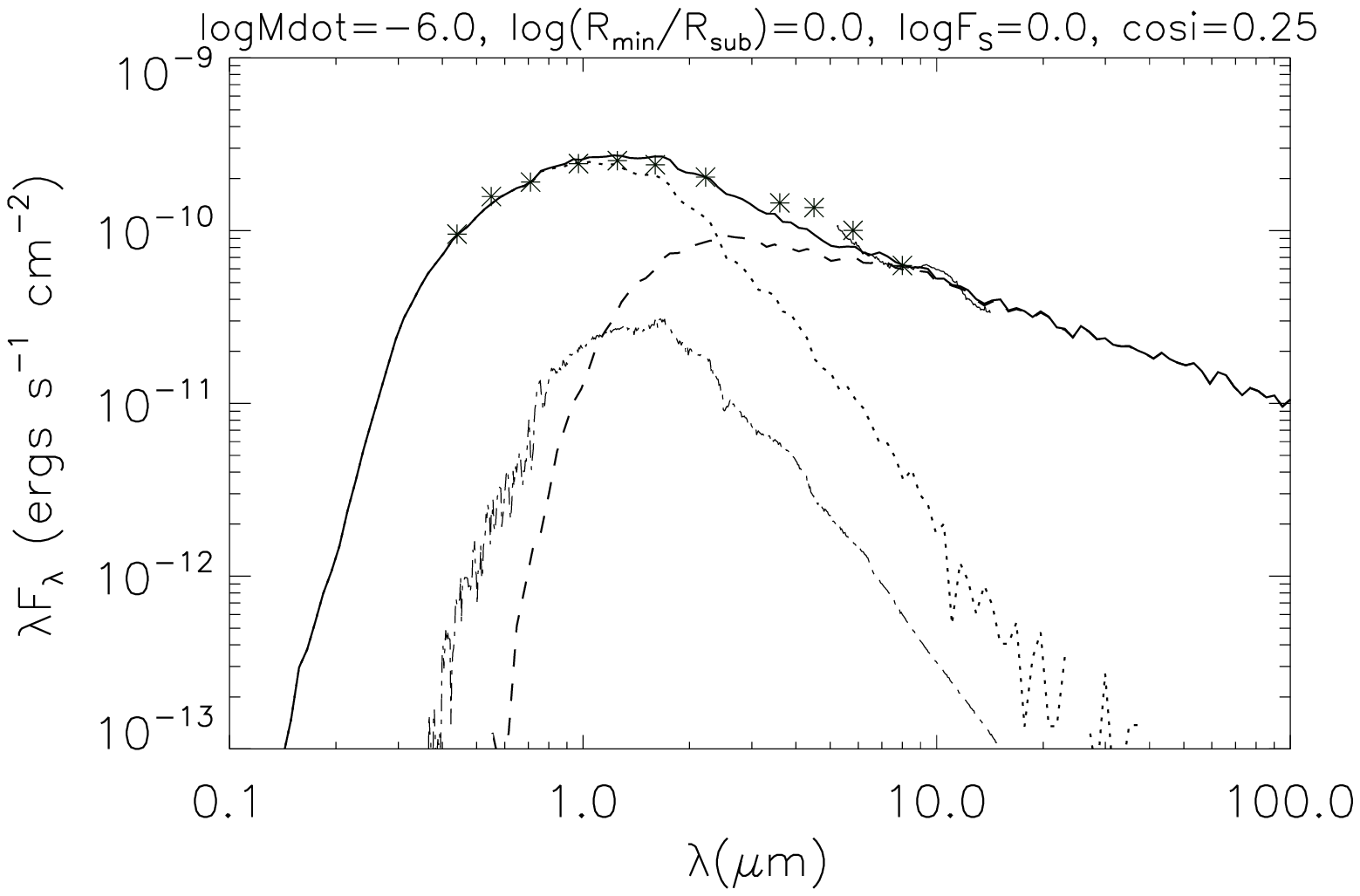}
\caption{The best-fit models to the ``quiescent'' SED (left) and to the peak outburst SED (right). The direct stellar contribution (including emission due to the hot spot) is shown as a dotted line (the $A_V = 1.5$~mag reddened stellar atmosphere model is shown as a dash-dotted line), the thermal disk spectrum as a dashed line.
\label{fig:sedfit}}
\end{figure*}

Figure~\ref{fig:sedfit} shows the best fits to the ``quiescent'' SED and to the peak of outburst SED, respectively. 
The total (bolometric) luminosity (including all components) varied from about $2.0 L_ \odot$ in quiescence (with a contribution of the stellar luminosity, $L=4 \pi R^2 \sigma T^4_{\rm eff}$, of $0.25 L_\odot$ and of the thermal disk luminosity of $0.9 L_\odot$) to $\approx 7.4 L_\odot$   at the peak of the outburst\footnote{{\citet{lorenzetti06} quote a peak luminosity of $25.4 L_\odot$; however, they used the peak magnitudes of a previous outburst to build the outburst peak SED and to derive the luminosity. The previous outburst had a peak magnitude of $J=8.5$, in contrast to the 2005--2006 outburst that peaked at $J=10.6$. In addition, our method derives the luminosity by matching the observed SED with a detailed radiative transfer model, while \citet{lorenzetti06} integrated the flux densities. The two above issues may explain the difference in derived luminosities.}} (the thermal disk luminosity increasing to about $1.7 L_\odot$). 

Thus the bolometric luminosity is dominated by the stellar direct emission, i.e., the stellar photospheric emission and the hotspot emission (the latter in particular during the outburst). We note, however, that we would need data at shorter wavelengths than $B$ to constrain better the hotspot emission during the outburst.

\section{Discussion}
\label{sect:discussion}

Our multi-wavelength campaign allows us to obtain several pieces of information on the outburst of V1118 Ori, as seen in the optical and infrared, and in the X-rays.
{Overall, the evolution of the X-ray flux during the outburst followed the evolution of the optical and near-infrared fluxes.}
The sampling of the X-ray observation does not allow us to determine whether the increase in X-ray flux in March 2006 is related to the short-term increase in optical/infrared flux observed about 45 days earlier. The thermal properties of V1118 Ori showed evidence of a plasma with temperature of a few MK to a few tens of MK. In our initial paper, we reported a change in the spectral properties from a predominantly hot (20 MK) plasma before the outburst (and possibly at the very beginning) to a lower temperature (8 MK) in February 2005 \citep{audard05b}. Our reanalysis confirms this finding and a gradual return to larger values in the later phases of the outburst. However, the low signal-to-noise ratio of the X-ray spectra taken during our monitoring campaign makes it difficult to constrain accurately the plasma temperature (in contrast, the February and March 2005 observations had a much better signal-to-noise ratio). The alternative quantile analysis showed evidence of less variations in the plasma temperature. Nevertheless, it is consistent with the detected change in temperature. While V1647 Ori showed a strong increase in X-ray flux during the outburst \citep{kastner04,grosso05}, V1118 Ori only showed moderate X-ray flux increase, and possibly an eventual decrease in flux after the outburst, if we assume that the 2002 flux level was representative of the pre-outburst X-ray flux. In this case, the impact of the mass infall onto  the stellar magnetosphere must have significantly affected the latter, leaving it in a lower X-ray flux state. It is, however, also possible that the 2002 flux was observed by chance in a high quiescent state, and that the 2007 X-ray flux is closer to the pre-outburst flux level. In any case, the changes in the X-ray fluxes during the outburst, and possibly the temperature changes all indicate a strong interplay between the material falling from the disk due to the increase in mass accretion rate and the stellar magnetosphere.

From the optical and infrared SED data and Monte-Carlo models using the code of \citet{whitney03a,whitney03b}, there is clear evidence that the thermal disk emission 
dominates in the mid-infrared in quiescence and during the outburst, while emission from a hotspot and
reprocessed emission dominates the outburst SED in the optical and near-infrared, and contributes to some extent before the outburst. However, we note that this
conclusion depends on our choice of using disk models that stop at the sublimation radius, i.e., there is no disk contribution at shorter radii, where the
gas is optically thin. This could contribute to the near-infrared fluxes, but it should not contribute significantly to the optical fluxes. Therefore, we believe
that the addition of hotspot emission is necessary to fit the blue side of the V1118 Ori SED, during the outburst and probably before the outburst.

The increase in mass accretion rate during the outburst leads to the change in inner disk radius which moves closer to the star's surface.  Such an effect could
explain the changes in the X-ray emission (see above). From our simulations, we have determined that the mass accretion rate increased during the
outburst: consequently, the dust sublimation radius \emph{increased} from $\approx 17~R_\star = 0.1$~AU to $\approx 32~R_\star = 0.2$~AU, whereas
the inner disk radius (based on the infrared dust emission) moved from $0.4$~AU in quiescence to $0.2$~AU in outburst (although the optically thin gas
disk could be closer to the star), and the inner disk radius is consistent with the dust sublimation radius at the peak of the outburst.
We emphasize that such estimates of the inner disk radius are biased due to our use of dust thermal emission for the disk. The optically thin gas disk is
difficult  to detect and we would require gas tracers, e.g., H$_2$ or CO bandheads to properly determine the inner extent of the disk. In any case, 
the strong increase in fractional area of the hotspot suggests that large amounts of matter fell from the disk onto the star during
the outburst, incidentally interacting with the stellar magnetosphere.

Using their near-infrared spectrum taken in September 2005 (i.e., about at the end of the plateau phase) and a wind loss model, \citet{lorenzetti06}
derived a mass loss rate\footnote{\citet{lorenzetti06} derive the mass loss rate from the wind model of \citet{carr89} and his Figure 8. While their derived mass loss rate is accurate, they incorrectly give a CO luminosity of $6.3 \times 10^{-5} L_\odot$, while it should have been $7.9 \times 10^{-4}L_\odot$ based on the quoted line flux (D. Lorenzetti 2009, priv.~comm.). } of about $4 \times 10^{-8}$~\msunyr\ from \ion{H}{i}. Furthermore, little or no absorption was reported ($A_V < 2$), with no evidence of variations during the outburst. They also derived a mass loss rate of $(3-8) \times 10^{-7}$~\msunyr\  from the CO overtone emission, indicating an ionization fraction of $0.1-0.2$.  
Furthermore, \citet{lorenzetti07} observed V1118 Ori in September 2006, i.e., well after the outburst, and detected no emission line, while the 2.3~\mic\ CO band was seen in \emph{absorption}. Evidence of wind emission in H$\alpha$ (P Cyg profile) was also reported  by \citet{herbig08} in November 2005, in contrast with the more symmetric profile observed after the outburst in December 2006.  It is, therefore, likely that high wind loss rates occurred transiently due to the instability in the disk. 
We note that disk emission may provide a better fit to CO bands in young stars than mass loss emission in general \citep[e.g.,][]{chandler95}, but this may not be the case during the outburst of a young star such as V1118 Ori: our SED models including emission due to matter falling from the accretion disk onto the stellar photosphere clearly show that the outburst SED at wavelengths below $3$\mic\ is dominated by the hotspot emission, not by disk thermal emission (while it remains negligible outside the outburst and stellar photospheric emission dominates). Wind loss emission may thus be present in the optical and near-infrared.

We can compare the mass loss rates derived for the neutral gas by \citet{lorenzetti06} to our {derived mass accretion rates}: the SED modeling suggests a variation in the mass accretion rate from about $2.5 \times 10^{-7}$~\msunyr\  to $1.0 \times 10^{-6}$~\msunyr, i.e., {these mass accretion rates are comparable to the mass loss rates determined by \citet{lorenzetti06}}.
This is somewhat surprising, since mass outflow rates are expected to be about 0.1 times lower than mass accretion rates (\citealt{hartigan95}, with revised mass accretion rates, see \citealt{gullbring98,edwards09}). Our mass accretion rate in quiescence is larger by about a factor of 10 compared to those derived for classical T Tau stars  ($\approx 10^{-8}$\msunyr; e.g., \citealt{gullbring98}), and lower by a factor of 10 than the rates derived for embedded Class I stars ($\approx 5 \times 10^{-6}$\msunyr; \citealt{kenyon93}). Therefore, our derived mass accretion rates are consistent with placing V1118 Ori in a category between Class I and CTTS stars. We note also that the IRAC color indices of V1118 Ori in quiescence are [3.6]--[4.5]=0.65--0.75  and [5.8]--[8.0]=0.65--0.7  which places it formally in the region of Class II sources, but also close to the region for Class 0/I \citep[e.g.,][]{allen04}.
Furthermore, the $K$--[3.6] color of 1.52 in ``quiescence'' also place the star as a strong accretor. In contrast, $U-V = -0.05$ which formally would (incorrectly) place V1118 Ori as a weak-lined T Tauri star. In fact, using the \citet{gullbring98} method to derive mass accretion rates from the $U$-band luminosity excess, we find a mass accretion rate of $4.1 \times 10^{-9}$~\msunyr, much smaller than values derived by SED modeling, and also inconsistent with the strong infrared excess. While our derived mass accretion rates are model-dependent,  we are confident that our derived mass accretion rates are realistic, despite the difficulty to derive mass accretion rates from infrared excess \citep[e.g.,][]{kenyon87}.

Overall, the data provide ample evidence that a significant portion of matter fell from the disk onto the young star due to the increase in mass accretion rate. Over a duration of about  400 days, we estimate about $5 \times 10^{-7}~M_\odot = 0.18 M_\oplus$ was deposited (using an average of half the peak mass accretion rate), i.e., about twice the amount that would have been deposited over the same time range if using the quiescent mass accretion rate
The hotspot covering factor we derive at the peak of the outburst is also relatively large ($f_S > 0.1$), suggesting that a large fraction of the stellar magnetosphere may also have been influenced by the increase in mass accretion rate, which in turn supports the hypothesis that the stellar magnetosphere was modified during the outburst and the X-ray properties of the coronal plasma have changed \citet{audard05b}. Our SED modeling of the dust thermal emission suggests a change of the inner {dust} disk radius {from about 0.4~AU in quiescence to 0.2~AU} at the peak of the outburst (the optically thin gas could reach down to 0.2~AU or below before the outburst, although it would be unclear why dust would not go down to the sublimation radius, unless, e.g., disk clearing by a giant planet or significant grain growth making the dust optically thin would be present in the inner portion of the disk). Together with the increase in $f_S$, this provides evidence that the disk closed in during the outburst, disrupted the magnetosphere, and matter fell onto the stellar photosphere, producing strong hotspot emission that dominated the optical and near-infrared in outburst (while the thermal disk emission increased by a factor of 4 and dominated the mid-infrared emission).

While color-color and color-magnitude diagrams can be used in young stars to determine the level of reddening, such diagrams  might be difficult to use to measure any reddening during the outburst of a young star: in addition to reddening, the whole SED of a young star changes dramatically, increasing in flux, but also in shape: in the case of V1118 Ori, the optical and near-infrared fluxes increased much more than the mid-infrared spectrum. The former are dominated by stellar and hotspot emission, while the latter is dominated by thermal disk emission. Color-color diagrams show a clear blueing of the colors (Fig.~\ref{fig:colindcolind}), not due to clearing of any enshrouding material, but due to the strong increase in flux below $1.5$\mic\  (Fig.~\ref{fig:pseudospec2}). Such multi-wavelength data show the importance that a good coverage of the SED is necessary to understand the disk-star interactions in young stars.
Furthermore, Figures \ref{fig:colindmag} and \ref{fig:colindcolind} indicate that V1118 Ori returned (within uncertainties) to its ``initial'', pre-outburst conditions.

\section{Conclusions}
\label{sect:concl}

We have followed the young accreting star V1118 Ori before, during, and after its outburst in 2005--2006 with photometry in 
$UBVRIJHK$ and in X-rays with \xmm\ and \cxc.  We have further obtained \spitz\ photometry and spectroscopy during the outburst and after the outburst. 

The optical and infrared data showed significant variations: the data below $1$~\mic\ brightened by as much as  about $3-4$~mag, while the
infrared data brightened only up to $\approx 2$~mag. Monte-Carlo simulations of a star+disk+hotspot model suggested that the optical data were dominated by
hotspot emission at the peak of the outburst (while the stellar photosphere dominated before the outburst), while thermal disk emission
from optically thick dust dominated the infrared. The SED analysis showed an increase in mass accretion rate from $\dot{M} =2.5 \times 10^{-7}$~\msunyr\  in quiescence to
$\dot{M} = 1.0 \times 10^{-6}$~\msunyr\  at the peak of the outburst, and an increase of the hotspot fractional coverage area from $f_S \leq 10^{-4}$ to $f_S > 0.1$. The SED modeling suggests that the dust sublimation radius increased from $0.1$ to $0.2$~AU. Based on the thermal dust emission, the inner disk radius moved from about 4 times the dust sublimation radius (0.4~AU) to 0.2~AU ($=R_\mathrm{sub}$) at the peak of the outburst. But the gas inner disk radius could extend below the dust sublimation radius.
The inclination angle of the disk is constrained to values of $\cos i = 0.1 - 0.3$ with a peak at $0.25$, i.e., $i \approx 76^\circ$.

The \spitz\ IRS spectra showed the silicate feature at 10~\mic\ in emission during and after the outburst, with no apparent variation in shape or flux, indicating
that the outburst had little impact on the optically thin region where the silicate emission is originating. Emission lines were detected in the high-resolution module data
although the background subtraction proved to be difficult, especially for the outburst spectra. Follow-up observations, e.g., from the ground should confirm
the detection of lines of particular interest, e.g., [\ion{Ne}{ii}], [\ion{S}{iii}], and H$_2$ $0-0$.

The initial X-ray observations (Jan--Mar 05) indicated a change in the thermal structure of the coronal spectrum of V1118 Ori  {with little change in the X-ray flux \citep{audard05b}, while fading in the X-rays was already detected in September 2005
\citep{lorenzetti06}.} Our continuing monitoring with \xmm\ and \cxc\ provided the complete picture, confirmed the low plasma temperature measured in February 2005, and showed that V1118 Ori's X-ray flux was strongly correlated with the optical and infrared fluxes, although intrinsic variability (due to magnetic activity) was also observed. Compared to the 2002 flux, the increase in X-ray flux was not large during the outburst and the post-outburst X-ray observation in late 2007 showed a lower flux level than observed serendipitously in 2002, which could either indicate that the outburst significantly impacted the stellar magnetosphere to leave it in a lower state than before the outburst, or that the 2002 observation caught V1118 Ori in a quiescent high state. The X-ray spectra showed evidence of thermal variations of the corona during the outburst, but the signal-to-noise ratio of our data was not large (except in March 2006 during which we could detect a second temperature component, probably due to the increase in flux at the end of the observation, {likely contaminated by a flare}). After the apparent change to a cooler plasma temperature in the early phase of the outburst, we argue that the thermal properties of V1118 Ori's corona returned to its ``normal'', hotter values before the end of the outburst, but that the continuing mass infall affected significantly the magnetospheric configuration.

Our observations have shown the interplay between disk and stellar magnetosphere in young accreting stars, as demonstrated by the changes in the X-ray plasma emission due to the strong increase in mass infall from the disk onto the stellar photosphere. While our data do not directly help to determine the origin of the outbursts (thermal disk instabilities or instabilities due to close companions or giant planets in the disk), it is interesting that \citet{reipurth07} identified V1118 Ori as a close visual binary. Future observations should, ideally, take into account this binarity to determine which star is actually outbursting.

\begin{acknowledgements}

We acknowledge support by NASA through \cxc\ award DD5-6029X and through \xmm\ award NNG05GI96G
to Columbia University,  SAO GO6-7005D, SAO GO8-9020X, and NNX07AI37G to the University of Colorado.
M.~A. acknowledges support from a Swiss National Science Foundation Professorship (PP002--110504). 
G.~S. Stringfellow further acknowledged support from NASA programs NNX06AG44G, NNX07AT30G,
and \spitz\ through JPL contract 1265288.
The \cxc\ X-ray Observatory Center is operated by the Smithsonian Astrophysical 
Observatory for and on behalf of NASA under contract NAS8-03060. Based on observations obtained with 
\xmm, an ESA science mission with instruments and contributions directly funded by ESA Member States and NASA. 
This work is based in part on observations made with the \spitz\  Space Telescope, which is operated by the Jet Propulsion Laboratory, California Institute of Technology under a contract with NASA. Support for this work was provided by NASA.
Stony Brook's participation in SMARTS is made possible by support from the 
offices of the Provost and the Vice President for Research. We thank J. Allyn Smith, P. McGehee, 
J. Espinoza, and D. Gonzalez for doing the observations with the SMARTS telescopes.
We also thank H. Tannanbaum and N. Schartel
for granting \cxc\  DDT and \xmm\ TOO time to observe V1118 Ori.
We also thank N. Grosso for discussions on quantiles and outbursting sources, A. Carmona and F. Fontani for useful comments on the manuscript.
{Finally, we thank the referee, D. Lorenzetti, for comments that clarified aspects of the manuscript.}
\end{acknowledgements}


\clearpage

\Online
\appendix

\section{Details on the SMARTS data reduction}
\label{app:smarts}

\begin{table*}[t]
\centering
\caption{Comparison stars for the SMARTS photometry\label{tab:smartsref}}
\begin{tabular}{lrrrrll}
\hline\hline
 \noalign{\vskip .8ex}%
\makebox[50mm][c]{Comparison} & \multicolumn{1}{c}{$B$}  & \multicolumn{1}{c}{$V$}  & \multicolumn{1}{c}{$R$}  & \multicolumn{1}{c}{$I$} & \multicolumn{1}{c}{$\delta V$} & \multicolumn{1}{c}{$\delta I$}\\
 \noalign{\vskip .8ex}%
\hline\\
 \noalign{\vskip -2ex}%
IV Ori\dotfill       & 16.19$\pm$0.35 & 14.94$\pm$0.26& 13.97$\pm$0.22& 13.19$\pm$0.13 & \nodata &  \nodata \\
CSV 6218\dotfill     & 17.40$\pm$0.13 & 15.90$\pm$0.13& 14.74$\pm$0.13& 13.68$\pm$0.15& 0.20 &  0.36\\
V759 Ori\dotfill     & 16.81$\pm$0.12 & 15.44$\pm$0.09& 14.32$\pm$0.10& 13.14$\pm$0.13& 0.10 & 0.24\\
NGC 1977 311\dotfill & 14.32$\pm$0.11 & 13.24$\pm$0.09& 12.46$\pm$0.09& 11.73$\pm$0.12&  \nodata &  \nodata\\
V1175 Ori\dotfill    & 16.59$\pm$0.10 & 15.19$\pm$0.07& 14.19$\pm$0.09& 13.26$\pm$0.12& 0.08 & 0.14\\
JW 94\dotfill        & 20.82$\pm$0.35 & 19.10$\pm$0.17& 17.37$\pm$0.10& 15.45$\pm$0.13&0.10 & 0.29 \\
JW 100\dotfill       & 20.25$\pm$0.32& 18.48$\pm$0.15& 17.09$\pm$0.10&  15.44$\pm$0.12& 0.03 & 0.23\\
Par 1606\dotfill     & 17.18$\pm$0.10 & 15.61$\pm$0.07& 14.38$\pm$0.08& 13.22$\pm$0.12& 0.03 & 0.12\\
CXORRS J053447.9-053544\dotfill&  22.00$\pm$0.99&  19.79$\pm$0.32& 17.76$\pm$0.13&  15.79$\pm$0.12& 0.67 & 0.23\\
NSV2246\dotfill      & 11.32$\pm$0.26 & 10.30$\pm$0.17&  9.78$\pm$0.16&  9.25$\pm$0.13& 0.08 & 0.06\\
JW 33\dotfill        & 18.75$\pm$0.32 & 17.30$\pm$0.10& 16.16$\pm$0.09& 14.84$\pm$0.12& -0.03 & 0.15\\
\hline
 \noalign{\vskip .8ex}%
\multicolumn{7}{l}{Notes: $\delta V$ and $\delta I$ are the difference between these $V$, $I$
determinations and those published by \citet{hillenbrand97}.}
\end{tabular}
\end{table*}

{The ANDICAM optical detector is a Fairchild 447 2048$\times$2048 CCD with
30$\mu$m pixels. 
It is used in 2$\times$2 binning mode, yielding a nominal plate scale of
0.369 arcsec pixel$^{-1}$ and about a $6\arcmin \times 6\arcmin$  field. Readnoise is
6.5 e$^-$ RMS; gain is 2.3 e$^-$/DN. It is read with a single amplifier.
Filters are Johnson $B$ and $V$ and Kron-Cousins $R$ and $I$. }
We generally took 3 exposures in $B$ and $V$, and 2 images in $R_C$ and $I_C$,
and co-added these prior to analysis. {The IR channel detector is a Rockwell 1024$^2$ HgCdTe ``Hawaii'' array with
18 $\mu$m pixels binned 2$\times$2 to yield a $0\farcs 274$ pixel$^{-1}$
plate scale. The IR channel field is $2\farcm 4 \times 2\farcm 4$. The filters we used are
standard CIT/CTIO $JHK$.}  We flatten the images using
dome flats obtained every 3 nights.
We observe using a 3-point dither pattern and a $40\arcsec$ throw,
and use the median of the unshifted images as the local sky.

\subsection{Differential Photometry}

We initially determine relative magnitudes using differential photometry.
In the optical, we selected the 11 brightest stars in the $\approx 6\arcmin$
field of view (Table~\ref{tab:smartsref}; V372~Ori is overexposed).
Instrumental magnitudes are determined for the target and all the comparison stars, by summing the counts in a 5
pixel ($1\farcs 85$) aperture. Background is the median level in an annulus from
10-20 pixel ($3\farcs 69-7\farcs 38$) radius.
Since V1118~Ori is in a region of complex nebulosity,
we also tried modelling the background by taking radial cuts through the 
image and extrapolating the level at the position of the target. This affected
the magnitudes at less than 5\% at $B$, and even less at longer wavelengths.

In the near-infrared, we use 6 comparison stars in the field (again excluding
V372~Ori). We use a 5
pixel ($1\farcs 37$) aperture. Background is the median level in an annulus from
10-20 pixel ($2\farcs 74-5\farcs 48$) radius.

We construct a single comparison, the ``superstar" from a weighted
sum of all the
comparison stars. The differential magnitude is the difference between the
instrumental magnitudes of the target and the ``superstar".

As it happens, most of the stars in this field are known or suspected variables.
Fortunately, none are as variable as V1118~Ori.
We assume that, by summing all the
comparisons, the ``superstar" will be less variable than any single comparison.
We have constructed light curves using subsets of the comparison stars, and do
not see
any gross effects that can be attributed to variations of the comparisons.
Nonetheless we have not incorporated those stars with measured {standard deviations}
after absolute calibration (see below) 
$>$0.10~mag at $V$ into the optical ``superstar''.
We caution that there is no certainty that the comparison is
stable at the few percent level.

\subsection{Optical Absolute Calibration}
Although the SMARTS operations are not optimized for absolute photometry,
single observations are made of Landolt standard fields on each night
judged photometric to establish the photometric zero point. These can be used
to determine apparent magnitudes with an accuracy of better than 10\%.

On each night on which we have both observations of the target and of a
photometric standard field, we compute the zero-point offset (apparent
magnitude - instrumental magnitude).
We assume the standard atmospheric extinction correction for Cerro Tololo. 
Since the standard field
is observed only once per night, we cannot account for changes in transparency
through the night. We do not solve for reddening terms, so the photometric
solution consists of only the zero-point. 

We use the zero-point correction to determine the apparent magnitudes of all
the comparison stars in the field. We could do this on 71 nights. 
In doing so, we confirm that all the comparison stars are variable (Table~\ref{tab:smartsref}).
We compared these magnitudes to the $V, I_C$ magnitudes published by
\citet{hillenbrand97}. The median offsets are 0.08 and 0.23~mag at $V$ and $I_C$,
respectively. Given that all the stars are
variable members of the Orion Nebula population, this is acceptable agreement.

We apply this absolute calibration on all nights (including non-photometric
nights) to determine the apparent magnitudes of the target.
There are 14 nights where we overlap with the Villanova photometry.
On these nights there are systematic offsets, in the sense (SMARTS-Villanova),
of 0.16$\pm$0.10, 0.24$\pm$0.04, and 0.34$\pm$0.04 mag in
$V$, $R_C$, and $I_C$,
respectively. The cause for the systematic offsets may be the lack of a 
color term in the photometric solution, exacerbated by the fact
that most of our comparisons have $V-I_C$ colors $>$2~mag.
{To allow comparison between the different telescope data,
we have applied the above systematic offsets to the SMARTS magnitudes} and present
the nightly average data in Table~\ref{tab:smartsmag} (available online only).

\subsection{Near-infrared  Absolute Calibration}

For the $JHK$ absolute calibration, we compare the instrumental magnitudes
directly to the 2MASS magnitudes of the
6 comparison stars. To check for target variability, we have compared these
magnitudes to those published in the DENIS catalog, and by \citet{ali95} and
\citet{hillenbrand98}. Two of the six comparisons (2MASS J05344159-0534249
and 2MASS J05344219-0533036 appear variable, with variances $>$0.2~mag at $K$
(4 observations), and 2 others (JW~94 and JW~100) appear variable in the
optical (Table~\ref{tab:smartsref}).

Our ability to absolutely calibrate the photometry is ultimately limited by the
stability of the zero-point. We trust in regression to the mean, that by using
at the possible comparisons, on average the zero-point will be {more} stable that
determined from any single comparison. 
We do not include the systematic uncertainty in the zero-point in our error budgets. 
Overall, the $JHK$ fluxes compare well to those published by \citet{lorenzetti07}.
However, we have added a systematic correction of $+0.1$~mag for $J$ and $+0.2$~mag for $K$ to match
their photometry. {V1118 Ori is intrinsically variable, as are the comparison stars. The   
only reason for applying the systematic offsets to our data are to simplify 
comparison graphically with the Lorenzetti et al. data. We note that the offsets do not change
significantly the results of this paper, they merely shift the near-infrared data points by $0.1$ to $0.2$~mag, e.g.,  in Figures~\ref{fig:colindmag} and \ref{fig:colindcolind}.}

{\small
\begin{longtable}{crrrrrrrr}
\caption{SMARTS average nightly magnitudes.\label{tab:smartsmag}}\\
\hline\hline
 \noalign{\vskip .8ex}%
MJD 		& \multicolumn{1}{c}{$U$} & \multicolumn{1}{c}{$B$} & \multicolumn{1}{c}{$V$} & \multicolumn{1}{c}{$R$} & \multicolumn{1}{c}{$I$} & \multicolumn{1}{c}{$J$} & \multicolumn{1}{c}{$H$} & \multicolumn{1}{c}{$K$} \\
 \noalign{\vskip .8ex}%
\hline\\
 \noalign{\vskip -2ex}%
\endfirsthead
\caption{continued.}\\
\hline\hline
 \noalign{\vskip .8ex}%
MJD 		& \multicolumn{1}{c}{$U$} & \multicolumn{1}{c}{$B$} & \multicolumn{1}{c}{$V$} & \multicolumn{1}{c}{$R$} & \multicolumn{1}{c}{$I$} & \multicolumn{1}{c}{$J$} & \multicolumn{1}{c}{$H$} & \multicolumn{1}{c}{$K$} \\
 \noalign{\vskip .8ex}%
\hline\\
 \noalign{\vskip -2ex}%
\endhead
\hline
\endfoot
53012 & \multicolumn{1}{c}{\nodata} & \multicolumn{1}{c}{\nodata} & \multicolumn{1}{c}{\nodata} & \multicolumn{1}{c}{\nodata} & \multicolumn{1}{c}{\nodata} & $12.74\pm 0.07$ & $11.68\pm 0.03$ & $11.15\pm 0.04$ \\
53031 & \multicolumn{1}{c}{\nodata} & \multicolumn{1}{c}{\nodata} & \multicolumn{1}{c}{\nodata} & \multicolumn{1}{c}{\nodata} & \multicolumn{1}{c}{\nodata} & $12.72\pm 0.06$ & $11.38\pm 0.03$ & $11.06\pm 0.03$ \\
53038 & $18.17\pm 0.05$ & $18.37\pm 0.03$ & $17.45\pm 0.02$ & $16.32\pm 0.01$ & $14.93\pm 0.01$ & \multicolumn{1}{c}{\nodata} & \multicolumn{1}{c}{\nodata} & \multicolumn{1}{c}{\nodata} \\
53058 & $18.67\pm 0.11$ & \multicolumn{1}{c}{\nodata} & \multicolumn{1}{c}{\nodata} & \multicolumn{1}{c}{\nodata} & \multicolumn{1}{c}{\nodata} & \multicolumn{1}{c}{\nodata} & \multicolumn{1}{c}{\nodata} & $11.32\pm 0.03$ \\
53062 & $18.41\pm 0.08$ & $18.41\pm 0.03$ & $17.61\pm 0.02$ & $16.32\pm 0.02$ & $14.91\pm 0.01$ & \multicolumn{1}{c}{\nodata} & \multicolumn{1}{c}{\nodata} & \multicolumn{1}{c}{\nodata} \\
53092 & $18.32\pm 0.07$ & $18.35\pm 0.03$ & $17.47\pm 0.02$ & $16.18\pm 0.01$ & $14.86\pm 0.01$ & $12.54\pm 0.07$ & $11.23\pm 0.03$ & $11.02\pm 0.04$ \\
53103 & $17.70\pm 0.03$ & $17.91\pm 0.02$ & $17.06\pm 0.01$ & $15.98\pm 0.01$ & $14.72\pm 0.01$ & \multicolumn{1}{c}{\nodata} & \multicolumn{1}{c}{\nodata} & \multicolumn{1}{c}{\nodata} \\
53381 & \multicolumn{1}{c}{\nodata} & $15.62\pm 0.01$ & $14.70\pm 0.01$ & $13.78\pm 0.01$ & $12.89\pm 0.01$ & $12.23\pm 0.05$ & $11.12\pm 0.03$ & $10.56\pm 0.03$ \\
53382 & \multicolumn{1}{c}{\nodata} & $15.68\pm 0.01$ & $14.78\pm 0.01$ & $13.87\pm 0.02$ & $12.93\pm 0.01$ & $12.27\pm 0.05$ & $11.17\pm 0.03$ & $10.64\pm 0.03$ \\
53383 & \multicolumn{1}{c}{\nodata} & $15.61\pm 0.01$ & $14.70\pm 0.01$ & $13.79\pm 0.01$ & $12.87\pm 0.01$ & $11.81\pm 0.05$ & $11.12\pm 0.03$ & $10.53\pm 0.03$ \\
53384 & \multicolumn{1}{c}{\nodata} & $15.92\pm 0.05$ & $14.94\pm 0.02$ & $13.97\pm 0.02$ & $13.09\pm 0.03$ & \multicolumn{1}{c}{\nodata} & \multicolumn{1}{c}{\nodata} & \multicolumn{1}{c}{\nodata} \\
53385 & \multicolumn{1}{c}{\nodata} & $15.25\pm 0.01$ & $14.30\pm 0.01$ & $13.40\pm 0.01$ & $12.56\pm 0.01$ & $11.71\pm 0.04$ & $10.86\pm 0.02$ & $10.35\pm 0.03$ \\
53394 & \multicolumn{1}{c}{\nodata} & $14.36\pm 0.02$ & $13.48\pm 0.01$ & $12.71\pm 0.01$ & $11.90\pm 0.01$ & $11.18\pm 0.03$ & $10.43\pm 0.02$ & $9.85\pm 0.02$ \\
53395 & \multicolumn{1}{c}{\nodata} & $14.04\pm 0.02$ & $13.09\pm 0.01$ & $12.35\pm 0.01$ & $11.59\pm 0.01$ & $11.23\pm 0.03$ & $10.26\pm 0.02$ & $10.0\pm 0.02$ \\
53396 & \multicolumn{1}{c}{\nodata} & $14.32\pm 0.02$ & $13.41\pm 0.01$ & $12.59\pm 0.01$ & $11.80\pm 0.01$ & $11.07\pm 0.03$ & $10.30\pm 0.02$ & $9.77\pm 0.02$ \\
53397 & \multicolumn{1}{c}{\nodata} & $14.33\pm 0.01$ & $13.40\pm 0.01$ & $12.60\pm 0.01$ & $11.81\pm 0.01$ & $11.41\pm 0.03$ & $10.41\pm 0.02$ & $9.94\pm 0.02$ \\
53398 & \multicolumn{1}{c}{\nodata} & $14.51\pm 0.01$ & $13.58\pm 0.01$ & $12.76\pm 0.01$ & $11.96\pm 0.01$ & $11.23\pm 0.03$ & $10.41\pm 0.02$ & $9.97\pm 0.02$ \\
53399 & \multicolumn{1}{c}{\nodata} & $14.44\pm 0.01$ & $13.52\pm 0.01$ & $12.70\pm 0.01$ & $11.91\pm 0.01$ & $11.39\pm 0.03$ & $10.58\pm 0.02$ & $10.01\pm 0.02$ \\
53400 & \multicolumn{1}{c}{\nodata} & $14.66\pm 0.02$ & $13.75\pm 0.01$ & $12.95\pm 0.01$ & $12.07\pm 0.01$ & $11.81\pm 0.04$ & $10.60\pm 0.02$ & $9.90\pm 0.03$ \\
53401 & \multicolumn{1}{c}{\nodata} & $14.31\pm 0.05$ & $13.42\pm 0.02$ & $12.60\pm 0.02$ & $11.90\pm 0.02$ & $11.25\pm 0.05$ & $10.17\pm 0.03$ & $9.78\pm 0.03$ \\
53405 & \multicolumn{1}{c}{\nodata} & $14.73\pm 0.02$ & $13.75\pm 0.02$ & $12.94\pm 0.01$ & $12.12\pm 0.01$ & $11.72\pm 0.04$ & $10.64\pm 0.02$ & $10.09\pm 0.03$ \\
53407 & \multicolumn{1}{c}{\nodata} & $14.24\pm 0.01$ & $13.29\pm 0.01$ & $12.51\pm 0.01$ & $11.74\pm 0.01$ & $11.14\pm 0.03$ & $10.32\pm 0.02$ & $9.87\pm 0.02$ \\
53408 & \multicolumn{1}{c}{\nodata} & $14.21\pm 0.02$ & $13.35\pm 0.02$ & $12.59\pm 0.01$ & $11.84\pm 0.01$ & $11.17\pm 0.04$ & $10.41\pm 0.02$ & $9.90\pm 0.03$ \\
53411 & \multicolumn{1}{c}{\nodata} & $14.31\pm 0.02$ & $13.38\pm 0.02$ & $12.60\pm 0.01$ & $11.79\pm 0.01$ & $11.21\pm 0.03$ & $10.39\pm 0.02$ & $10.01\pm 0.03$ \\
53413 & \multicolumn{1}{c}{\nodata} & $14.30\pm 0.02$ & $13.32\pm 0.02$ & $12.50\pm 0.01$ & $11.69\pm 0.01$ & $11.23\pm 0.03$ & $10.34\pm 0.02$ & $9.92\pm 0.03$ \\
53414 & \multicolumn{1}{c}{\nodata} & $14.24\pm 0.02$ & $13.38\pm 0.02$ & $12.61\pm 0.01$ & $11.81\pm 0.01$ & $11.06\pm 0.03$ & $10.38\pm 0.02$ & $9.93\pm 0.03$ \\
53416 & \multicolumn{1}{c}{\nodata} & $14.17\pm 0.03$ & $13.47\pm 0.01$ & $12.71\pm 0.01$ & $11.95\pm 0.01$ & \multicolumn{1}{c}{\nodata} & \multicolumn{1}{c}{\nodata} & \multicolumn{1}{c}{\nodata} \\
53417 & \multicolumn{1}{c}{\nodata} & $14.09\pm 0.03$ & $13.25\pm 0.01$ & $12.52\pm 0.01$ & $11.77\pm 0.01$ & \multicolumn{1}{c}{\nodata} & \multicolumn{1}{c}{\nodata} & \multicolumn{1}{c}{\nodata} \\
53418 & \multicolumn{1}{c}{\nodata} & $14.08\pm 0.03$ & $13.25\pm 0.01$ & $12.49\pm 0.01$ & $11.75\pm 0.01$ & \multicolumn{1}{c}{\nodata} & \multicolumn{1}{c}{\nodata} & \multicolumn{1}{c}{\nodata} \\
53419 & \multicolumn{1}{c}{\nodata} & $14.25\pm 0.03$ & \multicolumn{1}{c}{\nodata} & $12.65\pm 0.01$ & $11.87\pm 0.01$ & \multicolumn{1}{c}{\nodata} & \multicolumn{1}{c}{\nodata} & \multicolumn{1}{c}{\nodata} \\
53420 & \multicolumn{1}{c}{\nodata} & $14.24\pm 0.03$ & $13.42\pm 0.01$ & $12.64\pm 0.01$ & $11.89\pm 0.01$ & \multicolumn{1}{c}{\nodata} & \multicolumn{1}{c}{\nodata} & \multicolumn{1}{c}{\nodata} \\
53421 & \multicolumn{1}{c}{\nodata} & $13.77\pm 0.04$ & $13.04\pm 0.02$ & $12.31\pm 0.01$ & $11.60\pm 0.01$ & $10.92\pm 0.03$ & $10.21\pm 0.02$ & $9.68\pm 0.03$ \\
53422 & \multicolumn{1}{c}{\nodata} & $13.78\pm 0.03$ & $12.88\pm 0.03$ & $12.15\pm 0.01$ & $11.42\pm 0.01$ & $10.69\pm 0.03$ & $10.04\pm 0.02$ & $9.58\pm 0.03$ \\
53423 & \multicolumn{1}{c}{\nodata} & $13.85\pm 0.11$ & $13.07\pm 0.05$ & $12.27\pm 0.04$ & $11.58\pm 0.03$ & \multicolumn{1}{c}{\nodata} & \multicolumn{1}{c}{\nodata} & \multicolumn{1}{c}{\nodata} \\
53424 & \multicolumn{1}{c}{\nodata} & $14.10\pm 0.05$ & $13.33\pm 0.02$ & $12.54\pm 0.01$ & $11.76\pm 0.01$ & \multicolumn{1}{c}{\nodata} & \multicolumn{1}{c}{\nodata} & \multicolumn{1}{c}{\nodata} \\
53425 & \multicolumn{1}{c}{\nodata} & $13.90\pm 0.05$ & $13.14\pm 0.02$ & $12.39\pm 0.01$ & $11.69\pm 0.01$ & \multicolumn{1}{c}{\nodata} & \multicolumn{1}{c}{\nodata} & \multicolumn{1}{c}{\nodata} \\
53426 & \multicolumn{1}{c}{\nodata} & $13.85\pm 0.05$ & $13.13\pm 0.02$ & $12.39\pm 0.01$ & $11.67\pm 0.01$ & \multicolumn{1}{c}{\nodata} & \multicolumn{1}{c}{\nodata} & \multicolumn{1}{c}{\nodata} \\
53427 & \multicolumn{1}{c}{\nodata} & $13.84\pm 0.04$ & $13.22\pm 0.01$ & $12.48\pm 0.01$ & $11.75\pm 0.01$ & \multicolumn{1}{c}{\nodata} & \multicolumn{1}{c}{\nodata} & \multicolumn{1}{c}{\nodata} \\
53428 & \multicolumn{1}{c}{\nodata} & $13.55\pm 0.06$ & $12.95\pm 0.03$ & $12.24\pm 0.02$ & $11.55\pm 0.01$ & $11.18\pm 0.03$ & $10.23\pm 0.01$ & $9.67\pm 0.03$ \\
53429 & \multicolumn{1}{c}{\nodata} & $14.05\pm 0.03$ & $13.29\pm 0.01$ & $12.54\pm 0.01$ & $11.78\pm 0.01$ & \multicolumn{1}{c}{\nodata} & \multicolumn{1}{c}{\nodata} & \multicolumn{1}{c}{\nodata} \\
53430 & \multicolumn{1}{c}{\nodata} & $13.83\pm 0.02$ & $13.06\pm 0.02$ & $12.31\pm 0.01$ & $11.60\pm 0.01$ & $11.23\pm 0.03$ & $10.20\pm 0.01$ & $9.80\pm 0.02$ \\
53431 & \multicolumn{1}{c}{\nodata} & $13.89\pm 0.02$ & $12.97\pm 0.01$ & $12.24\pm 0.01$ & $11.55\pm 0.01$ & \multicolumn{1}{c}{\nodata} & \multicolumn{1}{c}{\nodata} & \multicolumn{1}{c}{\nodata} \\
53432 & \multicolumn{1}{c}{\nodata} & $13.74\pm 0.03$ & $13.01\pm 0.01$ & $12.27\pm 0.01$ & $11.54\pm 0.01$ & \multicolumn{1}{c}{\nodata} & \multicolumn{1}{c}{\nodata} & \multicolumn{1}{c}{\nodata} \\
53433 & \multicolumn{1}{c}{\nodata} & $13.60\pm 0.02$ & $12.84\pm 0.01$ & $12.11\pm 0.01$ & $11.41\pm 0.01$ & \multicolumn{1}{c}{\nodata} & \multicolumn{1}{c}{\nodata} & \multicolumn{1}{c}{\nodata} \\
53458 & \multicolumn{1}{c}{\nodata} & $13.83\pm 0.03$ & $12.90\pm 0.02$ & $12.15\pm 0.01$ & $11.41\pm 0.01$ & $10.74\pm 0.03$ & $9.83\pm 0.01$ & $9.38\pm 0.02$ \\
53460 & \multicolumn{1}{c}{\nodata} & $13.95\pm 0.02$ & $13.03\pm 0.02$ & $12.26\pm 0.01$ & $11.49\pm 0.01$ & $10.85\pm 0.03$ & $10.04\pm 0.01$ & $9.39\pm 0.02$ \\
53463 & \multicolumn{1}{c}{\nodata} & $13.95\pm 0.03$ & $13.01\pm 0.03$ & $12.21\pm 0.01$ & $11.45\pm 0.01$ & $10.67\pm 0.03$ & $9.98\pm 0.01$ & $9.30\pm 0.02$ \\
53464 & \multicolumn{1}{c}{\nodata} & $13.99\pm 0.02$ & $13.04\pm 0.02$ & $12.27\pm 0.01$ & $11.51\pm 0.01$ & $10.60\pm 0.03$ & $9.93\pm 0.01$ & $9.43\pm 0.02$ \\
53467 & \multicolumn{1}{c}{\nodata} & $13.67\pm 0.02$ & $12.79\pm 0.02$ & $12.03\pm 0.01$ & $11.31\pm 0.01$ & $10.60\pm 0.03$ & $9.81\pm 0.01$ & $9.29\pm 0.02$ \\
53468 & \multicolumn{1}{c}{\nodata} & $13.81\pm 0.02$ & $12.90\pm 0.02$ & $12.12\pm 0.01$ & $11.38\pm 0.01$ & $10.80\pm 0.03$ & $9.81\pm 0.01$ & $9.30\pm 0.02$ \\
53469 & \multicolumn{1}{c}{\nodata} & $13.95\pm 0.02$ & \multicolumn{1}{c}{\nodata} & \multicolumn{1}{c}{\nodata} & \multicolumn{1}{c}{\nodata} & \multicolumn{1}{c}{\nodata} & $9.87\pm 0.01$ & \multicolumn{1}{c}{\nodata} \\
53470 & \multicolumn{1}{c}{\nodata} & \multicolumn{1}{c}{\nodata} & $13.00\pm 0.02$ & $12.19\pm 0.01$ & $11.46\pm 0.01$ & $10.90\pm 0.03$ & \multicolumn{1}{c}{\nodata} & $9.20\pm 0.02$ \\
53476 & \multicolumn{1}{c}{\nodata} & $14.17\pm 0.03$ & $13.23\pm 0.02$ & $12.48\pm 0.01$ & $11.69\pm 0.01$ & $10.87\pm 0.03$ & $10.08\pm 0.02$ & $9.50\pm 0.03$ \\
53477 & \multicolumn{1}{c}{\nodata} & $14.22\pm 0.03$ & $13.25\pm 0.02$ & $12.48\pm 0.01$ & $11.67\pm 0.01$ & $10.81\pm 0.03$ & $9.97\pm 0.02$ & $9.51\pm 0.02$ \\
53488 & \multicolumn{1}{c}{\nodata} & $13.86\pm 0.03$ & $12.94\pm 0.02$ & $12.18\pm 0.01$ & $11.42\pm 0.01$ & $10.45\pm 0.03$ & $9.84\pm 0.01$ & $9.42\pm 0.02$ \\
53498 & \multicolumn{1}{c}{\nodata} & $13.94\pm 0.14$ & $13.22\pm 0.04$ & $12.25\pm 0.02$ & $11.46\pm 0.01$ & $10.77\pm 0.03$ & $10.25\pm 0.01$ & $9.38\pm 0.02$ \\
53502 & \multicolumn{1}{c}{\nodata} & \multicolumn{1}{c}{\nodata} & $13.08\pm 0.06$ & \multicolumn{1}{c}{\nodata} & \multicolumn{1}{c}{\nodata} & \multicolumn{1}{c}{\nodata} & \multicolumn{1}{c}{\nodata} & $9.11\pm 0.02$ \\
53584 & \multicolumn{1}{c}{\nodata} & $14.28\pm 0.17$ & $13.22\pm 0.01$ & $12.39\pm 0.01$ & $11.61\pm 0.01$ & $10.82\pm 0.08$ & $9.98\pm 0.02$ & \multicolumn{1}{c}{\nodata} \\
53608 & \multicolumn{1}{c}{\nodata} & $14.46\pm 0.01$ & $13.49\pm 0.01$ & $12.67\pm 0.01$ & $11.86\pm 0.01$ & $10.93\pm 0.05$ & $10.13\pm 0.02$ & $9.57\pm 0.03$ \\
53618 & \multicolumn{1}{c}{\nodata} & $14.43\pm 0.01$ & $13.44\pm 0.01$ & $12.62\pm 0.01$ & $11.78\pm 0.01$ & $11.01\pm 0.06$ & $10.21\pm 0.02$ & $9.57\pm 0.03$ \\
53625 & \multicolumn{1}{c}{\nodata} & $14.89\pm 0.01$ & $13.94\pm 0.01$ & $13.09\pm 0.01$ & $12.21\pm 0.01$ & $11.44\pm 0.04$ & $10.44\pm 0.02$ & $9.83\pm 0.03$ \\
53632 & \multicolumn{1}{c}{\nodata} & $14.50\pm 0.02$ & $13.51\pm 0.01$ & $12.72\pm 0.01$ & $11.86\pm 0.01$ & $11.05\pm 0.05$ & $10.20\pm 0.02$ & $9.51\pm 0.03$ \\
53637 & \multicolumn{1}{c}{\nodata} & $14.62\pm 0.02$ & $13.54\pm 0.04$ & \multicolumn{1}{c}{\nodata} & \multicolumn{1}{c}{\nodata} & \multicolumn{1}{c}{\nodata} & $10.24\pm 0.03$ & $9.30\pm 0.07$ \\
53640 & \multicolumn{1}{c}{\nodata} & $14.06\pm 0.01$ & $13.09\pm 0.01$ & $12.31\pm 0.01$ & $11.52\pm 0.01$ & $11.01\pm 0.04$ & $10.02\pm 0.02$ & $9.46\pm 0.03$ \\
53650 & \multicolumn{1}{c}{\nodata} & $14.03\pm 0.01$ & $13.12\pm 0.01$ & $12.31\pm 0.01$ & $11.56\pm 0.01$ & $10.74\pm 0.04$ & $10.04\pm 0.02$ & $9.43\pm 0.03$ \\
53653 & \multicolumn{1}{c}{\nodata} & $14.34\pm 0.01$ & $13.43\pm 0.01$ & $12.62\pm 0.01$ & $11.77\pm 0.01$ & $11.39\pm 0.04$ & $10.48\pm 0.02$ & $10.02\pm 0.03$ \\
53657 & \multicolumn{1}{c}{\nodata} & $14.15\pm 0.01$ & $13.25\pm 0.01$ & $12.45\pm 0.01$ & $11.65\pm 0.01$ & $10.93\pm 0.04$ & $10.12\pm 0.01$ & $9.46\pm 0.02$ \\
53663 & \multicolumn{1}{c}{\nodata} & $14.67\pm 0.01$ & $13.73\pm 0.01$ & $12.88\pm 0.01$ & $12.02\pm 0.01$ & $11.23\pm 0.04$ & $10.36\pm 0.02$ & $9.66\pm 0.03$ \\
53670 & \multicolumn{1}{c}{\nodata} & $14.76\pm 0.01$ & $13.82\pm 0.01$ & $13.01\pm 0.01$ & $12.16\pm 0.01$ & $11.73\pm 0.05$ & $10.86\pm 0.02$ & $10.14\pm 0.03$ \\
53674 & \multicolumn{1}{c}{\nodata} & $14.35\pm 0.01$ & $13.38\pm 0.01$ & $12.60\pm 0.01$ & $11.83\pm 0.01$ & $11.05\pm 0.04$ & $10.18\pm 0.02$ & $9.47\pm 0.03$ \\
53678 & \multicolumn{1}{c}{\nodata} & $14.65\pm 0.01$ & $13.70\pm 0.01$ & $12.86\pm 0.01$ & $12.02\pm 0.01$ & $11.30\pm 0.04$ & $10.30\pm 0.02$ & $9.74\pm 0.03$ \\
53687 & \multicolumn{1}{c}{\nodata} & $14.45\pm 0.01$ & $13.52\pm 0.01$ & $12.69\pm 0.01$ & $11.87\pm 0.01$ & $11.05\pm 0.04$ & $10.26\pm 0.02$ & $9.57\pm 0.03$ \\
53693 & \multicolumn{1}{c}{\nodata} & $14.54\pm 0.01$ & $13.63\pm 0.01$ & $12.81\pm 0.01$ & $12.00\pm 0.01$ & $11.21\pm 0.04$ & $10.33\pm 0.02$ & $9.63\pm 0.03$ \\
53700 & \multicolumn{1}{c}{\nodata} & $14.63\pm 0.01$ & $13.70\pm 0.01$ & $12.87\pm 0.01$ & $12.05\pm 0.01$ & $11.27\pm 0.04$ & $10.41\pm 0.02$ & $9.69\pm 0.03$ \\
53705 & \multicolumn{1}{c}{\nodata} & $14.96\pm 0.01$ & $13.98\pm 0.01$ & $13.13\pm 0.01$ & $12.27\pm 0.01$ & $11.44\pm 0.05$ & $10.53\pm 0.02$ & $9.81\pm 0.03$ \\
53710 & \multicolumn{1}{c}{\nodata} & $15.35\pm 0.01$ & $14.51\pm 0.01$ & \multicolumn{1}{c}{\nodata} & $12.75\pm 0.01$ & $11.75\pm 0.07$ & $10.95\pm 0.02$ & $10.24\pm 0.03$ \\
53712 & \multicolumn{1}{c}{\nodata} & $15.64\pm 0.01$ & $14.74\pm 0.01$ & $13.81\pm 0.01$ & $12.89\pm 0.01$ & \multicolumn{1}{c}{\nodata} & $10.91\pm 0.02$ & \multicolumn{1}{c}{\nodata} \\
53714 & \multicolumn{1}{c}{\nodata} & $15.50\pm 0.01$ & $14.63\pm 0.01$ & $13.73\pm 0.01$ & $12.84\pm 0.01$ & $11.84\pm 0.06$ & $10.89\pm 0.02$ & $10.08\pm 0.04$ \\
53715 & \multicolumn{1}{c}{\nodata} & $15.43\pm 0.01$ & $14.53\pm 0.01$ & $13.66\pm 0.01$ & $12.79\pm 0.01$ & $11.80\pm 0.05$ & $10.85\pm 0.02$ & $9.97\pm 0.03$ \\
53716 & \multicolumn{1}{c}{\nodata} & $15.69\pm 0.01$ & $14.77\pm 0.01$ & $13.85\pm 0.01$ & $12.93\pm 0.01$ & $11.95\pm 0.06$ & $10.93\pm 0.02$ & $10.12\pm 0.04$ \\
53717 & \multicolumn{1}{c}{\nodata} & $15.54\pm 0.01$ & $14.61\pm 0.01$ & $13.68\pm 0.01$ & $12.76\pm 0.01$ & $11.84\pm 0.06$ & $10.87\pm 0.02$ & $10.08\pm 0.04$ \\
53718 & \multicolumn{1}{c}{\nodata} & $15.36\pm 0.01$ & $14.44\pm 0.01$ & $13.54\pm 0.01$ & $12.65\pm 0.01$ & $11.72\pm 0.05$ & $10.80\pm 0.02$ & $9.95\pm 0.03$ \\
53719 & \multicolumn{1}{c}{\nodata} & $15.08\pm 0.02$ & $14.14\pm 0.01$ & $13.26\pm 0.01$ & $12.38\pm 0.01$ & $11.48\pm 0.05$ & $10.60\pm 0.02$ & $9.83\pm 0.03$ \\
53720 & \multicolumn{1}{c}{\nodata} & $15.54\pm 0.02$ & $14.63\pm 0.01$ & $13.69\pm 0.01$ & $12.75\pm 0.01$ & $11.87\pm 0.05$ & $10.94\pm 0.02$ & $10.22\pm 0.03$ \\
53721 & \multicolumn{1}{c}{\nodata} & $15.36\pm 0.02$ & $14.41\pm 0.01$ & $13.49\pm 0.01$ & $12.61\pm 0.01$ & $11.73\pm 0.05$ & $10.78\pm 0.02$ & $10.06\pm 0.03$ \\
53722 & \multicolumn{1}{c}{\nodata} & $15.25\pm 0.01$ & $14.30\pm 0.01$ & $13.42\pm 0.01$ & $12.54\pm 0.01$ & $11.69\pm 0.06$ & $10.73\pm 0.02$ & $10.07\pm 0.04$ \\
53723 & \multicolumn{1}{c}{\nodata} & $14.97\pm 0.01$ & $14.03\pm 0.01$ & $13.20\pm 0.01$ & $12.37\pm 0.01$ & $11.50\pm 0.05$ & $10.66\pm 0.02$ & $9.93\pm 0.03$ \\
53724 & \multicolumn{1}{c}{\nodata} & $14.96\pm 0.01$ & $14.03\pm 0.01$ & $13.16\pm 0.01$ & $12.35\pm 0.01$ & $11.51\pm 0.05$ & $10.64\pm 0.02$ & $9.89\pm 0.03$ \\
53725 & \multicolumn{1}{c}{\nodata} & $14.84\pm 0.01$ & $13.91\pm 0.01$ & $13.06\pm 0.01$ & $12.26\pm 0.01$ & $11.36\pm 0.05$ & $10.57\pm 0.02$ & $9.83\pm 0.03$ \\
53727 & \multicolumn{1}{c}{\nodata} & $15.00\pm 0.01$ & $14.07\pm 0.01$ & $13.23\pm 0.01$ & $12.41\pm 0.01$ & $11.49\pm 0.05$ & $10.66\pm 0.02$ & $9.90\pm 0.03$ \\
53728 & \multicolumn{1}{c}{\nodata} & $15.00\pm 0.01$ & $14.03\pm 0.01$ & $13.17\pm 0.01$ & $12.32\pm 0.01$ & $11.54\pm 0.05$ & $10.62\pm 0.02$ & $9.93\pm 0.03$ \\
53748 & \multicolumn{1}{c}{\nodata} & $16.01\pm 0.02$ & $15.16\pm 0.02$ & $14.26\pm 0.01$ & $13.35\pm 0.01$ & $12.15\pm 0.07$ & $11.08\pm 0.03$ & $10.36\pm 0.04$ \\
53749 & \multicolumn{1}{c}{\nodata} & $16.08\pm 0.02$ & $15.21\pm 0.02$ & $14.32\pm 0.01$ & $13.42\pm 0.01$ & $12.19\pm 0.07$ & $11.11\pm 0.03$ & $10.34\pm 0.04$ \\
53750 & \multicolumn{1}{c}{\nodata} & $16.20\pm 0.02$ & $15.36\pm 0.02$ & $14.42\pm 0.01$ & $13.51\pm 0.01$ & $11.85\pm 0.08$ & $11.20\pm 0.03$ & $10.10\pm 0.04$ \\
53751 & \multicolumn{1}{c}{\nodata} & $16.16\pm 0.02$ & $15.26\pm 0.01$ & $14.31\pm 0.01$ & $13.37\pm 0.01$ & $12.19\pm 0.06$ & $11.04\pm 0.03$ & $10.22\pm 0.04$ \\
53752 & \multicolumn{1}{c}{\nodata} & $16.10\pm 0.01$ & $15.20\pm 0.01$ & $14.26\pm 0.01$ & $13.32\pm 0.01$ & $12.03\pm 0.06$ & $11.02\pm 0.02$ & $10.20\pm 0.04$ \\
53755 & \multicolumn{1}{c}{\nodata} & $15.51\pm 0.01$ & $14.59\pm 0.01$ & $13.68\pm 0.01$ & $12.77\pm 0.01$ & $11.83\pm 0.06$ & $10.68\pm 0.02$ & $9.79\pm 0.03$ \\
53757 & \multicolumn{1}{c}{\nodata} & $15.68\pm 0.01$ & $14.73\pm 0.01$ & $13.79\pm 0.01$ & $12.85\pm 0.01$ & $11.94\pm 0.06$ & $10.81\pm 0.03$ & $9.94\pm 0.04$ \\
53759 & \multicolumn{1}{c}{\nodata} & $15.17\pm 0.01$ & $14.28\pm 0.01$ & $13.40\pm 0.01$ & $12.53\pm 0.01$ & $11.70\pm 0.05$ & $10.70\pm 0.02$ & $9.81\pm 0.03$ \\
53762 & \multicolumn{1}{c}{\nodata} & $14.98\pm 0.01$ & $14.07\pm 0.01$ & $13.21\pm 0.01$ & $12.39\pm 0.01$ & $11.54\pm 0.05$ & $10.64\pm 0.02$ & $9.89\pm 0.03$ \\
53765 & \multicolumn{1}{c}{\nodata} & $15.48\pm 0.01$ & $14.64\pm 0.01$ & $13.73\pm 0.01$ & $12.86\pm 0.01$ & $11.77\pm 0.06$ & $10.91\pm 0.02$ & $10.18\pm 0.04$ \\
53767 & \multicolumn{1}{c}{\nodata} & $15.74\pm 0.01$ & $14.89\pm 0.01$ & $13.94\pm 0.01$ & $13.08\pm 0.01$ & $11.88\pm 0.07$ & $11.03\pm 0.03$ & $10.07\pm 0.05$ \\
53770 & \multicolumn{1}{c}{\nodata} & $15.81\pm 0.01$ & $14.87\pm 0.01$ & $13.95\pm 0.01$ & $13.06\pm 0.01$ & $12.09\pm 0.06$ & $11.06\pm 0.03$ & $10.17\pm 0.04$ \\
53773 & \multicolumn{1}{c}{\nodata} & $16.00\pm 0.01$ & $15.11\pm 0.01$ & $14.17\pm 0.01$ & $13.24\pm 0.01$ & $12.12\pm 0.07$ & $11.03\pm 0.03$ & $10.21\pm 0.04$ \\
53776 & \multicolumn{1}{c}{\nodata} & $15.73\pm 0.02$ & $14.82\pm 0.01$ & $13.90\pm 0.01$ & $12.99\pm 0.01$ & $11.90\pm 0.06$ & $10.91\pm 0.02$ & $10.20\pm 0.04$ \\
53779 & \multicolumn{1}{c}{\nodata} & $15.88\pm 0.02$ & $15.00\pm 0.01$ & $14.01\pm 0.01$ & $13.11\pm 0.01$ & $11.97\pm 0.07$ & $11.00\pm 0.03$ & $10.33\pm 0.04$ \\
53782 & \multicolumn{1}{c}{\nodata} & $15.71\pm 0.01$ & $14.81\pm 0.01$ & $13.84\pm 0.01$ & $12.98\pm 0.01$ & $11.79\pm 0.07$ & $11.01\pm 0.03$ & $10.30\pm 0.04$ \\
53785 & \multicolumn{1}{c}{\nodata} & $15.72\pm 0.01$ & $14.85\pm 0.01$ & $13.95\pm 0.01$ & $13.05\pm 0.01$ & $11.98\pm 0.07$ & $10.96\pm 0.03$ & $10.31\pm 0.04$ \\
53789 & \multicolumn{1}{c}{\nodata} & $15.83\pm 0.01$ & $14.95\pm 0.01$ & $14.00\pm 0.01$ & $13.09\pm 0.01$ & $12.04\pm 0.06$ & $10.99\pm 0.03$ & $10.24\pm 0.04$ \\
53790 & \multicolumn{1}{c}{\nodata} & $15.98\pm 0.01$ & $15.15\pm 0.01$ & $14.19\pm 0.01$ & $13.25\pm 0.01$ & $12.09\pm 0.07$ & $11.01\pm 0.03$ & $10.36\pm 0.04$ \\
53792 & \multicolumn{1}{c}{\nodata} & $16.16\pm 0.01$ & $15.27\pm 0.01$ & $14.31\pm 0.01$ & $13.38\pm 0.01$ & $12.08\pm 0.07$ & $11.07\pm 0.03$ & $10.41\pm 0.04$ \\
53796 & \multicolumn{1}{c}{\nodata} & \multicolumn{1}{c}{\nodata} & \multicolumn{1}{c}{\nodata} & \multicolumn{1}{c}{\nodata} & \multicolumn{1}{c}{\nodata} & $12.08\pm 0.07$ & $11.08\pm 0.03$ & $10.44\pm 0.04$ \\
53797 & \multicolumn{1}{c}{\nodata} & $16.23\pm 0.01$ & $15.38\pm 0.01$ & $14.40\pm 0.01$ & $13.45\pm 0.01$ & $12.35\pm 0.07$ & $11.14\pm 0.03$ & $10.64\pm 0.04$ \\
53798 & \multicolumn{1}{c}{\nodata} & $16.48\pm 0.01$ & $15.60\pm 0.01$ & $14.59\pm 0.01$ & $13.61\pm 0.01$ & $12.37\pm 0.07$ & $11.30\pm 0.03$ & $10.63\pm 0.04$ \\
53801 & \multicolumn{1}{c}{\nodata} & $16.29\pm 0.01$ & $15.47\pm 0.01$ & $14.49\pm 0.01$ & $13.54\pm 0.01$ & $12.31\pm 0.07$ & $11.23\pm 0.03$ & $10.58\pm 0.04$ \\
53804 & \multicolumn{1}{c}{\nodata} & $16.45\pm 0.02$ & $15.57\pm 0.02$ & $14.62\pm 0.01$ & $13.66\pm 0.01$ & $12.26\pm 0.07$ & $11.24\pm 0.03$ & $10.46\pm 0.05$ \\
53808 & \multicolumn{1}{c}{\nodata} & $16.12\pm 0.02$ & $15.33\pm 0.02$ & $14.39\pm 0.01$ & $13.47\pm 0.01$ & $12.13\pm 0.07$ & $11.14\pm 0.03$ & $10.57\pm 0.05$ \\
53811 & \multicolumn{1}{c}{\nodata} & $16.46\pm 0.02$ & $15.64\pm 0.02$ & $14.59\pm 0.01$ & $13.57\pm 0.01$ & $12.08\pm 0.08$ & $11.23\pm 0.03$ & $10.65\pm 0.05$ \\
53814 & \multicolumn{1}{c}{\nodata} & $16.60\pm 0.01$ & $15.78\pm 0.02$ & $14.74\pm 0.01$ & $13.75\pm 0.01$ & $12.42\pm 0.07$ & $11.35\pm 0.03$ & $10.78\pm 0.05$ \\
53817 & \multicolumn{1}{c}{\nodata} & $16.49\pm 0.01$ & $15.62\pm 0.02$ & $14.63\pm 0.01$ & $13.67\pm 0.01$ & $12.27\pm 0.08$ & $11.25\pm 0.03$ & $10.66\pm 0.05$ \\
53819 & \multicolumn{1}{c}{\nodata} & $17.10\pm 0.02$ & $16.30\pm 0.02$ & \multicolumn{1}{c}{\nodata} & \multicolumn{1}{c}{\nodata} & \multicolumn{1}{c}{\nodata} & $11.37\pm 0.03$ & $10.80\pm 0.05$ \\
53820 & \multicolumn{1}{c}{\nodata} & \multicolumn{1}{c}{\nodata} & \multicolumn{1}{c}{\nodata} & $15.16\pm 0.01$ & $14.07\pm 0.02$ & $12.49\pm 0.07$ & \multicolumn{1}{c}{\nodata} & \multicolumn{1}{c}{\nodata} \\
53823 & \multicolumn{1}{c}{\nodata} & $16.67\pm 0.02$ & $15.86\pm 0.02$ & $14.82\pm 0.01$ & $13.82\pm 0.01$ & $12.34\pm 0.08$ & $11.24\pm 0.03$ & $10.83\pm 0.05$ \\
53826 & \multicolumn{1}{c}{\nodata} & $16.82\pm 0.03$ & $16.07\pm 0.02$ & $14.99\pm 0.02$ & $13.94\pm 0.02$ & $12.29\pm 0.08$ & $11.23\pm 0.03$ & $10.69\pm 0.05$ \\
53829 & \multicolumn{1}{c}{\nodata} & $16.97\pm 0.02$ & $16.13\pm 0.02$ & $15.04\pm 0.01$ & $14.00\pm 0.02$ & $12.51\pm 0.08$ & $11.36\pm 0.03$ & $10.69\pm 0.05$ \\
53832 & \multicolumn{1}{c}{\nodata} & $16.88\pm 0.03$ & $16.06\pm 0.02$ & $14.95\pm 0.01$ & $13.95\pm 0.02$ & $12.33\pm 0.08$ & $11.21\pm 0.03$ & $10.59\pm 0.05$ \\
53835 & \multicolumn{1}{c}{\nodata} & $16.97\pm 0.02$ & $16.09\pm 0.02$ & $15.01\pm 0.01$ & $13.98\pm 0.02$ & $12.29\pm 0.07$ & $11.28\pm 0.03$ & $10.68\pm 0.05$ \\
53839 & \multicolumn{1}{c}{\nodata} & $17.38\pm 0.03$ & $16.53\pm 0.03$ & $15.41\pm 0.02$ & $14.27\pm 0.02$ & $12.50\pm 0.09$ & $11.37\pm 0.03$ & $10.81\pm 0.05$ \\
53843 & \multicolumn{1}{c}{\nodata} & $17.41\pm 0.02$ & $16.55\pm 0.03$ & $15.42\pm 0.02$ & $14.27\pm 0.02$ & $12.41\pm 0.08$ & $11.30\pm 0.03$ & $10.78\pm 0.05$ \\
53847 & \multicolumn{1}{c}{\nodata} & $17.29\pm 0.05$ & $16.68\pm 0.05$ & $15.52\pm 0.03$ & $14.35\pm 0.02$ & $12.36\pm 0.08$ & $11.27\pm 0.03$ & $10.62\pm 0.06$ \\
53848 & \multicolumn{1}{c}{\nodata} & $17.55\pm 0.03$ & $16.69\pm 0.03$ & $15.53\pm 0.02$ & $14.38\pm 0.02$ & $12.47\pm 0.08$ & $11.32\pm 0.03$ & $10.77\pm 0.05$ \\
53851 & \multicolumn{1}{c}{\nodata} & $17.69\pm 0.15$ & $16.71\pm 0.09$ & $15.75\pm 0.05$ & $14.48\pm 0.04$ & $12.48\pm 0.08$ & $11.42\pm 0.03$ & $10.85\pm 0.05$ \\
53854 & \multicolumn{1}{c}{\nodata} & $17.80\pm 0.05$ & $16.98\pm 0.07$ & $15.76\pm 0.04$ & $14.53\pm 0.04$ & $12.42\pm 0.13$ & $11.41\pm 0.04$ & $10.52\pm 0.07$ \\
53860 & \multicolumn{1}{c}{\nodata} & $18.50\pm 0.29$ & $17.23\pm 0.13$ & $16.33\pm 0.08$ & $14.79\pm 0.05$ & $12.60\pm 0.10$ & $11.60\pm 0.04$ & $11.06\pm 0.06$ \\
53861 & \multicolumn{1}{c}{\nodata} & $18.29\pm 0.29$ & $17.85\pm 0.28$ & $16.22\pm 0.09$ & $14.84\pm 0.05$ & $12.70\pm 0.08$ & $11.60\pm 0.03$ & $10.82\pm 0.03$ \\
53956 & \multicolumn{1}{c}{\nodata} & \multicolumn{1}{c}{\nodata} & \multicolumn{1}{c}{\nodata} & \multicolumn{1}{c}{\nodata} & $14.89\pm 0.08$ & $12.73\pm 0.18$ & $11.83\pm 0.12$ & $10.91\pm 0.17$ \\
54111 & \multicolumn{1}{c}{\nodata} & $18.56\pm 0.04$ & $17.58\pm 0.03$ & $16.26\pm 0.03$ & $14.81\pm 0.02$ & $12.78\pm 0.10$ & $11.60\pm 0.04$ & $11.16\pm 0.06$ \\
54456 & \multicolumn{1}{c}{\nodata} & $16.50\pm 0.01$ & $15.61\pm 0.01$ & $14.67\pm 0.01$ & $13.74\pm 0.01$ & $12.42\pm 0.08$ & $11.41\pm 0.03$ & $10.96\pm 0.05$ \\
54465 & \multicolumn{1}{c}{\nodata} & $16.85\pm 0.01$ & $15.99\pm 0.01$ & $14.93\pm 0.01$ & $13.95\pm 0.01$ & $12.48\pm 0.09$ & $11.45\pm 0.03$ & $10.98\pm 0.06$ \\
54466 & \multicolumn{1}{c}{\nodata} & $16.75\pm 0.01$ & $15.91\pm 0.01$ & $14.84\pm 0.01$ & $13.85\pm 0.01$ & $12.45\pm 0.08$ & $11.40\pm 0.03$ & $10.93\pm 0.05$ \\
54468 & \multicolumn{1}{c}{\nodata} & $16.79\pm 0.01$ & $15.92\pm 0.01$ & $14.86\pm 0.01$ & $13.90\pm 0.01$ & $12.51\pm 0.09$ & $11.42\pm 0.03$ & $10.96\pm 0.05$ \\
54469 & \multicolumn{1}{c}{\nodata} & $16.57\pm 0.01$ & $15.71\pm 0.01$ & $14.69\pm 0.01$ & $13.74\pm 0.01$ & $12.35\pm 0.07$ & $11.33\pm 0.02$ & $10.77\pm 0.04$ \\
54471 & \multicolumn{1}{c}{\nodata} & $16.66\pm 0.01$ & $15.79\pm 0.01$ & $14.84\pm 0.01$ & $13.86\pm 0.01$ & $12.36\pm 0.07$ & $11.31\pm 0.03$ & $10.81\pm 0.04$ \\
54472 & \multicolumn{1}{c}{\nodata} & $16.54\pm 0.01$ & $15.67\pm 0.01$ & $14.71\pm 0.01$ & $13.76\pm 0.01$ & $12.33\pm 0.07$ & $11.30\pm 0.02$ & $10.78\pm 0.04$ \\
54473 & \multicolumn{1}{c}{\nodata} & $16.78\pm 0.01$ & $15.92\pm 0.01$ & $14.91\pm 0.01$ & $13.95\pm 0.01$ & $12.31\pm 0.08$ & $11.38\pm 0.03$ & $10.81\pm 0.05$ \\
54480 & \multicolumn{1}{c}{\nodata} & $16.80\pm 0.01$ & $16.12\pm 0.01$ & $15.11\pm 0.01$ & $14.08\pm 0.01$ & $11.75\pm 0.06$ & $11.33\pm 0.03$ & $10.78\pm 0.04$ \\
54484 & \multicolumn{1}{c}{\nodata} & $16.94\pm 0.01$ & $16.11\pm 0.01$ & $15.16\pm 0.01$ & $14.15\pm 0.01$ & $12.47\pm 0.07$ & $11.43\pm 0.02$ & $10.83\pm 0.04$ \\
54487 & \multicolumn{1}{c}{\nodata} & $16.94\pm 0.02$ & $16.10\pm 0.01$ & $15.12\pm 0.01$ & $14.08\pm 0.01$ & $12.45\pm 0.07$ & $11.37\pm 0.03$ & $10.87\pm 0.04$ \\
54490 & \multicolumn{1}{c}{\nodata} & $17.08\pm 0.02$ & $16.19\pm 0.02$ & $15.18\pm 0.02$ & $14.15\pm 0.01$ & $12.51\pm 0.09$ & $11.43\pm 0.03$ & $10.80\pm 0.05$ \\
54496 & \multicolumn{1}{c}{\nodata} & $16.76\pm 0.01$ & $15.91\pm 0.01$ & $14.92\pm 0.01$ & $13.90\pm 0.01$ & $12.38\pm 0.07$ & $11.39\pm 0.03$ & $10.77\pm 0.04$ \\
54499 & \multicolumn{1}{c}{\nodata} & $16.80\pm 0.01$ & $15.98\pm 0.01$ & $14.97\pm 0.01$ & $13.98\pm 0.01$ & $12.49\pm 0.08$ & $11.41\pm 0.03$ & $10.80\pm 0.04$ \\
54514 & \multicolumn{1}{c}{\nodata} & $17.71\pm 0.04$ & $16.86\pm 0.03$ & $15.71\pm 0.03$ & $14.52\pm 0.01$ & $12.66\pm 0.09$ & $11.51\pm 0.04$ & $11.08\pm 0.05$ \\
54522 & \multicolumn{1}{c}{\nodata} & $17.82\pm 0.03$ & $16.95\pm 0.02$ & $15.75\pm 0.02$ & $14.55\pm 0.01$ & $12.62\pm 0.08$ & $11.52\pm 0.03$ & $11.06\pm 0.05$ \\
54549 & \multicolumn{1}{c}{\nodata} & $18.42\pm 0.04$ & $17.55\pm 0.03$ & $16.18\pm 0.03$ & $14.81\pm 0.02$ & $12.69\pm 0.09$ & $11.62\pm 0.03$ & $11.21\pm 0.05$ \\
54553 & \multicolumn{1}{c}{\nodata} & $18.35\pm 0.03$ & $17.48\pm 0.02$ & $16.18\pm 0.02$ & $14.79\pm 0.01$ & $12.65\pm 0.08$ & $11.66\pm 0.03$ & $11.19\pm 0.05$ \\
\end{longtable}

}

\clearpage

\begin{table}
\caption{Villanova magnitudes.\label{tab:villmag}}
\centering
\begin{tabular}{cccc}
\hline\hline
 \noalign{\vskip .8ex}%
MJD 		& $V$ & $R$ & $I$\\
 \noalign{\vskip .8ex}%
\hline\\
 \noalign{\vskip -2ex}%
53398 & $13.66\pm 0.05$ & $12.81\pm 0.06$ & $12.00\pm 0.02$ \\
53399 & $13.55\pm 0.02$ & $12.71\pm 0.03$ & $11.91\pm 0.01$ \\
53401 & $13.62\pm 0.07$ & $12.75\pm 0.05$ & $11.95\pm 0.04$ \\
53402 & $13.70\pm 0.06$ & $12.81\pm 0.05$ & $12.02\pm 0.01$ \\
53403 & $13.81\pm 0.06$ & $12.92\pm 0.05$ & $12.11\pm 0.02$ \\
53406 & $13.64\pm 0.03$ & $12.78\pm 0.05$ & $11.97\pm 0.02$ \\
53418 & $13.58\pm 0.33$ & $12.49\pm 0.09$ & $11.71\pm 0.04$ \\
53428 & $13.09\pm 0.07$ & $12.28\pm 0.04$ & $11.53\pm 0.03$ \\
53435 & $12.77\pm 0.02$ & $11.95\pm 0.01$ & $11.18\pm 0.05$ \\
53436 & $12.57\pm 0.02$ & $11.78\pm 0.02$ & $11.05\pm 0.01$ \\
53444 & $13.23\pm 0.03$ & $12.37\pm 0.03$ & $11.57\pm 0.03$ \\
53445 & $13.21\pm 0.03$ & $12.36\pm 0.03$ & $11.56\pm 0.03$ \\
53621 & $13.58\pm 0.04$ & $12.68\pm 0.03$ & $11.86\pm 0.02$ \\
53625 & $13.99\pm 0.03$ & $13.07\pm 0.05$ & $12.20\pm 0.02$ \\
53634 & $13.52\pm 0.06$ & $12.64\pm 0.03$ & $11.80\pm 0.04$ \\
53641 & $13.10\pm 0.04$ & $12.26\pm 0.02$ & $11.50\pm 0.02$ \\
53659 & $13.49\pm 0.04$ & $12.62\pm 0.03$ & $11.79\pm 0.03$ \\
53674 & $13.44\pm 0.02$ & $12.60\pm 0.02$ & $11.82\pm 0.02$ \\
53707 & $14.09\pm 0.03$ & $13.17\pm 0.04$ & $12.34\pm 0.03$ \\
53711 & $14.82\pm 0.09$ & $13.89\pm 0.08$ & $12.99\pm 0.03$ \\
53712 & $14.74\pm 0.13$ & $13.81\pm 0.06$ & $12.89\pm 0.04$ \\
53718 & $14.45\pm 0.12$ & $13.57\pm 0.11$ & $12.69\pm 0.05$ \\
53748 & $15.12\pm 0.22$ & $14.20\pm 0.16$ & $13.28\pm 0.11$ \\
53751 & $15.12\pm 0.34$ & $14.36\pm 0.19$ & $13.41\pm 0.09$ \\
53754 & $14.49\pm 0.16$ & $13.55\pm 0.04$ & $12.68\pm 0.04$ \\
53755 & $14.61\pm 0.04$ & $13.69\pm 0.06$ & $12.78\pm 0.03$ \\
53762 & $14.16\pm 0.08$ & $13.30\pm 0.08$ & $12.44\pm 0.06$ \\
53773 & $14.99\pm 0.08$ & $14.11\pm 0.07$ & $13.19\pm 0.05$ \\
53774 & $14.78\pm 0.11$ & $13.92\pm 0.07$ & $13.00\pm 0.02$ \\
53785 & $14.83\pm 0.12$ & $13.94\pm 0.08$ & $13.06\pm 0.06$ \\
53790 & $15.03\pm 0.07$ & $14.12\pm 0.06$ & $13.18\pm 0.03$ \\
53793 & $15.39\pm 0.14$ & $14.48\pm 0.10$ & $13.52\pm 0.03$ \\
53795 & $14.87\pm 0.08$ & $14.09\pm 0.05$ & $13.19\pm 0.03$ \\
53800 & $15.23\pm 0.10$ & $14.37\pm 0.08$ & $13.42\pm 0.05$ \\
53802 & $15.16\pm 0.08$ & $14.24\pm 0.05$ & $13.36\pm 0.03$ \\
\hline
\end{tabular}
\end{table}


\twocolumn
\section{Details on the \spitz\ data reduction}
\label{app:spitz}
\subsection{IRAC}
 We used the MOsaicker and Point source EXtractor (MOPEX) software release of June 2007. We started from the Basic Calibrated Data (BCD) individual images to produce mosaics with a (native) pixel size of 1\farcs 22.
For the data taken in sub-array mode (PIDs 3716 and 41019), we collapsed the 3-dimensional image BCDs into 2-dimensional BCDs by using, for each pixel, the median of  the 64 planes. For the uncertainty BCD files, we used the standard deviation of the 64 planes of the BCD file instead of using the input uncertainties. This method allows the removal of most particle hits and the ingestion of the collapsed BCDs in MOPEX. We then used aperture photometry, centered on V1118 Ori, using an extraction radius of 10 pixels (12\farcs 2), and a concentric annulus of radii 12 and 15 pixels (we used radii of 10 and 12.5 pixels for the IRAC4 band at 8.0~$\mu$m due to the strength of the background intensity at this wavelength for radii larger than 12.5 pixels. Indeed, the background is dominated by emission due to the nearby Herbig Ae star V372 Ori; this effect is less prominent in the other IRAC bands, but the presence of the Herbig star limited us to outer radii of 15 pixels). No aperture correction is needed for an extraction circle radius of 10 pixels (according to the IRAC data handbook v3.0).

\subsection{MIPS}

We used the BCD files where V1118 Ori was on the detector as input files for the MOPEX pipeline. A native pixel size of $2\farcs 5$ was used to create the mosaic, and we used an extraction radius of 13\arcsec\ centered on V1118 Ori to derive the aperture photometry. We used an annulus of radii 15\arcsec\ and 20\arcsec\ for the background, and finally used an aperture correction factor of 1.17 (from Table~3.12 of the MIPS data handbook, v.~3.3.1). We also investigated the background with a different method, i.e., to calculate the background in a nearby region using a circle with radius 13\arcsec. Indeed, the MIPS24 background near V1118 Ori is highly inhomogeneous. 
Notice that, while the uncertainty in the flux density is of the order 0.5 mJy, we estimate the true uncertainty, mainly due to the inhomogeneity of the background and the difficulty to find a representative background region, to be closer to 15-20 mJy.

\subsection{IRS}
{We extracted the IRS  spectra using the \textit{Spitzer} IRS Custom Extraction (SPICE) v2.1.2 software and the post-BCD \emph{co-added}  2-D images {(pipeline version S15.3.0 for SL/SH and S17.2.0 for LH/LL for the outburst data; S18.5.0 for all post-outburst spectra)}.
We investigated in detail the methodology to subtract the background in the IRS spectra.} 
Indeed, for the low-resolution modules, we have investigated different techniques to estimate the background contamination. First, we used the standard technique that takes the image from nod A, subtracts the nod B image, and then the net (A--B) spectrum of V1118 Ori is extracted with SPICE. The same is then done for the net (B--A) spectrum, and both net spectra are averaged. Unfortunately, this technique is unsatisfactory due to the inhomogeneity of the background emission along the cross-dispersion axis of the slit. A second technique consisted in using the images taken for the SL2\footnote{For clarity, we explain the second technique for SL2 only in the main text. For the SL1 data, simply permute SL2 and SL1 in the text.} nods and use the source-free region in the SL1 image for the background of the SL1 nods. This method removes any contamination by V1118 Ori in the SL1 image, but the downside is that the region of the sky covered by the SL1 slit (during the SL2 observations) are away from V1118 Ori. Thus, the measured background may not represent the background near V1118 Ori. A third technique consisted in using a background region near V1118 Ori obtained during the observed nod. This technique allows to get a better estimate for the background near V1118 Ori, but has the disadvantage that V1118 Ori may contaminate the background, especially at longer wavelengths, since the standard extraction width increases with increasing wavelength, to take into account the increasing size of the point spread function of a point source. For the February 2005 observation, technique \#2 proved better for SL1, while technique \#3 was better for SL2. We used the PAH emission at 6.2 and 11.3~$\mu$m (coming mostly from the diffuse
background emission) to check that the background emission was adequately removed. For the March 2005 observation, we used technique \#3 for both SL slits, as technique \#2 gave an excess in the PAH at 11.3~$\mu$m in the SL1 slit. For the post-outburst observations, we used technique \#3 for both SL slits and also for the LL slits (in this case, the background subtraction is more difficult for $\lambda > 30$~mic). 

For the high-resolution modules, we used the standard approach to average the spectra from both nods using the full slit aperture. For the post-outburst spectra, we used the accompanying background spectra to obtain the background-subtracted spectra. This procedure worked nicely for the SH module (i.e., up to $19.35$~\mic), as demonstrated by the good subtraction of the PAH emission at 11.3~\mic. Notice that the
H$_2$  $0-0$ $S(1)$ $\lambda 17.0348$~\mic\ line was apparently completely subtracted, while the H$_2$  $0-0$ $S(2)$ $\lambda 12.2786$~\mic\ line is faintly detected, We also
detect [\ion{Ne}{ii}] $\lambda 12.8136$~\mic\  ($^2P_{1/2}$ -- $^2P_{3/2}$) an [\ion{S}{iii}] $\lambda 18.7130$~\mic\ ($^3P_2$ -- $^3P_1$) in the background-subtracted spectrum (we will come back to this issue below). Furthermore, the silicate feature and continuum flux up to $14$~\mic\  are consistent with the SL flux, suggesting that the background subtraction was not too far off. 
For the LH module, we subtracted only 95\% of the background flux for the post-outburst data. Indeed, at the wavelengths covered by the LH, the background dominates the emission and is 
highly position-dependent. Although our background pointing was close to V1118 Ori, there is considerable inhomogeneity in the diffuse emission of the Orion nebula, as observed in the 8.0 and 24~\mic\ images. {Figure~\ref{fig:mipsirs} shows the MIPS 24~$\mu$m images together with the ``field-of-views" of the IRS slits during the first  outburst observation and in late 2008.}
We used the [\ion{Si}{ii}] $\lambda 34.815$~\mic\  ($^2P_{3/2}$ -- $^2P_{1/2}$) as the proxy for background subtraction for the LH module, and noticed that a scaling factor of 0.95 was better than 1.0.
We detect, however, excess flux in the [\ion{S}{iii}] $\lambda 33.481$~\mic\ ($^3P_1$ -- $^3P_0$) line and faint excess in the H$_2$  $0-0$ $S(0)$ $\lambda 28.2188$~\mic\  line (in 
contrast to the S(1) line in the SH spectrum). The SH and LH spectra relatively well at 19.4~\mic, suggesting again that the background subtraction was adequate.

We provide, however, a word of caution about the accuracy of the continuum flux level (especially above 30~\mic) and in the reality of the detected flux excess in some lines, especially in the LH wavelength region.
Indeed, we have scaled the background LH spectrum using the  [\ion{Si}{ii}] line. However, we remind that V1118 Ori is located at the center of the Orion nebula, where strong ionization and excitation of the
interstellar medium takes place. Strong variations of the diffuse background emission take place, and they may differ for the continuum emission and for the line emission (mostly  [\ion{Si}{ii}] ,  [\ion{S}{iii}], [\ion{Ne}{ii}], and H$_2$). Therefore, using a specific emission line to check that the background subtraction is adequate may result in an incorrect subtraction
of the continuum emission of the nebula and even in the subtraction of another emission line!  Nevertheless, the similarities of the continuum shape in the low-resolution and high-resolution modules 
(at least below 35~\mic) for the post-outburst data, the presence of line excess in [\ion{S}{iii}] in the LL spectrum and in both SH and LH spectra are suggestions that the background subtraction was overall accurate.
The presence of  [\ion{Ne}{ii}] in the SH module but not in the SL module demands, however, deeper analysis to confirm the detection of the line. We degraded the SH and LH spectra down to SL and LL resolution (Fig.~\ref{fig:irsconv}, available online only). The continuum shapes are well matched. But the degraded SH data covering the [\ion{Ne}{ii}] line shows a strong line which is not compatible with the SL data at this wavelength. This strongly suggests that the SH background [\ion{Ne}{ii}] line flux was too faint compared to the line flux near V1118 Ori and that the measured SH line flux is of background origin (from the Orion nebula or from diffuse emission close to V1118 Ori).  In the case of H$_2$ at 12.28~\mic, the degraded SH spectrum is consistent with the SL spectrum: the contrast between the line and the continuum is too faint to confirm the presence of this line in the SL spectrum.  The same comment applies to the other molecular hydrogen line at 28.2~\mic. Thus, we cannot confirm that this molecule is detected in V1118 Ori's spectrum. On the other hand the [\ion{S}{iii}] lines in the degraded SH and LH spectra have similar peak flux densities as in the LL spectrum, indicating that these lines originate from the immediate vicinity of V1118 Ori. In summary, we believe in the detection of the [\ion{S}{iii}] lines, while we have doubts for [\ion{Ne}{ii}] and the H$_2$ lines, but we provide nevertheless the line fluxes from the high-resolution module spectra in Table~\ref{tab:irsline}. In any case, in view of \spitz\ spatial resolution, higher spatial resolution 
observations would be required to confirm the origin of the excess line emission.

We have further investigated whether the post-outburst background observations could be used for the outburst SH and LH data. In principle, the zodiacal light dominates the continuum background emission
below about $40$~\mic, while the interstellar medium emission dominates above. The zodi contribution varies as a function of time. We have used the Spitzer Planning Observations Tool (SPOT) to determine the contribution at the three different
epochs of the IRS observations and found that they were of similar level (about 20-25 MJy~sr$^{-1}$ from 15 to 24~\mic). For the SH spectrum, we preferred to use the post-outburst spectrum without scaling factor,
since the PAH feature and the H$_2$ line {S(2) were well canceled (the S(1) line is slightly oversubtracted)}. For the LH spectrum, we used a scaling factor equivalent to $0.85$, i.e., about the value of the ratio of the zodiacal light contribution at $24$~\mic\  at
in February-March 2005 and November 2008. This ratio also cancels out relatively well the [\ion{Si}{ii}] line {while the H$_2$ S(0) line still remains detected}. With this procedure, the resulting spectrum shows an {\em increase} of the flux with longer wavelengths, perhaps coming from an envelope, but this increase is not detected in the post-outburst data, casting some doubt on the accuracy of the continuum shape in the SH and LH outburst spectra above 14~\mic. Since we have no outburst LL data to confirm this shape, we prefer to err on the safe side and claim that the high-resolution continuum
shape for  $\lambda > 14$~\mic\  during the outburst is unreliable. The strengths of the emission lines may also be affected by improper background reduction; indeed the flux levels are generally larger than in the post-outburst spectrum (except for [\ion{Ne}{ii}], but in this case, the line is probably not originating in the direct vicinity of V1118 Ori). Line fluxes from the high-resolution module data in outburst are also given in Table~\ref{tab:irsline}.

\onlfig{1}{
\begin{figure}[!ht]
\centering
\includegraphics[angle=0,width=\linewidth]{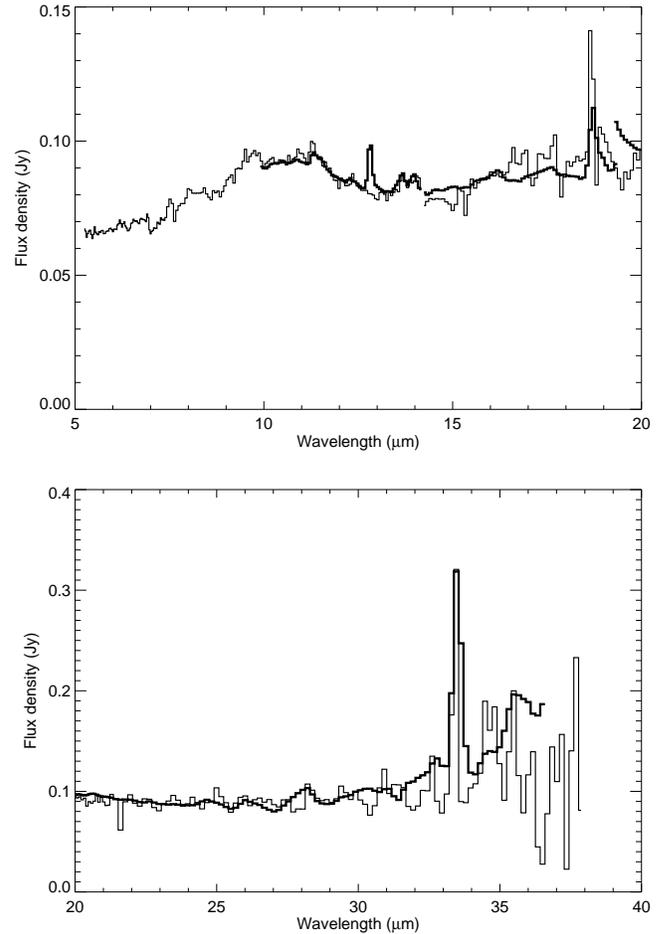}
\caption{\spitz\ IRS high-resolution {post-outburst} spectra (thick curve) degraded to the resolution of the SL and LL modules. The latter data are shown as well (thin curve). Notice the good agreement in the continuum and for most lines, except for [\ion{Ne}{ii}].
\label{fig:irsconv}}
\end{figure}
}

\onlfig{2}{
\begin{figure}[!hb]
\centering
\includegraphics[angle=0,width=\linewidth]{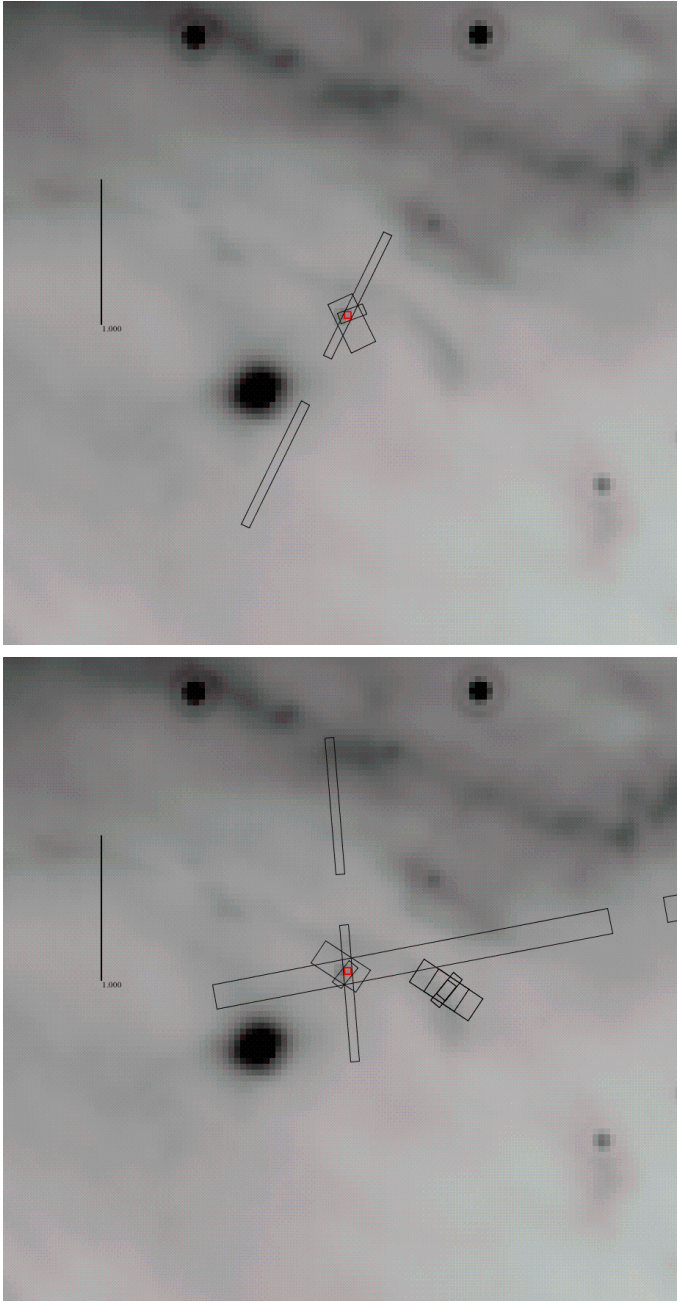}
\caption{\spitz\ MIPS 24~\mic\  image taken in March 2004 with overlays of the IRS modules during the observations on 2005 February 18 (top) and on 2008 November 14 after the outburst (bottom). Note that we only show the first 
position of the two nods, for clarity. Both nods for the background SH/LH observations are shown in the bottom figure. A $1\arcmin$ scale is shown on both North-orientated images. The image scale is linear from 0 to 250~\Mjysr.
\label{fig:mipsirs}}
\end{figure}
}


\begin{thebibliography}{}
\bibitem[{\'A}brah{\'a}m et al.(2009)]{abraham09} {\'A}brah{\'a}m, P., Juh\'asz, A., Dullemond, C.~P., et al.\ 2009, Nature, 459, 224

\bibitem[Ali \& Depoy(1995)]{ali95} Ali, B. \& Depoy, D.L. 1995, AJ, 109, 709

\bibitem[Allen et al.(2004)]{allen04} Allen, L. E., Calvet, N., D'Alessio, P., et al.\ 2004, ApJS, 154, 363

\bibitem[Argiroffi et al.(2007)]{argiroffi07} Argiroffi, C., Maggio, A., \& Peres, G.\ 2007, \aap, 465, L5 

\bibitem[Armitage et al.(2001)]{armitage01} Armitage, P.~J., Livio, M., \& Pringle, J.~E.\ 2001, \mnras, 324, 705 

\bibitem[Arnaud(1996)]{arnaud96} Arnaud, K.~A. 1996, in ASP Conf. Ser. 101,
 Astronomical Data Analysis Software and Systems V, ed. G. Jacoby \& J. Barnes
 (San Francisco: ASP), 1

\bibitem[Aspin(2008)]{aspin08b} Aspin, C.\ 2008b, IAU Circ, 8969

\bibitem[Aspin et al.(2008)]{aspin08a} Aspin, C., Beck, T.~L., \& Reipurth, B.\ 2008a, AJ, 135, 423

\bibitem[Aspin et al.(2009)]{aspin09} Aspin, C., Reipurth, B., Beck, T.~L., et al.\ 2009, \apjl, 692, L67 

\bibitem[Audard et al.(2005a)]{audard05a} Audard, M., Skinner, S.~L., Smith, K.~W., et al.\ 2005a,  in Proc. 13th Cool Stars Workshop, ed. F. Favata et al. (ESA), 411

\bibitem[Audard et al.(2005b)]{audard05b} Audard, M.,  G\"udel, M., Skinner, S.~L., et al.\ 2005b, ApJ, 635, L81

\bibitem[Bell \& Lin(1994)]{bell94} Bell, K.~R., \& Lin, D.~N.~C.\ 1994, \apj, 427, 987 

\bibitem[Bessell \& Brett(1988)]{bessell88} Bessell, M. S., \& Brett, J. M. 1988, PASP, 100, 1134

\bibitem[Bonnell \& Bastien(1992)]{bonnell92} Bonnell, I., \& Bastien, P.\ 1992, \apjl, 401, L31 

\bibitem[Brice{\~ n}o et al.(2004)]{briceno04} Brice{\~ n}o, C., Vivas, A.~K., Hern\'andez, J., et al.\ 2004, ApJ, 606, L123 

\bibitem[Calvet \& Gullbring(1998)]{calvet98} Calvet, N., \& Gullbring, E.\ 1998, \apj, 509, 802 

\bibitem[Carr(1989)]{carr89} Carr, J.~S.\ 1989, \apj, 345, 522 

\bibitem[Chandler et al.(1995)]{chandler95} Chandler, C.~J., Carlstrom, J.~E., \& Scoville, N.~Z.\ 1995, \apj, 446, 793 

\bibitem[Clarke \& Syer(1996)]{clarke96} Clarke, C.~J., \& Syer, D.\ 1996, \mnras, 278, L23 

\bibitem[Drake(2005)]{drake05} Drake, J.~J.\ 2005, 13th 
Cambridge Workshop on Cool Stars, Stellar Systems and the Sun, 560, 519 

\bibitem[Edwards(2009)]{edwards09} Edwards, S.\ 2009, American 
Institute of Physics Conference Series, 1094, 29 

\bibitem[Fazio et al.(2004)]{fazio04} Fazio, G.~G., Hora, J.~L., Allen, L.~E., et al.\ 2004, \apjs, 154, 10 

\bibitem[Fitzpatrick(1999)]{fitzpatrick99} Fitzpatrick, E.~L.\ 1999, \pasp, 111, 63 

\bibitem[Garcia \& Parsamian(2000)]{garcia00} Garcia, J.~G., \& Parsamian, E.~S.\ 2000, Informational Bulletin on Variable Stars, 4925, 1 

\bibitem[Garcia \& Parsamian(2008)]{garcia08} Garcia, J.~G., \& Parsamian, E.~S.\ 2008, Information Bulletin on Variable Stars, 5829, 1 

\bibitem[Garcia et al.(2006)]{garcia06} Garcia, J.~G., Parsamian, E.~S.,  \& Velazquez, J.~C.\ 2006, Information Bulletin on Variable Stars, 5691, 1 

\bibitem[Grevesse \& Sauval(1998)]{grevesse98} Grevesse, N., \& Sauval, A.~J.\ 1998, Space Science Reviews, 85, 161

\bibitem[Grosso et al.(2005)]{grosso05} Grosso, N., Kastner, J.~H., Ozawa, H., et al.\ 2005, \aap, 438, 159 

\bibitem[G{\"u}del  \& Telleschi(2007)]{guedel07} G{\"u}del, M., \& Telleschi, A.\ 2007, \aap, 474, L25 

\bibitem[G{\"u}del et al.(2005)]{guedel05} G{\"u}del, M., Skinner, S.~L., Briggs, K.~R., et al.\ 2005, \apjl, 626, L53 

\bibitem[G{\"u}del et  al.(2008)]{guedel08} G{\"u}del, M., Skinner, S.~L., Audard, M., et al.\ 2008, \aap, 478, 797 

\bibitem[Gullbring et al.(1998)]{gullbring98} Gullbring, E., Hartmann, L., Brice\~{n}o, C., \& Calvet, N.\ 1998, ApJ, 492, 323

\bibitem[G\"unther \& Schmitt(2009)]{guenther09} G{\"u}nther, H.~M., \& Schmitt, J.~H.~M.~M.\ 2009, \aap, 494, 1041 

\bibitem[G{\"u}nther et al.(2006)]{guenther06} G{\"u}nther, H.~M., Liefke, C., Schmitt, J.~H.~M.~M., 
	et al.\ 2006, \aap, 459, L29

\bibitem[G{\"u}nther et al.(2007)]{guenther07} G{\"u}nther, H.~M., Schmitt,
J.~H.~M.~M., Robrade, J., \& Liefke, C.\ 2007, \aap, 466, 1111 

\bibitem[Hartigan et al.(1995)]{hartigan95} Hartigan, P., Edwards, S., \& Ghandour, L.\ 1995, \apj, 452, 736 

\bibitem[Hartmann \& Kenyon(1996)]{hartmann96} Hartmann, L., \& Kenyon, S.~J.\ 1996, ARA\&A, 34, 207 

\bibitem[Hartmann et al.(1993)]{hartmann93} Hartmann, L., Kenyon,  S., \& Hartigan, P.\ 1993, Protostars and Planets III, 497 

\bibitem[Hartmann et al.(1998)]{hartmann98} Hartmann, L., Calvet, N., Gullbring, E., \& D'Alessio, P.\ 1998, \apj, 495, 385 

\bibitem[Hauschildt et al.(1999)]{hauschildt99} Hauschildt, P.~H., Allard, F., \& Baron, E.\ 1999, ApJ 512, 377

\bibitem[Herbig(2008)]{herbig08} Herbig, G.~H.\ 2008, \aj, 135, 637 

\bibitem[Hillenbrand(1997)]{hillenbrand97} Hillenbrand, L.~A.\ 1997, AJ, 113, 1733

\bibitem[Hillenbrand et al.(1998)]{hillenbrand98} Hillenbrand, L.~A., Strom, S.~E., Calvet, N., et al.\ 1998, AJ, 116, 1816

\bibitem[Hong et al.(2004)]{hong04} Hong, J., Schlegel, E.~M., \& Grindlay, J.~E.\ 2004, \apj, 614, 508

\bibitem[Houck et al.(2004)]{houck04} Houck, J.~R., Roellig, T.~L., van Cleve, J., et al.\ 2004, \apjs, 154, 18 

\bibitem[Itagaki(2008)]{itagaki08} Itagaki, K.\ 2008, IAU Circ., 8968, 

\bibitem[Jansen et al.(2001)]{jansen01} Jansen, F., Lumb, D., Altieri, B., et al. 2001, A\&A, 365, L1

\bibitem[Jones(2008a)]{jones08a} Jones, A.~F.~A.~L.\ 2008a, Central Bureau Electronic Telegrams, 1217, 1 

\bibitem[Jones(2008b)]{jones08b} Jones, A.~F.~A.~L.\ 2008b, Central Bureau Electronic Telegrams, 1231, 1 

\bibitem[Kastner et al.(2002)]{kastner02} Kastner, J.~H., Huenemoerder, D.~P., Schulz, N.~S., et al.\  2002, ApJ, 567, 434 

\bibitem[Kastner et al.(2004)]{kastner04} Kastner, J.~H., Richmond, M., Grosso, N., et al.\ 2004, Nature, 430, 429 

\bibitem[Kastner et al.(2005)]{kastner05} Kastner, J.~H., Franz, G., Grosso, N., et al.\ 2005, ApJS, 160, 511

\bibitem[Kastner et al.(2006a)]{kastner06a} Kastner, J.~H., Richmond, M., Grosso, N., et al.\ 2006a, ApJ, 648, L43

\bibitem[Kastner et al.(2006b)]{kastner06b} Kastner, J.~H., Richmond, M., Simon, T., et al.\ 2006b, Central Bureau Electronic Telegrams, 760, 1 

\bibitem[Kenyon \& Hartmann(1987)]{kenyon87} Kenyon, S.~J., \& Hartmann, L.\ 1987, ApJ, 323, 714

\bibitem[Kenyon et al.(1988)]{kenyon88} Kenyon, S.~J., Hartmann, L., \& Hewett, R.\ 1988, \apj, 325, 231 

\bibitem[Kenyon et al.(1993)]{kenyon93} Kenyon, S.~J., Calvet, N., \& Hartmann, L.\ 1993, \apj, 414, 676

\bibitem[Kessler-Silacci et al.(2005)]{kessler05} Kessler-Silacci, J.~E., Hillenbrand, L.~A., Blake, G.~A., 
\& Meyer, M.~R.\ 2005, \apj, 622, 404 

\bibitem[Kessler-Silacci et al.(2006)]{kessler06} Kessler-Silacci, J.~E., Augereau, J.-C., Dullemond, C.~P., et al.\ 2006, \apj, 639, 275 

\bibitem[Kospal et al.(2008)]{kospal08} K\'osp\'al, \'A., Nemeth, P., \'Abrah\'am, P., et al.\ 2008, Information Bulletin on Variable Stars, 5819, 1 

\bibitem[Lin \& Papaloizou(1985)]{lin85} Lin, D.~N.~C., \& Papaloizou, J.\ 1985, Protostars and Planets II, 981

\bibitem[Lodato \& Clarke(2004)]{lodato04} Lodato, G., \& Clarke, C.~J.\ 2004, \mnras, 353, 841 

\bibitem[Lorenzetti et al.(2006)]{lorenzetti06} Lorenzetti, D., Giannini, T., Calzoletti, L., et al.\ 2006, \aap, 453, 579 

\bibitem[Lorenzetti et al.(2007)]{lorenzetti07} Lorenzetti, D., Giannini, T., Larionov, V.~M., et al.\ 2007, \apj, 665, 1182 

\bibitem[Lorenzetti et al.(2009)]{lorenzetti09} Lorenzetti, D., Larionov, V.~M., Giannini, T., et al.\ 2009, \apj, 693, 1056 

\bibitem[Maggio et al.(2007)]{maggio07} Maggio, A., et al.\ 2007, ApJ, 660, 1462

\bibitem[Meyer et al.(1997)]{meyer97} Meyer, M. R., Calvet, N., \& Hillenbrand, L. A. 1997, AJ, 114, 288

\bibitem[McGehee et al.(2004)]{mcgehee04} McGehee, P.~M., Smith, J.~A., Henden, A.~A., et al.\ 2004, Apj, 616, 1058

\bibitem[McNeil(2004)]{mcneil04} McNeil, J.~W.\ 2004, IAU Circ., 616, 8284

\bibitem[Muench et al.(2008)]{muench08} Muench, A., Getman, K., Hillenbrand, L., 
\& Preibisch, T.\ 2008, Handbook of Star Forming Regions, Volume I: The Northern Sky ASP Monograph Publications, Vol.~4.~Edited by Bo Reipurth, p.483, 483 

\bibitem[Ness \& Schmitt(2005)]{ness05} Ness, J.-U., \& Schmitt, J.~H.~M.~M.\ 2005, \aap, 444, L41

\bibitem[Pfalzner et al.(2008)]{pfalzner08} Pfalzner, S., Tackenberg, J., \& Steinhausen, M.\ 2008, \aap, 487, L45 

\bibitem[Preibisch et al.(2005)]{preibisch05} Preibisch, T., Kim, Y.-C., Favata, F., et  al.\ 2005, ApJS, 160, 401 

\bibitem[Ram\'irez et al.(2004)]{ramirez04} Ram{\'{\i}}rez, S.~V., Rebull, L., Stauffer, J., et al.\ 2004, AJ, 128, 787 

\bibitem[Reipurth et al.(2007)]{reipurth07} Reipurth, B., 
Guimar{\~a}es, M.~M., Connelley, M.~S., \& Bally, J.\ 2007, \aj, 134, 2272 

\bibitem[Rieke et al.(1985)]{rieke85} Rieke, G.~H., Lebofsky,  M.~J., \& Low, F.~J.\ 1985, \aj, 90, 900 

\bibitem[Rieke et al.(2004)]{rieke04} Rieke, G.~H., Young, E.~T., Engelbracht, C.~W., et al.\ 2004, \apjs, 154, 25 

\bibitem[Robrade \& Schmitt(2006)]{robrade06} Robrade, J., \& Schmitt, J.~H.~M.~M. 2006, A\&A, 449, 737

\bibitem[Robrade \& Schmitt(2007)]{robrade07} Robrade, J., \& Schmitt, J.~H.~M.~M.\ 2007, \aap, 473, 229 

\bibitem[Robitaille et al.(2007)]{robitaille07} Robitaille, T., Whitney, B.~A., Indebetouw, R., \& Wood, K.\ 2007, ApJS, 169, 328

\bibitem[Sacco et al.(2008)]{sacco08} Sacco, G.~G., Argiroffi, C., Orlando, S., et al.\ 2008, \aap, 491, L17 

\bibitem[Schmitt et al.(2005)]{schmitt05} Schmitt, J.~H.~M.~M., Robrade, J., Ness, J.-U., et al.\ 2005, A\&A, 432, L35

\bibitem[Schneider \& Schmitt(2008)]{schneider08} Schneider, P.~C., \& Schmitt, J.~H.~M.~M.\ 2008, \aap, 488, L13 

\bibitem[Skinner et al.(2006)]{skinner06} Skinner, S.~L., Briggs, K.~R., G\"udel, M.\ 2006, \apj, 643, 995 

\bibitem[Skinner et al.(2009a)]{skinner09a} Skinner, S.~L., Sokal, K.~R., G{\"u}del, M., \& Briggs, K.~R.\ 2009, \apj, 696, 766 

\bibitem[Skinner et al.(2009b)]{skinner09b} Skinner, S.~L., Sokal, K.~R., Megeath, S.~T., et al.\ 2009, arXiv:0906.2428 

\bibitem[Smith et al.(2005)]{smith05} Smith, K.~W., Audard, M., G\"udel, M., et al.\ 2005, 
in Proc. 13th Cool Stars Workshop, ed. F. Favata et al. (ESA), 971

\bibitem[Smith et al.(2001)]{smith01} Smith, R.~K., Brickhouse, N.~S., Liedahl,
D.~A., Raymond, J.~C. 2001, ApJ, 556, L91

\bibitem[Stassun et al.(1999)]{stassun99} Stassun, K.~G., Mathieu, R.~D., Mazeh, T., \& Vrba, F.~J.\ 1999, AJ, 117, 2941

\bibitem[Stelzer \& Schmitt(2004)]{stelzer04} Stelzer, B., \& Schmitt, J.~H.~M.~M.\ 2004, A\&A, 418, 687 

\bibitem[Stelzer et al.(2009)]{stelzer09} Stelzer, B., Hubrig, S., Orlando, S., et al.\ 2009, A\&A, 499, 529 

\bibitem[Str\"uder et al.(2001)]{strueder01} Str\"uder, L., Aschenbach, B., Br\"auniger, H., et al. 2001, A\&A, 365, L18

\bibitem[Telleschi et al.(2007)]{telleschi07} Telleschi, A., G{\"u}del, M., Briggs, K.~R., et al.\ 2007, \aap, 468, 443 

\bibitem[Turner et al.(2001)]{turner01} Turner, M.~J.~L., Abbey, A., Arnaud, M., et al. 2001, A\&A, 365, L26

\bibitem[Venkat \& Anandarao(2008)]{venkat08} Venkat, V., \& Anandarao, B.~G.\ 2008, CBET, 1596

\bibitem[Vorobyov \& Basu(2005)]{vorobyov05} Vorobyov, E.~I., \& Basu, S.\ 2005, \apjl, 633, L137

\bibitem[Vorobyov \& Basu(2006)]{vorobyov06} Vorobyov, E.~I., \& Basu, S.\ 2006, \apj, 650, 956 

\bibitem[Weisskopf et al.(1996)]{weisskopf96} Weisskopf, M.~C.,  O'dell, S.~L., \& van Speybroeck, L.~P.\ 1996, \procspie, 2805, 2 

\bibitem[Werner et al.(2004)]{werner04} Werner, M.~W., Roellig, T.~L., Low, F.~J., et al.\ 2004, \apjs, 154, 1 

\bibitem[Whitney et al.(2003a)]{whitney03a} Whitney, B.~A., Wood,  K., Bjorkman, J.~E., \& Wolff, M.~J.\ 2003a, \apj, 591, 1049 

\bibitem[Whitney et al.(2003b)]{whitney03b} Whitney, B.~A., Wood,  K., Bjorkman, J.~E., \& Cohen, M.\ 2003b, \apj, 598, 1079 

\bibitem[Williams et al.(2005)]{williams05} Williams, P., Bembrick, C., Lee, S. 2005, IAUC, 8460, 3

\bibitem[Wood et al.(2002)]{wood02} Wood, K., Wolff, M.~J., Bjorkman, J.~E., \& Whitney, B. 2002, ApJ, 564, 887


\end{thebibliography}
\end{document}